\documentclass[12pt]{article}
 \usepackage{amsmath,amssymb,graphicx,ulem,mathrsfs,caption,color}
 \usepackage{bm}
 \usepackage{graphicx}
 \usepackage{subfigure}
 \usepackage{epstopdf}
 \usepackage{float}
 \usepackage{colordvi,multicol}
\usepackage{color}
  \usepackage{amsthm}
  \usepackage[margin=1in]{geometry}
 \allowdisplaybreaks
 \newcommand{\bbeta}{\boldsymbol{\beta}}
 \newtheorem{theorem}{Theorem}
\usepackage[symbol]{footmisc}

\begin{document}
  \title{\bf Functional varying index coefficient model for dynamic gene-environment interactions}
\date{\vspace{-12ex}}
\maketitle
\begin{center}
  \author{Jingyi Zhang$^{1,2}$, Xu Liu$^{3}$, Honglang Wang$^{4}$ and Yuehua Cui$^{1}$}\footnote{To whom correspondence should be addressed: cuiy@msu.edu}\\
$^{1}${\it Department of Statistics and Probability, Michigan State University, East Lansing, MI 48824}\\
 $^{2}${\it Amazon Lab126, Sunnyvale, CA 94089}\\
$^{3}${\it School of Statistics and Management, Shanghai University of Finance and Economics, Shanghai, China 200433} \\
$^{4}${\it Department of Mathematical Sciences, Indiana University-Purdue University Indianapolis, Indianapolis, IN 46202}\\
\end{center}

{\bf Running title}: Functional varying index coefficient model for G$\times$E interaction


\begin{abstract}
Rooted in genetics, human complex diseases are largely influenced by environmental factors. Existing literature has shown the power of integrative gene-environment interaction analysis by considering the joint effect of environmental mixtures on a disease risk. In this work, we propose a functional varying index coefficient model for longitudinal measurements of a phenotypic trait together with multiple environmental variables, and assess how the genetic effects on a longitudinal disease trait are nonlinearly modified by a mixture of environmental influences. We derive an estimation procedure for the nonparametric functional varying index coefficients under the quadratic inference function and penalized spline framework. Theoretical results such as estimation consistency and asymptotic normality of the estimates are established. In addition, we propose a hypothesis testing procedure to assess the significance of the nonparametric index coefficient function. We evaluate the performance of our estimation and testing procedure through Monte Carlo simulation studies. The proposed method is illustrated by applying to a real data set from a pain sensitivity study in which SNP effects are nonlinearly modulated by the combination of dosage levels and other environmental variables to affect patients' blood pressure and heart rate.
\end{abstract}

\section{Introduction}
It has been broadly recognized that gene-environment (G$\times$E) interaction plays important roles in human complex diseases. A growing number of scientific researches have confirmed the role of G$\times$E interaction in many human diseases, such as Parkinson disease (Ross and Smith, 2007) and type 2 diabetes (Zimmet et al., 2001). G$\times$E interaction is defined as how genotypes influence phenotypes differently under different environmental conditions (Falconer, 1952). It also refers to the genetic sensitivity to environmental changes. Usually, G$\times$E has been investigated based on a single environment exposure model. Evidence from epidemiological studies has suggested that disease risk can be modified by simultaneous exposures to multiple environmental factors. The effect of simultaneous exposure is larger than the simple addition of the effects of factors acting alone (e.g., Carpenter et al., 2002; Sexton and Hattis, 2007). This motivated us to assess the combined effect of environmental mixtures, and how they as a whole, interact with genes to affect a disease risk (Liu et al. 2016). In our previous models, we proposed a varying multi-index coefficient model (VMICM) to capture the nonlinear interaction between a gene and environmental mixtures (Liu et al. 2016). The method was extended for any univariate trait distribution in a generalized linear model framework (Liu et al. 2017).

In biomedical studies, longitudinal traits are often observed, with repeated measures of the same subject over time. The increased power of a longitudinal design to detect genetic associations over cross-sectional designs has been evaluated (Sitlani et al. 2015; Furlotte et al. 2015; Xu et al. 2014). With longitudinal disease traits, one can study the dynamic gene effect over time. Coupling with longitudinal measure of environmental exposures, one can study how genes respond to the dynamic change of environmental factors to affect a disease trait. This motivates us to extend the VMICM model to a longitudinal trait.

To explore time-dependent effects in longitudinal data analysis, some nonparametric and semiparametric models such as varying coefficient models have been proposed, for example, Hoover et al. (1998), Wu, Chiang, and Hoover (1998), Fan and Zhang (2000), Martinussen and Scheike (2001), Chiang, Rice, and Wu (2001), Huang, Wu, and Zhou (2002), Ma and Song (2015). However, these methods do not fit to our purpose. In order to capture the dynamic nonlinear G$\times$E interaction with combined effect of environmental factors for longitudinal data, we propose a functional varying index coefficient model (FVICM) for correlated response, i.e.,
\begin{equation}\label{FVICM}
    Y_{ij}=m_{0}(\bm{\beta}_{0}^{T}\textbf{X}_{ij})+m_{1}(\bm{\beta}_{1}^{T}\textbf{X}_{ij})G_{i}+\varepsilon_{ij},
\end{equation}
where $Y_{ij}$ is the response variable which measures the risk of certain disease on the $i$th subject at the $j$th time point, where $i=1,\cdots,N$, $j=1,\cdots,n_{i}$; $\textbf{X}_{ij}$ is a $p$-dimensional vector of environmental variables, which can be either time-dependent or time invariant; $G_{i}$ denotes the genetic variable; $\varepsilon_{ij}$ is an error term with mean 0 and some correlation structure; $m_{0}(\cdot)$ and $m_{1}(\cdot)$ are unknown functions; $\bm{\beta}_{0}$ and $\bm{\beta}_{1}$ are $p$-dimensional vectors of index loading coefficients. For model identifiability, we have the constraints $\|\bm{\beta}_{0}\|=\|\bm{\beta}_{1}\|=1$ and restrict the first elements of $\bm{\beta}_{0}$ and $\bm{\beta}_{1}$ be positive.

Qu et al. (2000) proposed the quadratic inference function (QIF) for longitudinal data analysis, as an improvement of the generalized estimation equation (GEE) approach introduced by Liang and Zeger (1986). The QIF approach avoids estimating the nuisance correlation parameters by assuming that the inverse of the correlation matrix can be approximated by a linear combination of several basis matrices. Qu et al. (2000) found that the QIF estimator could be generally more efficient than the GEE estimator. Qu and Li (2006) applied the QIF method to the varying coefficient model for longitudinal data. Bai et al. (2009) developed an estimating procedure for single index models with longitudinal data based on the QIF method. Motivated by that, in this paper, we extend the QIF method to the FVICM model for dynamic G$\times$E interactions.

Our goal in this work is to develop a set of statistical estimation and hypothesis testing procedure for model (\ref{FVICM}). We first approximate the varying index coefficient function by penalized splines (Ruppert and Carroll, 2000) and then extend the QIF approach to our model in order to estimate the index loading coefficients and the penalized spline coefficients. Under certain regularity conditions, we establish the consistency and asymptotic normality of the resulting estimators.

Another goal of this work is to test the linearity of the G$\times$E interaction effect. This is of particular interest in our model setting since if the G$\times$E interaction is linear, a simple linear regression model should be fit, and fitting any higher order nonlinear functions would be unnecessary. With a mixed effects model representation of the penalized spline approximations (Speed, 1991; Ruppert, Wand, and Carroll, 2003; Wand, 2003), we can transform the problem of testing an unknown function into testing some fixed effects and a variance component in a linear mixed effects model setup with multiple variance components, which will be evaluated in this study.

This work is organized as follows: in Section $\ref{estimation}$, we propose an estimation procedure under the FVICM model, and further establish the consistency and asymptotic normality of the proposed estimator in Section $\ref{theory}$. In Section $\ref{practical}$, we discuss some practical issues to implement the proposed estimation procedures. In Section $\ref{testing}$, a pseudo-likelihood ratio test procedure with a linear mixed effects model representation is illustrated. We assess the finite sample performance of the proposed procedure with Monte Carlo simulations in Section $\ref{simulation}$ and illustrate the proposed method by an analysis of a pain sensitivity data set in Section $\ref{realdata}$, followed by discussions in Section $\ref{discussion}$. Technical details are rendered in Appendix.


\section{Quadratic inference function for FVICM with longitudinal data}\label{estimation}
For longitudinal data, suppose the response $y_{ij}$, $p$-dimensional covariate vector $\bm{x}_{ij}$, and SNP variable $G_{i}$ are observed from the $i$th observation at the $j$th time point. SNP variable $\{G_{i}, i=1,...,N\}$ does not change over time. Assume the model satisfies
\begin{equation*}
 E(y_{ij}|\bm{x}_{ij},G_{i}) = m_{0}(\bm{\beta}_{0}^{T}\bm{x}_{ij})+m_{1}(\bm{\beta}_{1}^{T}\bm{x}_{ij})G_{i}.
\end{equation*}
We can approximate the unknown coefficient functions $m_{0}(u_{0})$ and $m_{1}(u_{1})$ by a $q$-degree truncated power spline basis, i.e.
\begin{equation*}
    m_{0}(u_{0})=m_{0}(u_{0},\bm{\beta})\approx\textbf{B}(u_{0})^{T}\bm{\gamma}_{0},
\end{equation*}
\begin{equation*}
  m_{1}(u_{1})=m_{1}(u_{1},\bm{\beta})\approx\textbf{B}(u_{1})^{T}\bm{\gamma}_{1},
\end{equation*}
where $\bm{\beta}=(\bm{\beta}_{0}^{T},\bm{\beta}_{1}^{T})^{T}$, $\textbf{B}(u)=(1,u,u^{2},\cdots,u^{q},(u-\kappa_{1})_{+}^{q},\cdots,(u-\kappa_{K})_{+}^{q})^{T}$ is a $q$-degree truncated power spline basis with $K$ knots $\kappa_{1},\cdots,\kappa_{K}$. $\bm{\gamma}_{0}$ and $\bm{\gamma}_{1}$ are $(q+K+1)$-dimensional vectors of spline coefficients.
Let $\bm{\gamma}=(\bm{\gamma}_{0}^{T},\bm{\gamma}_{1}^{T})^{T}$.

For longitudinal data, the conditional variance-covariance matrix of the response need to be modelled. The method of generalized estimation equation (GEE) is often applied to estimate the unknowns. The GEE is defined as,
\begin{equation*}
    \sum_{i=1}^{N}\dot{\bm{\mu}}_{i}^{T}\textbf{V}_{i}^{-1}(\textbf{y}_{i}-\bm{\mu}_{i})=0,
\end{equation*}
where $\textbf{V}_{i}$ is the covariance matrix of $\textbf{y}_{i}$, $\textbf{y}_{i}=(y_{i1},...,y_{in_{i}})^{T}$, $\bm{\mu}_{i}=E(\textbf{y}_{i})$ is the mean function and $\dot{\bm{\mu}}_{i}$ is the first derivative of $\bm{\mu}_{i}$ with respect to the parameters. Based on the spline approximation, the mean function can be written as
\begin{equation*}
    \bm{\mu}_{i}=\bm{\mu}_{i}(\bm{\theta})=\left[
                                  \begin{array}{c}
                                      \bm{\mu}_{i1}(\bm{\theta}) \\
                                       \vdots \\
                                      \bm{\mu}_{in_{i}}(\bm{\theta}) \\
                                  \end{array}
                                  \right]                      =       \left[
                                                                        \begin{array}{c}
                                                                         \textbf{B}^{T}(\bm{\beta}_{0}^{T}\textbf{x}_{i1})\bm{\gamma}_{0}+ \textbf{B}^{T}(\bm{\beta}_{1}^{T}\textbf{x}_{i1})\bm{\gamma}_{1}G_{i}\\
                                                                          \vdots \\
                                                                         \textbf{B}^{T}(\bm{\beta}_{0}^{T}\textbf{x}_{in_{i}})\bm{\gamma}_{0}+
                                                                         \textbf{B}^{T}(\bm{\beta}_{1}^{T}\textbf{x}_{in_{i}})\bm{\gamma}_{1}G_{i} \\
                                                                        \end{array}
                                                                        \right],
\end{equation*}
and the first derivative of $\bm{\mu}_{i}$ is
\begin{equation*}
   \dot{\bm{\mu}}_{i}=\left[
                         \begin{array}{cccc}
                           \textbf{B}_{d}^{T}(\bm{\beta}_{0}^{T}\textbf{x}_{i1})\bm{\gamma}_{0}\textbf{x}_{i1}^{T} & \textbf{B}_{d}^{T}(\bm{\beta}_{1}^{T}\textbf{x}_{i1})\bm{\gamma}_{1}G_{i}\textbf{x}_{i1}^{T} & \textbf{B}^{T}(\bm{\beta}_{0}^{T}\textbf{x}_{i1}) & \textbf{B}^{T}(\bm{\beta}_{1}^{T}\textbf{x}_{i1})G_{i} \\
                           \vdots & \vdots & \vdots & \vdots \\
                           \textbf{B}_{d}^{T}(\bm{\beta}_{0}^{T}\textbf{x}_{in_{i}})\bm{\gamma}_{0}\textbf{x}_{in_{i}}^{T} & \textbf{B}_{d}^{T}(\bm{\beta}_{1}^{T}\textbf{x}_{in_{i}})\bm{\gamma}_{1}G_{i}\textbf{x}_{in_{i}}^{T} & \textbf{B}^{T}(\bm{\beta}_{0}^{T}\textbf{x}_{in_{i}}) & \textbf{B}^{T}(\bm{\beta}_{1}^{T}\textbf{x}_{in_{i}})G_{i} \\
                         \end{array}
                       \right],
\end{equation*}
where $\textbf{B}_{d}(u)=\frac{\partial \textbf{B}(u)}{\partial u}=(0,1,2u,\cdots,qu^{q-1},q(u-\kappa_{1})_{+}^{q-1},\cdots,q(u-\kappa_{K})_{+}^{q-1})$, $\bm{\theta}=(\bm{\beta}^{T},\bm{\gamma}^{T})^{T}$.

When $\textbf{V}_{i}$ is unknown, Liang and Zeger (1986) suggested that $\textbf{V}_{i}$ can be simplified as $\textbf{V}_{i}=\textbf{A}_{i}^{1/2}\textbf{R}(\rho)\textbf{A}_{i}^{1/2}$ with $\textbf{A}_{i}$ being a diagonal matrix of marginal variances and $\textbf{R}(\rho)$ being a common working correlation matrix with a small number of nuisance parameters $\rho$. When $\rho$ is consistently estimated, the GEE estimators of the regression coefficients are consistent. When such consistent estimators for the nuisance parameters do not exist, Qu et al. (2000) suggested that the inverse of $\textbf{R}(\rho)$ can be represented by a linear combination of a class of basis matrices such as $\textbf{R}^{-1}(\rho)\approx a_{1}\textbf{M}_{1}+a_{2}\textbf{M}_{2}\cdots+a_{h}\textbf{M}_{h}$, where $\textbf{M}_{1}$ is the identity matrix and $\textbf{M}_{2},\cdots,\textbf{M}_{h}$ are symmetric matrices. The advantage of this method is that the estimation of nuisance parameters $a_{1},\cdots,a_{h}$ are not required. Following this idea, we define the estimation function as,
\begin{equation}\label{gN}
    \bar{g}_{N}(\bm{\theta})=\frac{1}{N}\sum_{i=1}^{N}g_{i}(\bm{\theta})=\frac{1}{N}\left[
                                                                                \begin{array}{c}
                                                                                  \sum_{i=1}^{N}\dot{\bm{\mu}}_{i}^{T}\textbf{A}_{i}^{-1/2}\textbf{M}_{1}\textbf{A}_{i}^{-1/2}(\textbf{y}_{i}-\bm{\mu}_{i}) \\
                                                                                  \vdots \\
                                                                                  \sum_{i=1}^{N}\dot{\bm{\mu}}_{i}^{T}\textbf{A}_{i}^{-1/2}\textbf{M}_{h}\textbf{A}_{i}^{-1/2}(\textbf{y}_{i}-\bm{\mu}_{i})\\
                                                                                \end{array}
                                                                              \right]
\end{equation}
Because the dimension of the estimation equation $\bar{g}_{N}$ is greater than the number of parameters, we cannot obtain the estimators by simply setting each element in $\bar{g}_{N}$ to be zero. Qu et al. (2000) introduced the Quadratic Inference Function (QIF) based on the generalized method of moments (Hansen, 1982). Thus, we can estimate the parameters by minimizing the QIF, which is defined as
\begin{equation}\label{QN}
    Q_{N}(\bm{\theta})= N\bar{g}_{N}^{T}\bar{C}_{N}^{-1}\bar{g}_{N},
\end{equation}
where $\bar{C}_{N}=\frac{1}{N}\sum_{i=1}^{N}g_{i}g_{i}^{T}$ is a consistent estimator for $\text{var}(g_{i})$.
By minimizing the quadratic inference function, we can obtain the estimation of the parameters
\begin{equation*}
    \widehat{\bm{\theta}}=\arg\min_{\bm{\theta}}Q_{N}(\bm{\theta}).
\end{equation*}
To overcome the well known over-parameterization issue, Qu et al. (2000) further proposed the penalized quadratic inference function
\begin{equation}\label{target}
    N^{-1}Q_{N}(\bm{\theta})+ \lambda \bm{\theta}^{T}\textbf{D}\bm{\theta},
\end{equation}
where $\textbf{D}$ is a diagonal matrix with element 1 if the corresponding parameters are spline coefficients associated with the knots and 0 otherwise, i.e., $\textbf{D} = \text{diag}(\textbf{0}^{T}_{(2p+q+1)\times1}, \textbf{1}^{T}_{K\times1}, \textbf{0}^{T}_{(q+1)\times1}, \textbf{1}^{T}_{K\times1})$.
Then the estimator is given by
\begin{equation}\label{thetahat}
    \widehat{\bm{\theta}}=\arg\min_{\bm{\theta}}(N^{-1}Q_{N}(\bm{\theta})+ \lambda \bm{\theta}^{T}\textbf{D}\bm{\theta}).
\end{equation}

\section{Asymptotic properties}\label{theory}
In this section, we establish the asymptotic properties for the penalized quadratic inference function estimators with fixed knots.
Assume $\bm{\theta}_{0}$ is the parameter satisfying $E_{\bm{\theta}_{0}}(g_{i})=0$. Theorem \ref{thm1} provides the consistency of the resulting estimators. We show the asymptotic normality of the estimators in Theorem \ref{thm2}. The theoretical results are similar to those provided in Qu and Li (2006). The difference is that we have constraints for the index loading parameters in our model, i.e. $\|\bm{\beta}_{0}\|$=$\|\bm{\beta}_{1}\|$=1, and $\beta_{01}>0$, $\beta_{11}>0$. To handle the constraints, we do the reparameterization as $\beta_{l1}=\sqrt{1-\|\bm{\beta}_{l,-1}\|^{2}_{2}}$ with $\bm{\beta}_{l,-1}=(\beta_{l2}, ..., \beta_{lp})^{T}$ for $l$=1, 2 (Yu and Ruppert, 2002; Cui et al., 2011; Ma and Song, 2015). Then the parameters space of $\bm{\beta}_{l}$, $l$=1,2, becomes $[\{(\sqrt{1-\|\bm{\beta}_{l,-1}\|^{2}_{2}}, \beta_{l2}, ..., \beta_{lp})^{T}\}: \|\bm{\beta}_{l,-1}\|^{2}_{2}<1].$
Let
\begin{equation*}
  \textbf{J}_{l}=\frac{\partial \bm{\beta}_{l}}{\partial \bm{\beta}_{l,-1}^{T}}=\left(
                                                                                  \begin{array}{c}
                                                                                    -\bm{\beta}_{l,-1}^{T}/\sqrt{1-\|\bm{\beta}_{l,-1}\|^{2}_{2}} \\
                                                                                    \textbf{I}_{p-1} \\
                                                                                  \end{array}
                                                                                \right)
\end{equation*}
be the Jacobian matrix of dimension $p\times(p-1)$. Denote $\bm{\beta}_{-1}=(\bm{\beta}_{0,-1}^{T}, \bm{\beta}_{1,-1}^{T})^{T}$, and $\bm{\theta}^{*}=(\bm{\beta}_{-1}, \bm{\gamma})^{T}$. From $\bm{\theta}$ to $\bm{\theta}^{*}$, we have Jacobian matrix $\textbf{J}=\text{diag}(\textbf{J}_{0}, \textbf{J}_{1}, \textbf{I}_{q+K+1}, \textbf{I}_{q+K+1})$.

\begin{theorem}\label{thm1}
Suppose the assumptions (A1)-(A6) in the Appendix hold, and the smoothing parameter $\lambda_{N}=o(1)$, then the estimator $\widehat{\bm{\theta}}$, which is obtained by minimizing the penalized quadratic function in (\ref{target}), exists and converges to the true parameters $\bm{\theta}_{0}$ in probability.
\end{theorem}

\begin{theorem}\label{thm2}
Suppose the assumptions (A1)-(A6) in the Appendix hold, and the smoothing parameter $\lambda_{N}=o(N^{-1/2})$, then the estimator $\widehat{\bm{\theta}}$ obtained by minimizing the penalized quadratic function in (\ref{target}) is asymptotically normally distributed, i.e.,
\begin{equation*}
  \sqrt{N}(\widehat{\bm{\theta}}-\bm{\theta}_{0})\xrightarrow{d}N(\textbf{0},\textbf{J}(\textbf{G}_{0}^{T}\textbf{C}_{0}^{-1}\textbf{G}_{0})^{-1}\textbf{J}^{T}),
\end{equation*}
where $\textbf{G}_{0}$ and $\textbf{C}_{0}$ are given in the Appendix.
\end{theorem}

\section{Practical implementation}\label{practical}
In this section, we discuss some practical issues when we implement the proposed method.
\subsection{Algorithm for estimation}
A two-step iterative Newton-Raphson algorithm is applied when we estimate the index loading parameters and the varying spline coefficients.
The algorithm of the estimation procedure can be summarized in the following steps.
\begin{description}
  \item[Step 0] Choose initial values for $\bm{\beta}$ and $\bm{\gamma}$. Denote them by $\bm{\beta}^{(old)}$ and $\bm{\gamma}^{(old)}$.
  \item[Step 1] Estimate $\bm{\gamma}^{(new)}$ by
  \begin{equation*}
    \bm{\gamma}^{(new)}=\arg\min_{\bm{\gamma}}(N^{-1}Q_{N}(\bm{\gamma},\bm{\beta}^{(old)})+ \lambda \bm{\gamma}^{T}\textbf{D}\bm{\gamma}.
\end{equation*}
The Newton-Raphson algorithm is used for the minimization.
  \item[Step 2] Estimate $\bm{\beta}^{(new)}$ by
  \begin{equation*}
    \bm{\beta}^{(new)}=\arg\min_{\bm{\beta}}Q_{N}(\bm{\beta},\bm{\gamma}^{(new)}).
\end{equation*}
Also use Newton-Raphson for minimization.
  \item[Step 3] Update $\bm{\beta}_{l}^{(old)}$ by $\bm{\beta}_{l}^{(old)}=\text{sign}(\beta_{l1}^{(new)})\bm{\beta}_{l}^{(new)}/\|\bm{\beta}_{l}^{(new)}\|_{2}$, $l=1,2$. Update $\bm{\gamma}^{(old)}$ by setting $\bm{\gamma}^{(old)}=\bm{\gamma}^{(new)}$.
  \item[Step 4] Repeat Steps 1-3 until convergence.
\end{description}

\subsection{Model selection}
It is important to determine the order and number of knots in the spline approximation since too many knots in the model might overfit the data. Under the assumption $E(g)=0$ ($g$ is the estimation function in (\ref{gN}) for a single observation) and the number of estimating equations is larger than the number of parameters, we have $Q(\widehat{\bm{\theta}})\rightarrow\chi^{2}_{r-k}$ in distribution (Hansen, 1982), where $r$ is the dimension of $\bar{g}_{N}(\bm{\theta})$, $k$ is the dimension of $\bm{\theta}$, $\widehat{\bm{\theta}}$ is the estimator by minimizing the QIF when certain order and number of knots are chosen. This asymptotic property of the QIF provides a goodness-of-fit test, which can be useful to determine the order and number of knots to be selected in our model.

However, it is also possible that the goodness-of-fit tests fail to reject several different models which may not be nested. Since $Q(\widehat{\bm{\theta}})$ is  asymptotically chi-square distributed, we can use BIC to penalize $Q(\widehat{\bm{\theta}})$ for the difference of the numbers of estimating equations and parameters. In particular, the BIC criterion for a model with $r$ estimating equation and $k$ parameters is defined as
\begin{equation*}
  Q(\widehat{\bm{\theta}})+(r-k)\ln N.
\end{equation*}
The model with minimum BIC would be considered better. If we choose $h$ basis matrices in (\ref{gN}), then $r-k$ = $hk-k$ = $(h-1)k$. As will be discussed in Section \ref{basis}, we usually use $h$=2 in our setting. Thus, the BIC criterion is actually
\begin{equation*}
  Q(\widehat{\bm{\theta}})+ k\ln N,
\end{equation*}
where $k$ is the number of parameters in the model.

In our simulation and real data application, we search the optimal order and the number of knots over a set of combinations of $q$ and $K$ using BIC. Knots are evenly distributed in the range of $u (=\bm{\beta}^{T}\textbf{X})$.

\subsection{Choice of the basis for the inverse of the correlation matrix}\label{basis}
Qu and Li (2006) offered several choices of basis matrixes. For exchangeable working correlation, $\textbf{M}_{1}$ is an identity matrix and $\textbf{M}_{2}$ has 0 on the diagonal and 1 off-diagonal. If the working correlation is AR(1), we can set $\textbf{M}_{2}$ to have 1 on its two subdiagonals and 0 elsewhere. Prior information on correlation can help us to determine the choice of appropriate basis matrices. The effect of choosing different basis matrices is discussed in Qu and Li (2006) through simulation studies. Qu and Lindsay (2003) also proposed an adaptive estimation method to approximate the correlation empirically when there is no prior information available.

\subsection{Choice of the tuning parameter}
Since the penalized spline is used to approximate the unknown functions, we need to determine the tuning parameter $\lambda$ involved in the method. As Qu and Li (2006) suggested, we can extend the generalized cross-validation (Ruppert, 2002) to the penalized QIF and define the generalized cross-validation statistic as
\begin{equation*}
    \text{GCV}(\lambda)=\frac{N^{-1}Q_{N}}{(1-N^{-1}\text{df})^{2}}
\end{equation*}
where $\text{df}=\text{tr}[(\ddot{Q}_{N}+2N\lambda D)^{-1}\ddot{Q}_{N}]$ is the effective degree of freedom, $Q_{N}$ is defined in (\ref{QN}) and $\ddot{Q}_{N}$ is the second derivative of $Q_{N}$. The desirable choice of tuning parameter $\lambda$ is
\begin{equation*}
    \widehat{\lambda}=\arg\min_{\lambda}\text{GCV}(\lambda).
\end{equation*}
In the implementation of GCV, the golden search method can be applied in order to reduce the computational time.


\section{Hypothesis test}\label{testing}

\subsection{Linear mixed model representation for FVICM model}
In our proposed FVICM model (\ref{FVICM}), it is of interest to test the unspecified coefficient function. In particular, we are interested in testing whether a linear function is good enough to describe the G$\times$E interaction.
Given $\bm{\beta}$, let $u_{0}=\bm{\beta}_{0}^{T}\mathbf{X}$, $u_{1}=\bm{\beta}_{1}^{T}\mathbf{X}$, with the truncated power spline basis, the coefficient function can be modeled by
\begin{equation*}
  m_{1}(u_{1})=\gamma_{10}+\gamma_{11}u_{1}+\gamma_{12}u_{1}^{2}+\cdots+\gamma_{1q}u_{1}^{q}+\sum_{k=1}^{K}b_{1k}(u_{1}-\kappa_{k})^{q}_{+}.
\end{equation*}
Note that under the current model setup, we cannot assess the zero effect of the nonparametric function $m_1(\cdot)$ since under the null hypothesis of $m_1(\cdot)=0$, the index loading parameters $\bbeta_1$ are not identifiable, unless we impose the constraint that $\bbeta_1=\bbeta_0=\bbeta$. This constraint, however, is practically unrealistic.
Thus, our goal is to test the linearity of $m_{1}(u_{1})$, which is equivalent to test
\begin{equation*}
  H_{0}: \gamma_{12}=\cdots=\gamma_{1q}=0, b_{11}=\cdots=b_{1K}=0.
\end{equation*}
If the above $H_0$ is rejected, we conclude there exists a nonlinear relationship. Otherwise, we assume a linear relationship and fit $m_1(\cdot)$ with a linear function and further test the zero effect of the linear relationship. Let $\textbf{w}_{0ij}=(1,u_{0ij},\cdots,u_{0ij}^{q})^{T}$, $\textbf{z}_{0ij}=\big((u_{0ij}-\kappa_{1})_{+}^{q} ,\cdots, (u_{0ij}-\kappa_{K})_{+}^{q}\big)^{T}$, $\bm{\tilde{\gamma}}_{0}=(\gamma_{00},\cdots,\gamma_{0q})^{T}$, $\textbf{b}_{0}=(b_{01},\cdots,b_{0K})^{T}$, $\textbf{w}_{1ij}=(1,u_{1ij},\cdots,u_{1ij}^{q})^{T}$, $\textbf{z}_{1ij}=\big((u_{1ij}-\kappa_{11})_{+}^{q} ,\cdots, (u_{1ij}-\kappa_{1K})_{+}^{q}\big)^{T}$,$\textbf{b}_{1}=(b_{11},\cdots,b_{1K})^{T}$, $\bm{\tilde{\gamma}}_{1}=(\gamma_{10},\cdots,\gamma_{1q})^{T}$, then we have
$m_{0}(u_{0ij})=\textbf{w}_{0ij}^{T}\bm{\tilde{\gamma}}_{0}+\textbf{z}_{0ij}^{T}\textbf{b}_{0}$, and
  $m_{1}(u_{1ij})=\textbf{w}_{1ij}^{T}\bm{\tilde{\gamma}}_{1}+\textbf{z}_{1ij}^{T}\textbf{b}_{1}$.

We further define $\textbf{Y}_{i}=(y_{i1},\cdots,y_{in_{i}})^{T}$, $\textbf{W}_{0i}=(\textbf{w}_{0i1},\cdots,\textbf{w}_{0in_{i}})^{T}$, $\textbf{W}_{1i}=(\textbf{w}_{1i1} G_{i},\cdots,\textbf{w}_{1in_{i}}G_{i})^{T}$, $\textbf{Z}_{0i}=(\textbf{z}_{0i1},\cdots,\textbf{z}_{0in_{i}})^{T}$, and
 $\textbf{Z}_{1i}=(\textbf{z}_{1i1}G_{i},\cdots,\textbf{z}_{1in_{i}}G_{i})^{T}$, then a linear mixed model (LMM) representation (Wang and Chen, 2012) can be obtained as,
\begin{equation}\label{lmm}
  \textbf{Y}_{i} = \textbf{1}_{i}a_{i}+\textbf{W}_{0i}\bm{\tilde{\gamma}}_{0} + \textbf{W}_{1i}\bm{\tilde{\gamma}}_{1} + \textbf{Z}_{0i}\textbf{b}_{0} + \textbf{Z}_{1i}\textbf{b}_{1} + \bm{\epsilon}_{i}, ~~i=1,\cdots,n,
\end{equation}
where $\textbf{b}_{l}\sim N(\textbf{0},\sigma_{\textbf{b}_{l}}^{2}\textbf{I}_{K}), ~l=0,1$, $\bm{\epsilon}_{i}\sim N(\textbf{0},\sigma_{\epsilon}^{2}\textbf{I})$, and the random incept effects $a_{i}$ are assumed to be independent as $N(0,\sigma_{a}^{2})$. 


With the LMM representation, testing the linearity of the varying index coefficients is equivalent to test some fixed effects and a variance component in model $(\ref{lmm})$. To be specific, we want to test
\begin{equation}\label{hypothesis}
  H_{0}: \gamma_{12}=\cdots=\gamma_{1q}=0 ~\text{and} ~\sigma_{\textbf{b}_{1}}^{2}=0.
\end{equation}

\subsection{Likelihood ratio test (LRT) and pseudo-LRT in LMM}\label{LRT}
\subsubsection{LRT for one variance component}\label{LRTintro}

Crainiceanu and Ruppert (2004) proposed the likelihood ratio test in linear mixed effect models with one variance component. Consider a LMM with one variance component
\begin{equation}\label{LMMonecomp}
  \textbf{Y}=\textbf{X}\bm{\beta}+\textbf{Z}\textbf{b}+\bm{\epsilon}, ~~E\left[\begin{array}{c}
                                                             \textbf{b} \\
                                                             \bm{\epsilon}
                                                           \end{array}\right] = \left[\begin{array}{c}
                                                             \mathbf{0}_{L}\\
                                                             \mathbf{0}_{n}
                                                           \end{array}\right],
                                                           ~~\text{Cov}\left[\begin{array}{c}
                                                             \textbf{b} \\
                                                             \bm{\epsilon}
                                                           \end{array}\right]=\left[\begin{array}{cc}
                                                                                \sigma_{b}^{2}\bm{\Sigma} & \mathbf{0} \\
                                                                                \mathbf{0} & \sigma_{\epsilon}^{2}\textbf{I}_{n}
                                                                              \end{array}\right],
\end{equation}
where $\bm{\beta}$ is a $p$-dimensional vector of fixed effect coefficients, $\textbf{b}$ is a $L$-dimensional vector of random effects, $\mathbf{0}_{L}$ is a $L$-dimensional vector of zeros, $\bm{\Sigma}$ is a known $L\times L$ symmetric positive definite matrix. Let $\lambda=\sigma_{b}^{2}/\sigma_{\epsilon}^{2}$ be the signal-to-noise ratio and then the covariance matrix of $\textbf{Y}$ cab be written as $\text{Cov}(\textbf{Y})=\sigma_{\epsilon}^{2}\textbf{V}_{\lambda}$, where $\textbf{V}_{\lambda}=\textbf{I}_{n}+\lambda \textbf{Z}\bm{\Sigma} \textbf{Z}^{T}$.
Consider testing for the null hypothesis
\begin{equation}\label{H0onecomp}
  H_{0}: \beta_{p+1-p'}=0,\cdots,\beta_{p}=0, ~\sigma_{b}^{2}=0
\end{equation}
for $p'>0$.

The LRT statistic is defined as
\begin{equation*}
  LRT_{n}\propto2\Big\{\sup_{H_{A}}L(\bm{\beta},\lambda,\sigma_{\epsilon}^{2})-\sup_{H_{0}}L(\bm{\beta},\lambda,\sigma_{\epsilon}^{2})\Big\}.
\end{equation*}
If we substitute the parameters $\bm{\beta}$ and $\sigma_{\epsilon}^{2}$ with their profile estimators
\begin{equation*}
  \widehat{\bm{\beta}}(\lambda)=(\textbf{X}^{T}\textbf{V}^{-1}_{\lambda}\textbf{X})^{-1}\textbf{X}^{T}\textbf{V}^{-1}_{\lambda}\textbf{Y},
\end{equation*}
\begin{equation*}
 \widehat{\sigma}_{\epsilon}^{2}(\lambda)=\frac{\{\textbf{Y}-\textbf{X}\widehat{\bm{\beta}}(\lambda)\}^{T}V_{\lambda}^{-1}\{\textbf{Y}-\textbf{X}\widehat{\bm{\beta}}(\lambda)\}}{n},
\end{equation*}
for fixed $\lambda$, we obtain the LRT statistic
\begin{equation}\label{LRT}
  LRT_{n}=\sup_{\lambda\geq0}\{n\log(\textbf{Y}^{T}\textbf{S}_{0}\textbf{Y})-n\log(\textbf{Y}^{T}\textbf{P}^{T}_{\lambda}\textbf{V}_{\lambda}^{-1}\textbf{P}_{\lambda}\textbf{Y})-\log|\textbf{V}_{\lambda}|\},
\end{equation}
where $\textbf{P}_{\lambda}=\textbf{I}_{n}-\textbf{X}(\textbf{X}^{T}\textbf{V}^{-1}_{\lambda}\textbf{X})^{-1}\textbf{X}^{T}\textbf{V}^{-1}_{\lambda}$, $\textbf{X}_{0}$ denotes the design matrix of fixed effects under the null hypothesis, and $\textbf{S}_{0}=\textbf{I}_{n}-\textbf{X}_{0}(\textbf{X}_{0}^{T}\textbf{X}_{0})^{-1}\textbf{X}_{0}^{T}$.

Theorem 1 in Crainiceanu and Ruppert (2004) provides the distribution of LRT statistic $(\ref{LRT})$. Let $\mu_{s}$ be the eigenvalues of $\bm{\Sigma}^{1/2}\textbf{Z}^{T}\textbf{P}_{0}\textbf{Z}\bm{\Sigma}^{1/2}$, $\xi_{s}$ be the eigenvalues of $\bm{\Sigma}^{1/2}\textbf{Z}^{T}\textbf{Z}\bm{\Sigma}^{1/2}$, $s=1,\cdots,L$, then
\begin{equation}\label{dLRT}
  LRT_{n}\buildrel d \over =n\Bigg(1+\frac{\sum_{1}^{p'}u^{2}_{s}}{\sum_{1}^{n-p}w^{2}_{s}}\Bigg)+\sup_{\lambda\geq0}f_{n}(\lambda),
\end{equation}
where $u_{s}\buildrel iid \over \sim N(0,1)$ for $s=1,\cdots,L$, $w_{s}\buildrel iid \over \sim N(0,1)$ for $s=1,\cdots,n-p$, and
\begin{equation*}
  f_{n}(\lambda)=n\log\bigg\{1+\frac{N_{n}(\lambda)}{D_{n}(\lambda)}\bigg\}-\sum_{s=1}^{L}\log(1+\lambda\mu_{s}),
\end{equation*}
with
\begin{equation*}
  N_{n}(\lambda)=\sum_{s=1}^{L}\frac{\lambda\mu_{s}}{1+\lambda\mu_{s}}w^{2}_{s},
\end{equation*}
\begin{equation*}
  D_{n}(\lambda)=\sum_{s=1}^{L}\frac{w^{2}_{s}}{1+\lambda\mu_{s}}+\sum_{s=L+1}^{n-p}w^{2}_{s}.
\end{equation*}
The distribution in ($\ref{dLRT}$) only depends on the eigenvalues $\mu_{s}$ and $\xi_{s}$. Based on the spectral decomposition, simulation from this distribution can be done very rapidly. Detailed algorithm for this simulation can be found in Crainiceanu and Ruppert (2004).

\subsubsection{Pseudo-LRT for multiple variance components}
For a LMM with multiple variance components
  \begin{equation}\label{LMMmulticomp}
  \textbf{Y}=\textbf{X}\bm{\beta}+\textbf{Z}\textbf{b}_{s}+\cdots+\textbf{Z}\textbf{b}_{S}+\bm{\epsilon},
\end{equation}
\begin{equation*}
  \textbf{b}_{s}\sim N(\textbf{0},\sigma_{s}^{2}\bm{\Sigma}_{s}), ~s=1,\cdots,S, ~~ \bm{\epsilon}\sim N(\textbf{0},\sigma_{\epsilon}^{2}\textbf{I}_{n}),
\end{equation*}
where $\textbf{b}_{s}$, $s=1,\cdots,S$ are random effects and $S>1$. Suppose we are interested in testing
\begin{equation*}
  H_{0}: \beta_{p+1-p'}=0,\cdots,\beta_{p}=0, ~\sigma^{2}_{s}=0.
\end{equation*}


Greven et al. (2008) proposed to approximate the distribution of LRT for the model (\ref{LMMmulticomp}) based on the pseudo-likelihood ratio test theory (Liang and Self, 1996) by using a pseudo-outcome. In the framework of model (\ref{LMMmulticomp}), 
$\textbf{b}_{l}, l\neq s$, are nuisance random parameters. We can define the pseudo-outcome as
\begin{equation*}
  \widetilde{\textbf{Y}}=\textbf{Y}-\sum_{l\neq s}\textbf{Z}_{l}\widehat{\textbf{b}}_{l},
\end{equation*}
where $\widehat{\textbf{b}}_{l}$ are the best linear unbiased predictors (BLUP) of nuisance random effects $\textbf{b}_{l}, l\neq s$.
Then the model (\ref{LMMmulticomp}) can be reduced to
\begin{equation}\label{reduced}
  \widetilde{\textbf{Y}}=\textbf{X}\bm{\beta}+\textbf{Z}_{s}\textbf{b}_{s}+\bm{\epsilon}.
\end{equation}
Then the method for testing one variance component introduced by Crainiceanu and Ruppert (2004) can be applied to the model in (\ref{reduced}).

\subsection{Pseudo-LRT in FVICM model}
For the model in (\ref{lmm}), we can use the idea of Greven et al. (2008) and define the pseudo-outcome
\begin{equation*}
  \widetilde{\textbf{Y}}_{i} = \textbf{Y}_{i} - \textbf{Z}_{0i}\widehat{\textbf{b}}_{0} - \textbf{U}_{i}\widehat{a}_{i},~~i=1,\cdots,n,
\end{equation*}
where $\widehat{\textbf{b}}_{0}$ and $\widehat{a}_{i}$ are BLUPs of $\textbf{b}_{0}$ and $a_{i}$, respectively.
The reduced model using pseudo-outcome for model (\ref{lmm}) can be written as
\begin{equation}\label{pseudomodel}
  \widetilde{\textbf{Y}}_{i}=\textbf{W}_{0i}\bm{\tilde{\gamma}}_{0} + \textbf{W}_{1i}\bm{\tilde{\gamma}}_{1} + \textbf{Z}_{1i}\textbf{b}_{1} +\bm{\epsilon}_{i}. ~~i=1,\cdots,n.
\end{equation}
For the new model (\ref{pseudomodel}) using pseudo-response, we can apply the method for the single variance component model introduced in Section \ref{LRT} to test hypothesis (\ref{hypothesis}). Statistical significance can be assessed through the resampling approach described in section \ref{LRTintro}.

\section{Simulation study}\label{simulation}
\subsection{Simulation}
In this section, the finite sample performance of the proposed method is evaluated through Monte Carlo simulation studies. We generate three covariates $X_{1}, X_{2}, X_{3}$. For each subject $i$, $X_{1ij}, X_{2ij}, X_{3ij}$ are generated independently from uniform distribution $U(0,1)$. We set the minor allele frequency (MAF) as $p_{A}$=(0.1, 0.3, 0.5) and assume Hardy-Weinberg equilibrium. We use $AA$, $Aa$ and $aa$ to denote three different SNP genotypes, where allele $A$ is the minor allele. These genotypes are simulated from a multinomial distribution with frequencies $p_{A}^{2}$, $2p_{A}(1-p_{A})$ and $(1-p_{A})^{2}$, respectively. Variable $G$ takes value in the set \{0,1,2\}, corresponding to genotypes $\{aa,Aa,AA\}$ respectively. The error term $\epsilon_{i}=(\epsilon_{i1},\cdots,\epsilon_{in_{i}})$ are independently generated from the multivariate normal distribution $N(\textbf{0},0.1\textbf{R}(\rho))$. The true correlation structure $\textbf{R}(\rho)$ is assumed to be exchangeable with $\rho=0.5$ and 0.8.

We set $m_{0}(u_{0})=\cos(\pi u_{0})$ and $m_{1}(u_{1})=\sin[\pi(u_{1}-A)/(B-A)]$ with $A=\sqrt{3}/2-1.645/\sqrt{12}$ and $B=\sqrt{3}/2+1.645/\sqrt{12}$. The true parameters are $\bm{\beta}_{0}=(\sqrt{5},\sqrt{4},\sqrt{4})/\sqrt{13}$ and $\bm{\beta}_{1}=(1,1,1)/\sqrt{3}$. To simplify the simulation and save computational time, we consider the balanced case, which means each observation has the same number of time points. We draw 1000 data sets with sample size $N=200, 500$ and time points $n_{i}=T=10$. Since the true correlation structure is exchangeable, we set $\textbf{M}_{1}$ to be the identity matrix and $\textbf{M}_{2}$ to be 0 on the diagonal and 1 off-diagonal. The order and number of knots of the splines are chosen by using the BIC method.

\subsection{Performance of estimation}
Table $\ref{rho05}$ summarizes the results based on 1000 replications. In this table, the average bias (Bias), the standard deviation of the 1000 estimates (SD), the average of the estimated standard error (SE) based on the theoretical results, and the estimated coverage probability (CP) at the $95\%$ confidence level are reported. Note that the estimation of the loading parameter $\bm{\beta}_{1}$ improves (smaller Bias, SD and SE and CP more close to 95\%), as MAF $p_{A}$ increases, while the estimation of $\bm{\beta}_{0}$ show a opposite direction. This is because we have limited data information to estimate the marginal effects $m_0(\cdot)$ when $p_{A}$ increases. As the sample size increases, the performance of the estimation improves by showing smaller Bias, SD and SE.

\begin{table}[h!]\centering
\caption{Simulation results for $p_A = 0.1, 0.3, 0.5$ with sample size $N=200, 500$ and correlation $\rho$=0.5. }{
 {\footnotesize \setlength{\tabcolsep}{1.0mm}
\begin{tabular}{cccccccccccccccccccccccccccccccccccccc}
\hline
 &&&  \multicolumn{4}{c}{$p_A=0.1$} &    &  \multicolumn{4}{c}{$p_A=0.3$} && \multicolumn{4}{c}{$p_A=0.5$}  \\
  \cline{5-8}\cline{10-13}\cline{15-18}
   $N$&Param& True& & Bias  & SD& SE& CP && Bias & SD& SE& CP && Bias & SD& SE& CP
         \\
\hline
200 &$\beta_{01}$ &    0.620 & & 7.3E-04&    0.008&    0.008&    95.6 & &   1.7E-03&    0.009&    0.010&    96.2& & 1.5E-03&    0.011&    0.011&     95.0   \\
&$\beta_{02}$ &    0.555 & & -3.9E-04&    0.008&    0.009&     93.2 & &   -1.0E-03&    0.010&    0.010&     92.5& & -1.2E-03&    0.012&    0.011&      92.4  \\
&$\beta_{03}$ &    0.555 & & -6.2E-04&    0.008&    0.008&    94.4 & &   -1.2E-03&    0.010&    0.010&     94.2& & -8.5E-04&    0.012&    0.011&      93.0   \\
&$\beta_{11}$ &    0.577 & & -2.3E-05&    0.018&    0.020&     91.0 & &   -3.1E-04&    0.011&    0.011&    93.7& & -8.6E-04&    0.009&    0.009&     94.7   \\
&$\beta_{12}$ &    0.577 & & -6.3E-04&    0.018&    0.020&    91.3 & &   -3.0E-04&    0.011&    0.011&     94.3& & -6.8E-05&    0.009&    0.009&      93.8   \\
&$\beta_{13}$ &    0.577 & & -3.9E-04&    0.018&    0.020&    91.0 & &   2.8E-04&    0.011&    0.011&    94.8& &7.1E-04&    0.009&    0.009&     93.1   \\\\
500 &$\beta_{01}$ &    0.620 & & 7.5E-04&    0.005&    0.005&    95.5 & &   1.7E-03&    0.006&    0.006&    95.1& &1.6E-03&    0.007&    0.007&     95.8   \\
&$\beta_{02}$ &    0.555 & & -5.7E-04&    0.005&    0.005&     94.4 & &   -1.1E-03&    0.006&    0.006&     94.6& &-8.8E-04&    0.007&    0.007&      95.2   \\
&$\beta_{03}$ &    0.555 & & -3.4E-04&    0.005&    0.005&     93.9 & &   -8.7E-04&    0.006&    0.006&     94.1& &-1.1E-03&    0.007&    0.007&      94.7   \\
&$\beta_{11}$ &    0.577 & & 6.4E-04&    0.012&    0.012&     93.8 & &   -1.7E-04&    0.007&    0.007&     95.6& &-7.3E-04&    0.006&    0.006&     95.1   \\
&$\beta_{12}$ &    0.577 & & -6.0E-04&    0.012&    0.012&    93.6 & &   -1.5E-05&    0.007&    0.007&    96.1& &5.3E-04&    0.006&    0.006&      94.7   \\
&$\beta_{13}$ &    0.577 & & -4.1E-04&    0.012&    0.012&    94.6 & &   6.0E-05&    0.007&    0.007&    95.0& &1.1E-04&    0.006&    0.006&     95.6   \\
\hline
\end{tabular}}}
\label{rho05}
\end{table}

\begin{figure}[h!]\centering
 \centering
 \subfigure{
   \begin{minipage}[t]{0.32\textwidth}
         \includegraphics[width=5.5cm, height=4.5cm]{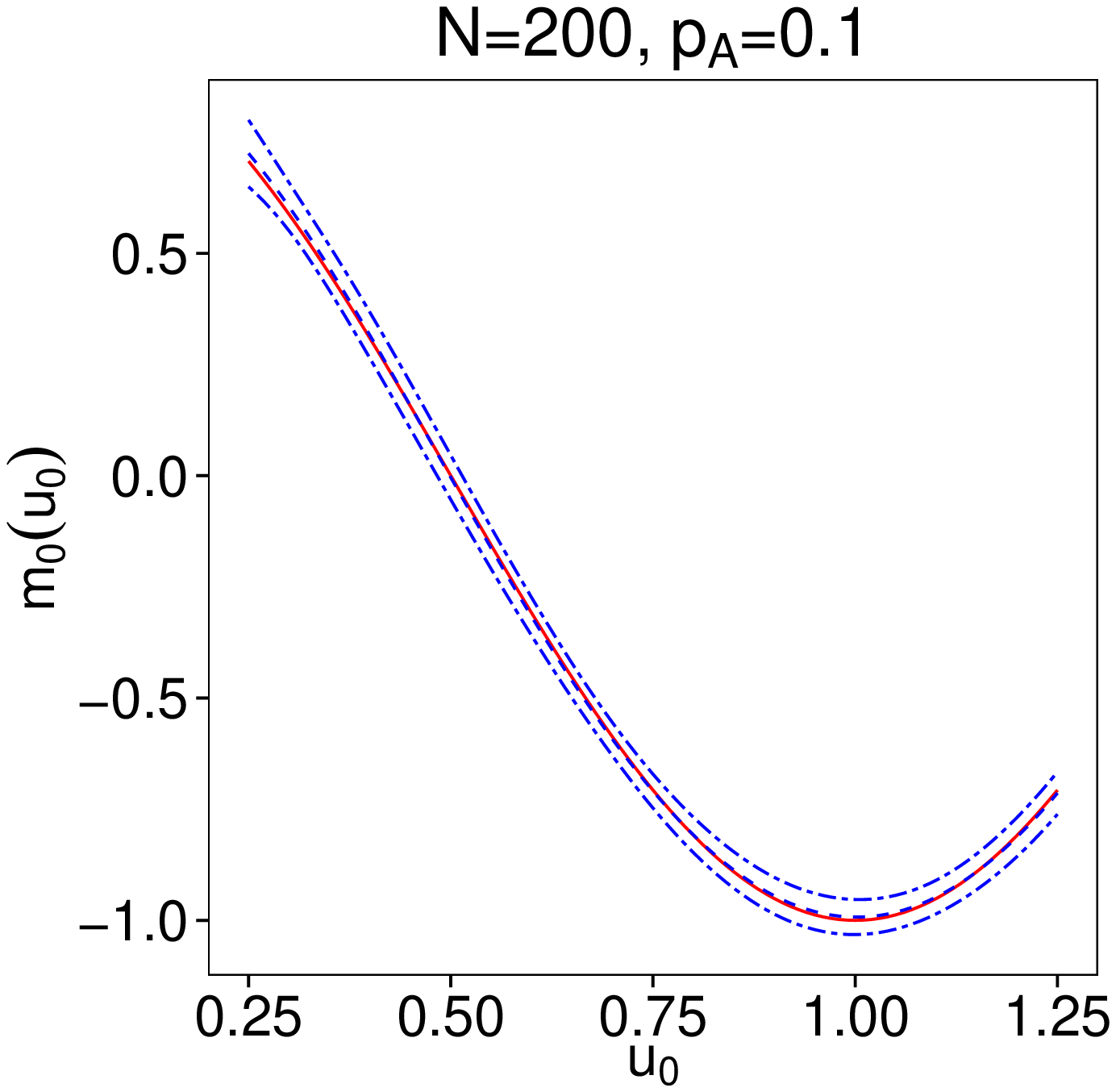}
   \end{minipage}
   \begin{minipage}[t]{0.32\textwidth}
         \includegraphics[width=5.5cm, height=4.5cm]{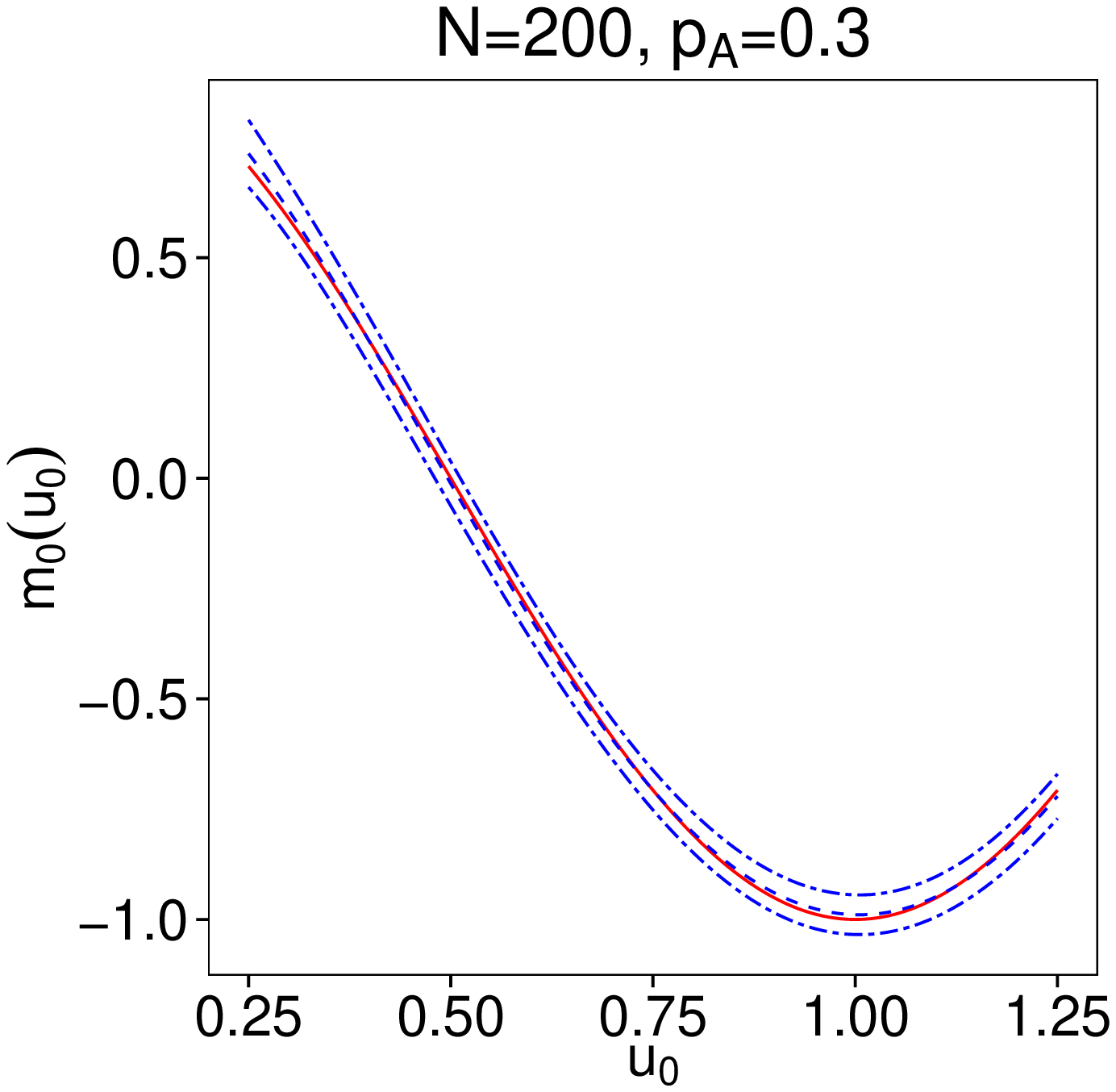}
   \end{minipage}
   \begin{minipage}[t]{0.32\textwidth}
         \includegraphics[width=5.5cm, height=4.5cm]{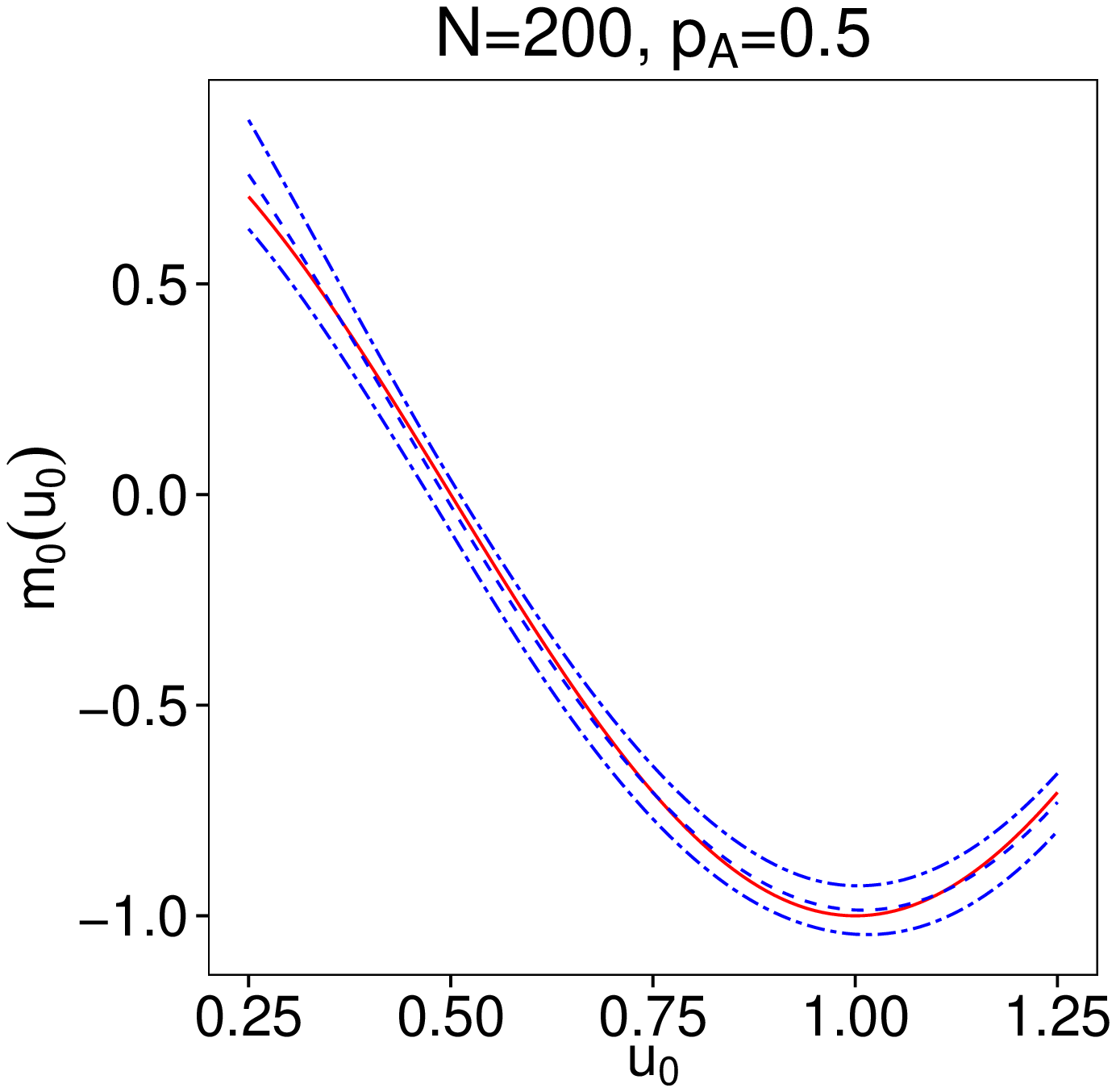}
   \end{minipage}
 }
 \subfigure{
   \begin{minipage}[t]{0.32\textwidth}
         \includegraphics[width=5.5cm, height=4.5cm]{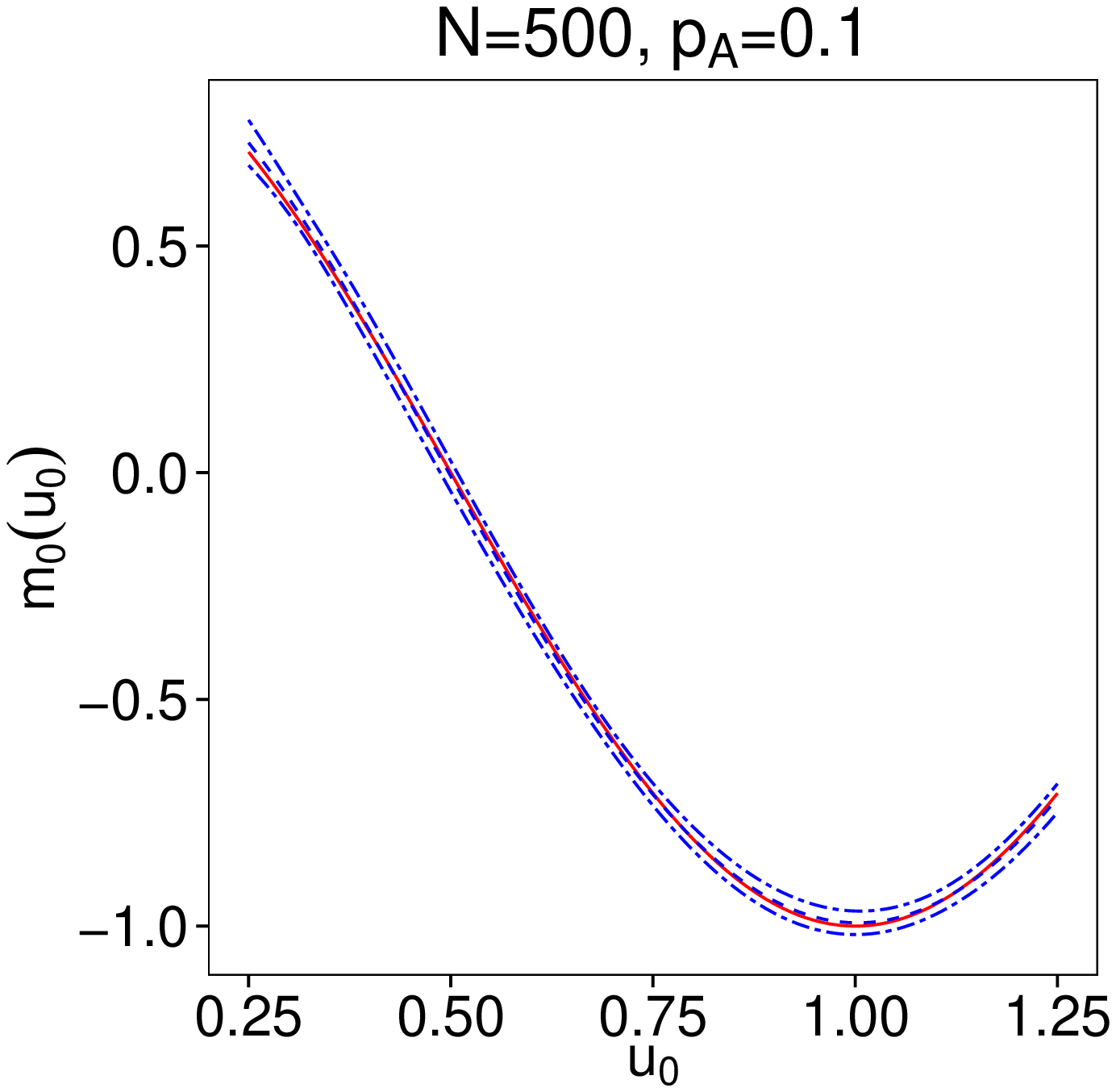}
   \end{minipage}
   \begin{minipage}[t]{0.32\textwidth}
         \includegraphics[width=5.5cm, height=4.5cm]{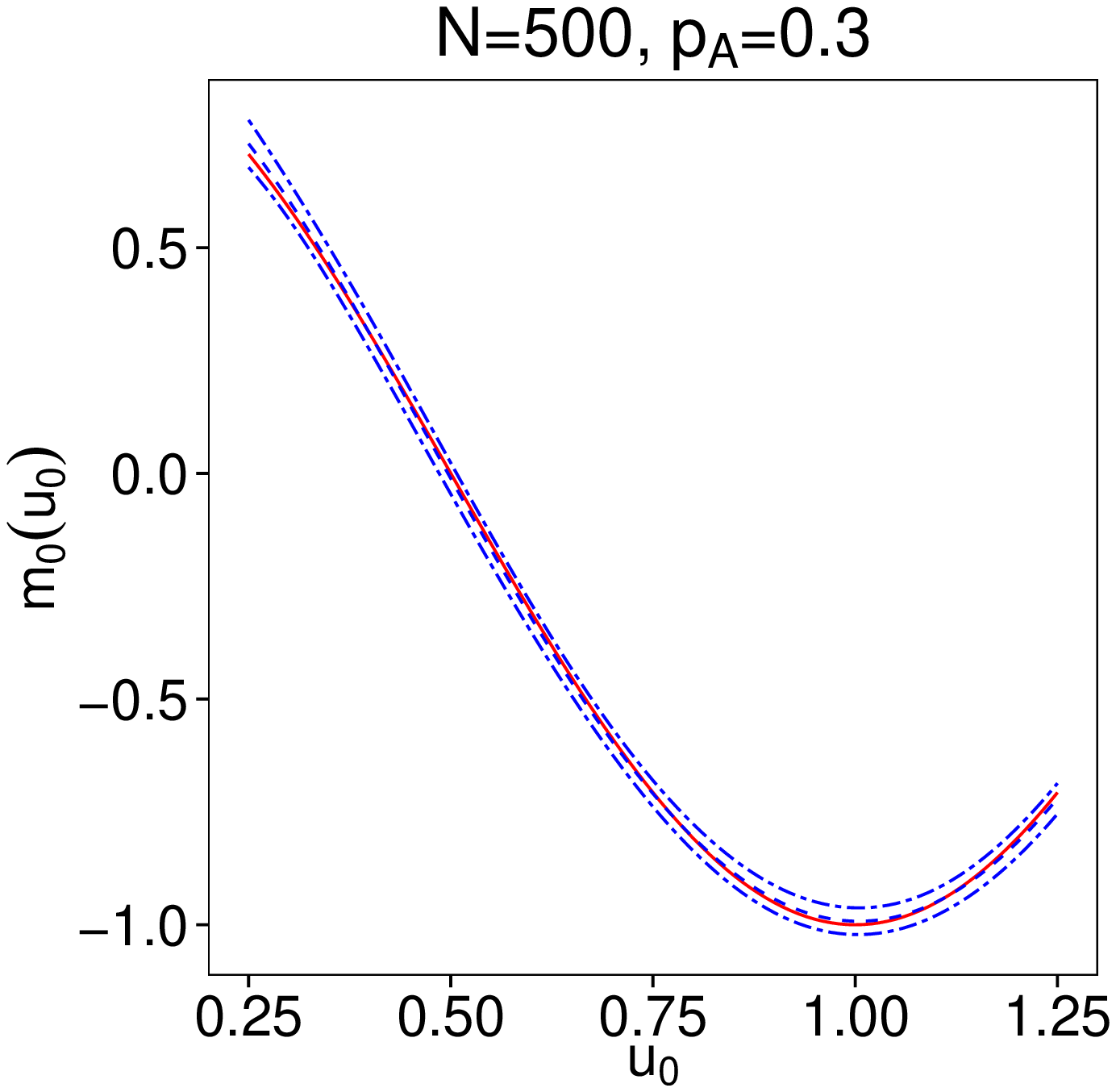}
   \end{minipage}
   \begin{minipage}[t]{0.32\textwidth}
         \includegraphics[width=5.5cm, height=4.5cm]{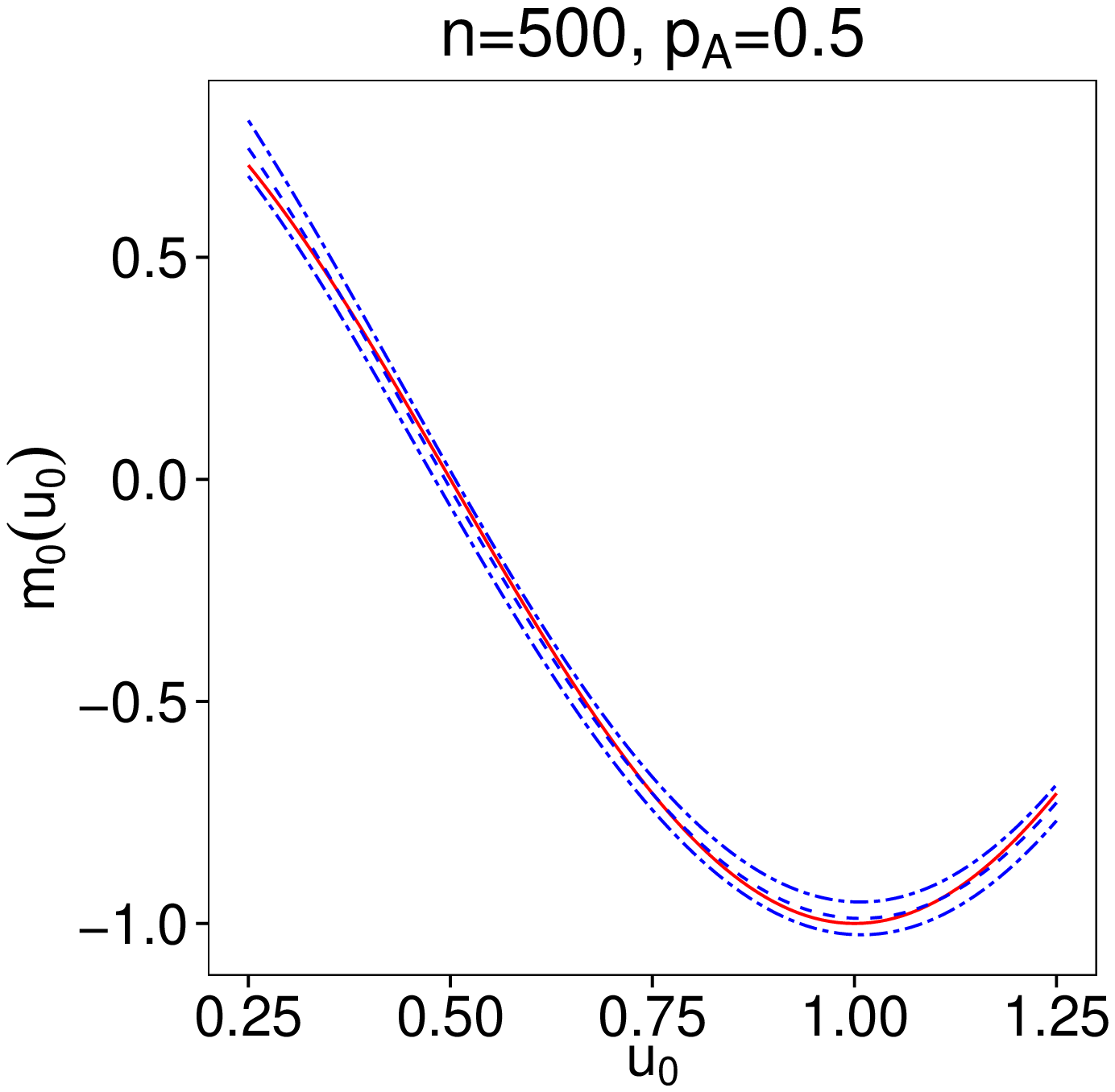}
   \end{minipage}
 }
 \caption{The estimation of function $m_{0}(\cdot)$ under different MAFs when $N$=200, 500 and $\rho$=0.5. The estimated and true functions are denoted by the solid and dashed lines respectively. The 95\% confidence band is denoted by the dotted-dash line.}
 \label{m0rho05}
 \end{figure}

\begin{figure}[h!]\centering
 \centering
 \subfigure{
   \begin{minipage}[t]{0.32\textwidth}
         \includegraphics[width=5.5cm, height=4.5cm]{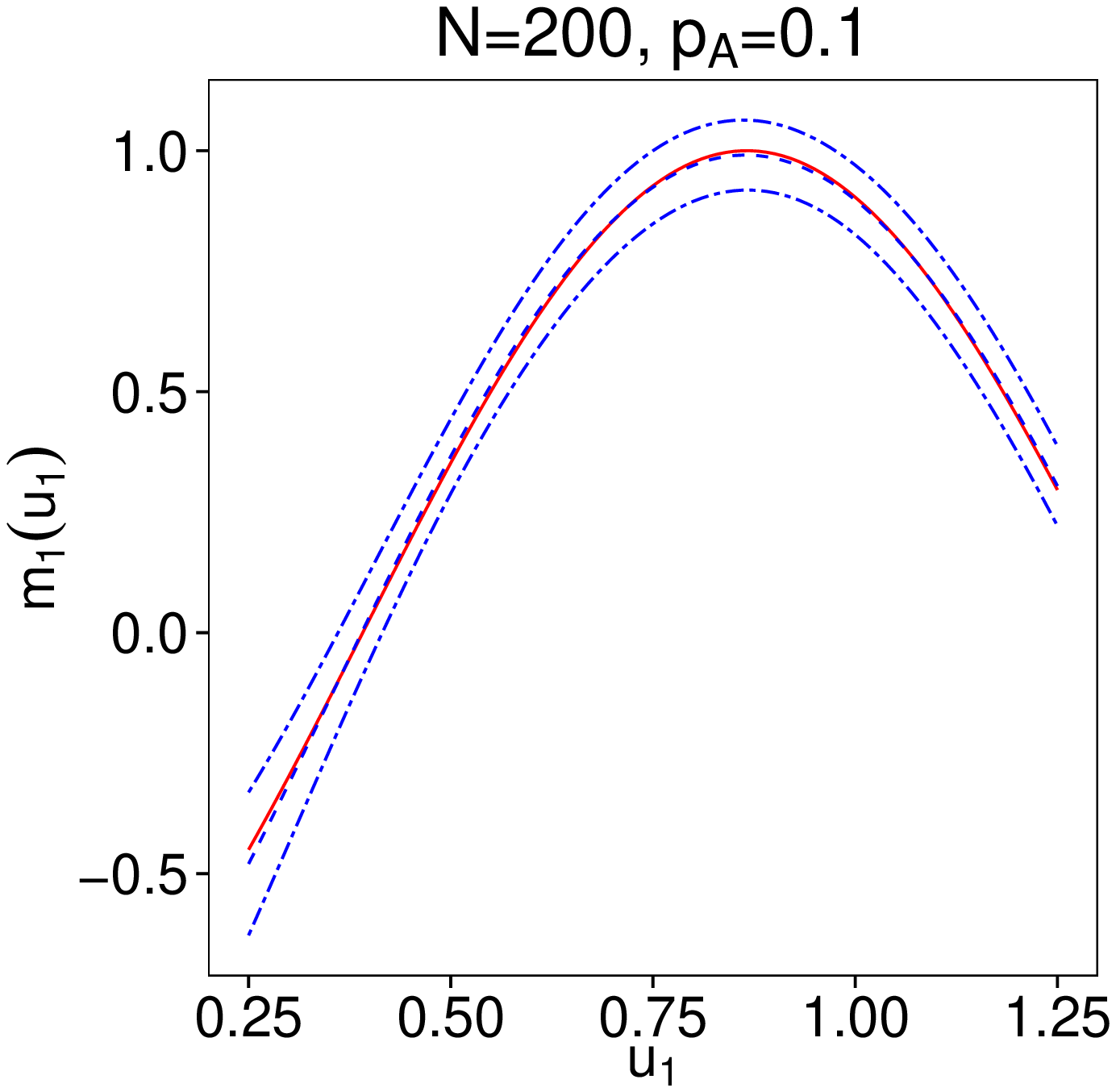}
   \end{minipage}
   \begin{minipage}[t]{0.32\textwidth}
         \includegraphics[width=5.5cm, height=4.5cm]{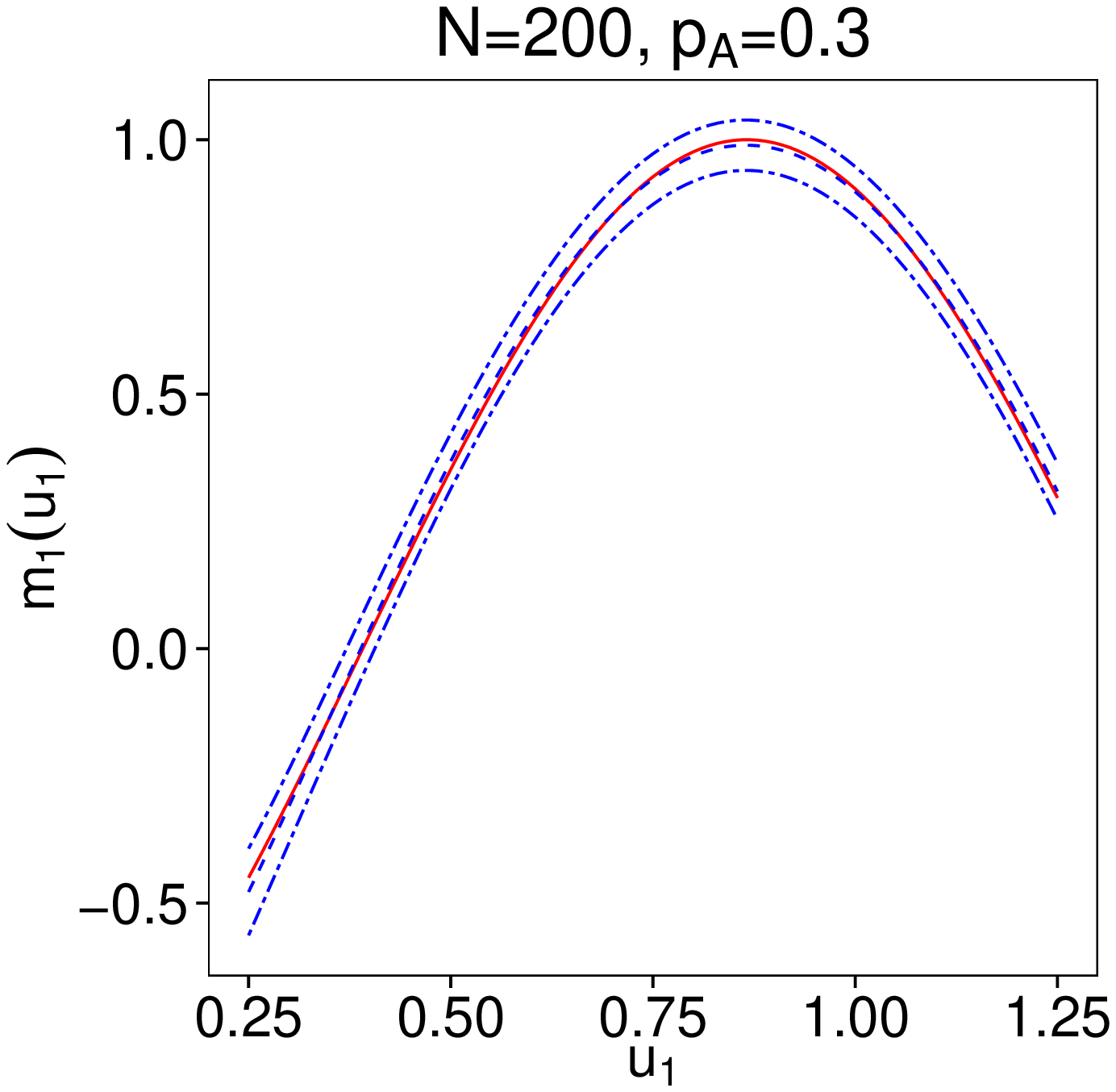}
   \end{minipage}
   \begin{minipage}[t]{0.32\textwidth}
         \includegraphics[width=5.5cm, height=4.5cm]{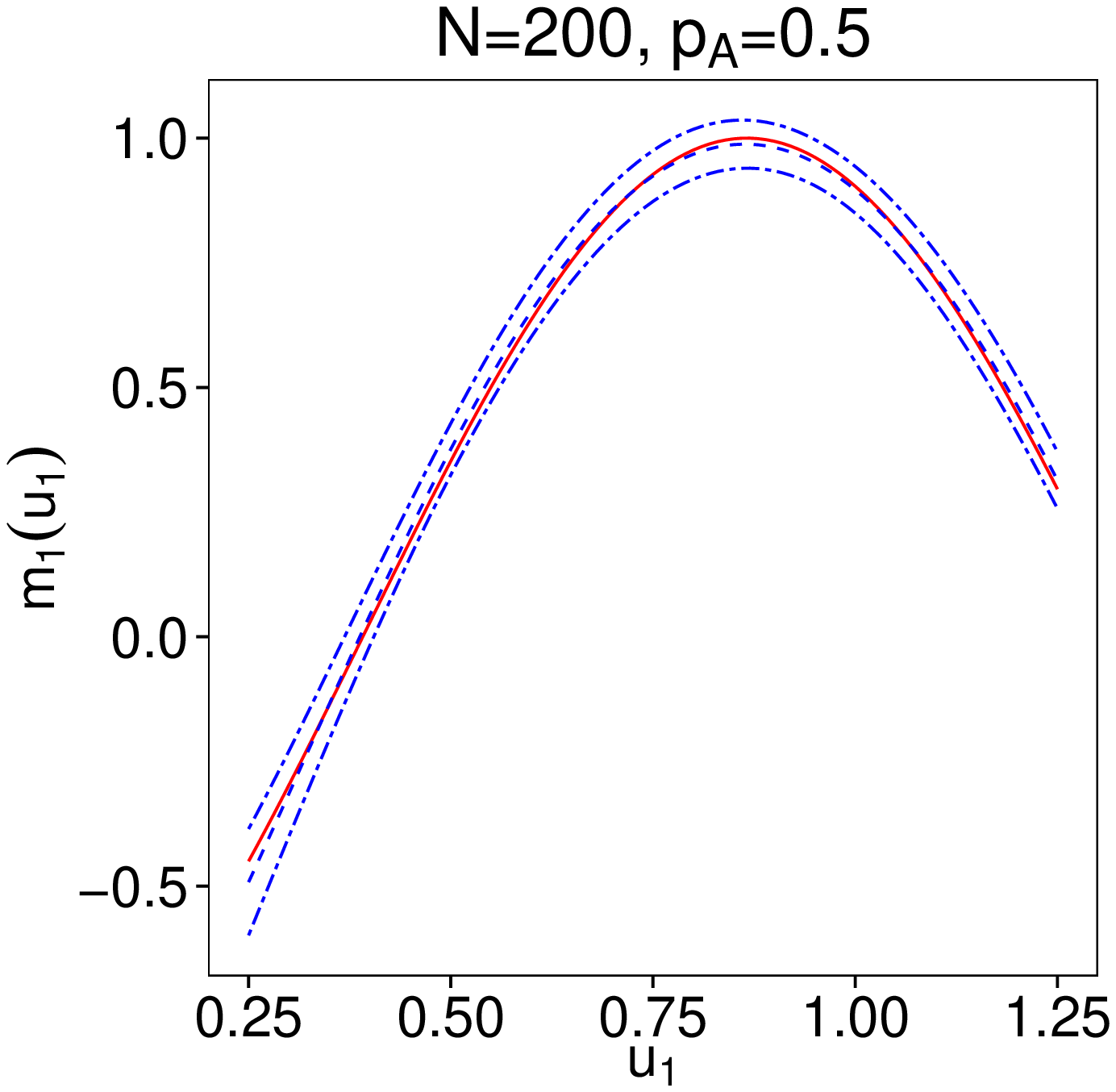}
   \end{minipage}
 }
 \subfigure{
   \begin{minipage}[t]{0.32\textwidth}
         \includegraphics[width=5.5cm, height=4.5cm]{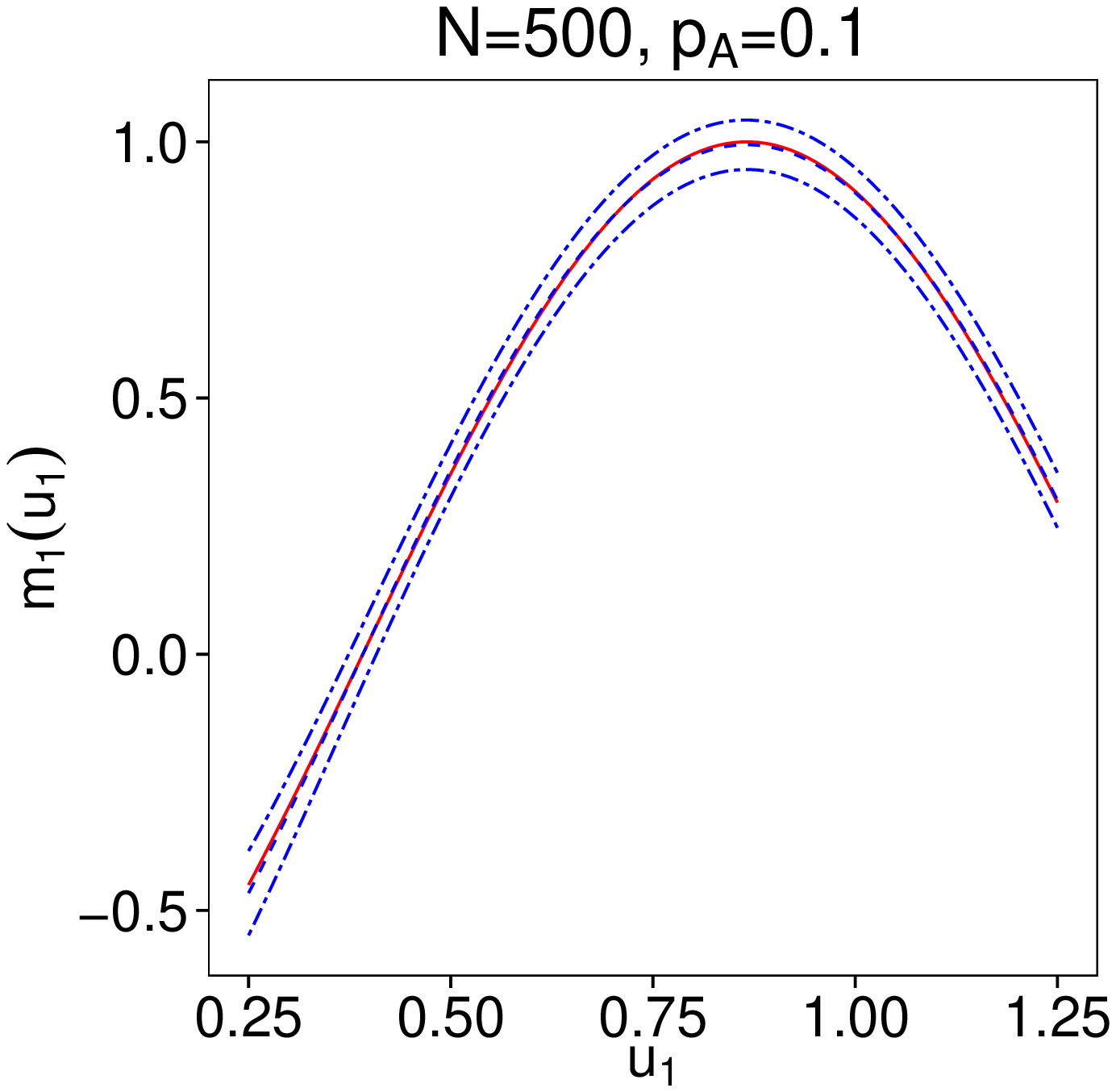}
   \end{minipage}
   \begin{minipage}[t]{0.32\textwidth}
         \includegraphics[width=5.5cm, height=4.5cm]{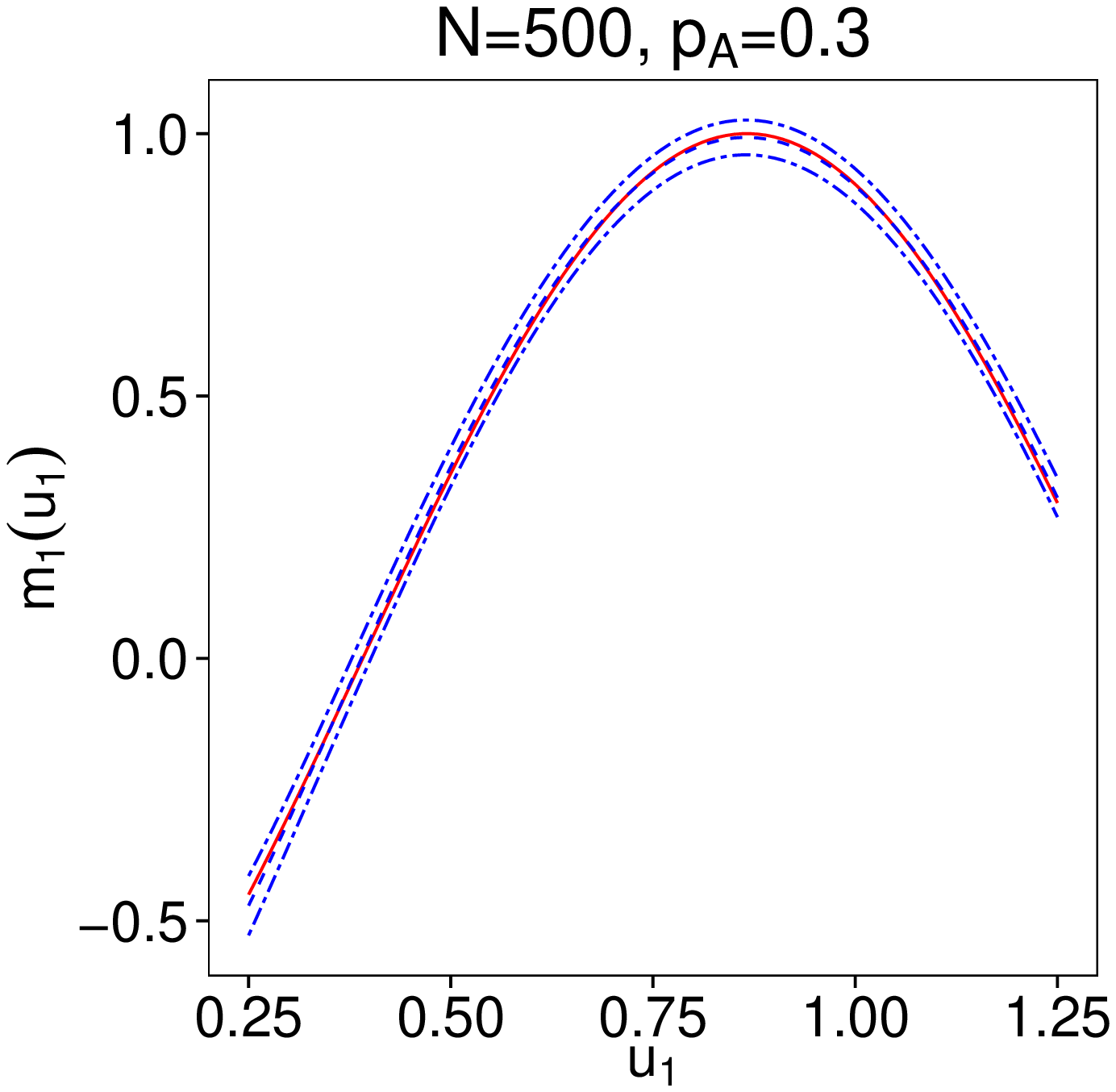}
   \end{minipage}
   \begin{minipage}[t]{0.32\textwidth}
         \includegraphics[width=5.5cm, height=4.5cm]{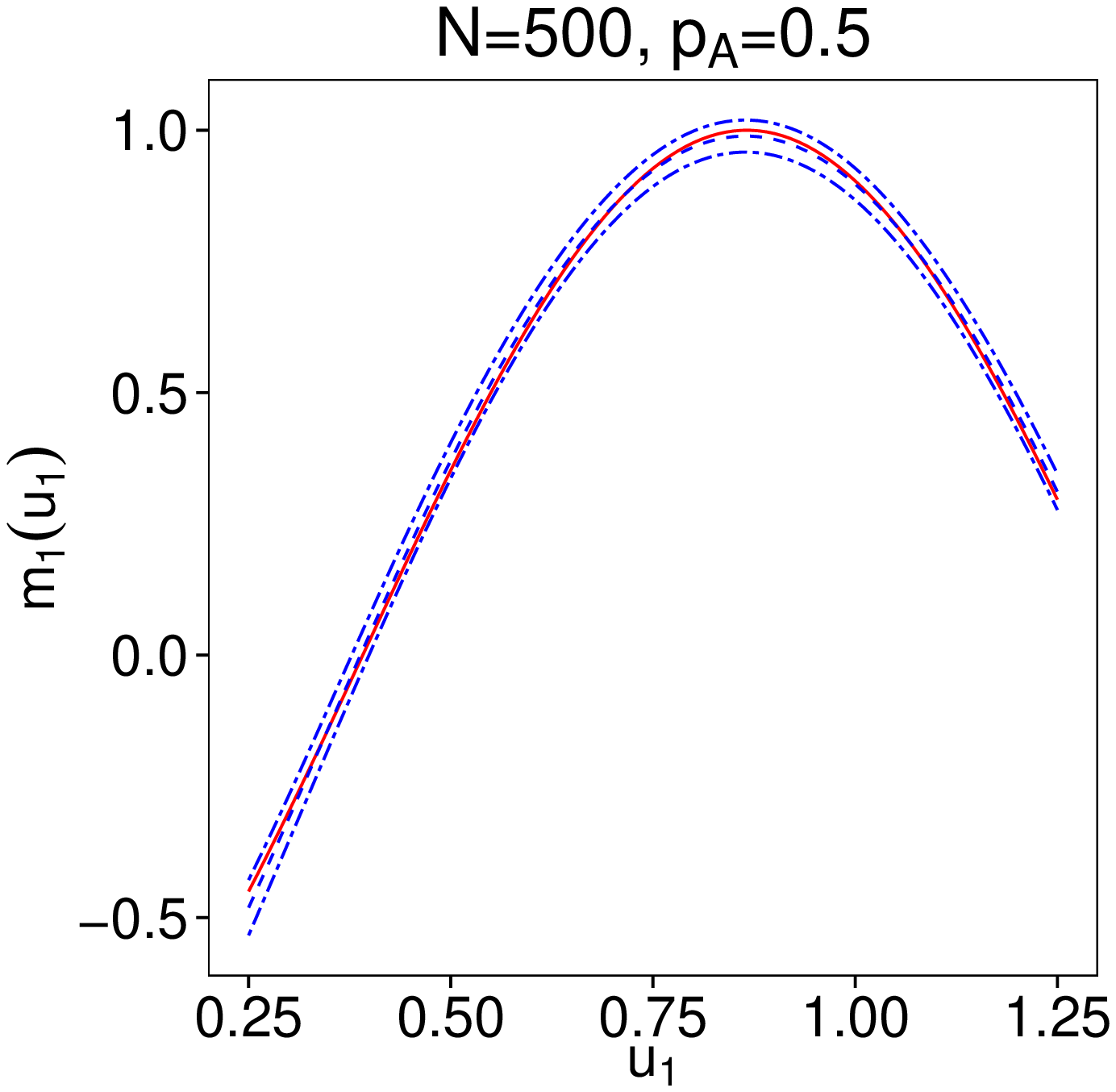}
   \end{minipage}
 }
 \caption{The estimation of function $m_{1}(\cdot)$ under different MAFs when $N$=200, 500 and $\rho$=0.5. The estimated and true functions are denoted by the solid and dashed lines respectively. The 95\% confidence band is denoted by the dotted-dash line.}
 \label{m1rho05}
 \end{figure}

The plots for the estimations of $m_{0}(u_{0})$ and $m_{1}(u_{1})$ under different sample sizes and MAFs are shown in Figure \ref{m0rho05} and Figure \ref{m1rho05}. The estimated and true functions are denoted by the solid and dashed lines, respectively. The $95\%$ confidence band is denoted by the dotted-dash line. The estimated curves almost overlap with the corresponding true curves as shown in the plots. The confidence bands are tight, especially under a large sample size. Note that the estimation for the interaction effects $m_{1}(u_{1})$ improves as MAF $p_{A}$ increases, while the estimation for the marginal effects $m_{0}(u_{0})$ show a opposite direction, which coincides with the results for the parametric estimation in Table \ref{rho05}.

\begin{figure}[H]\centering
\centering
\subfigure{
  \begin{minipage}[t]{0.32\textwidth}
        \includegraphics[width=5.5cm, height=4.5cm]{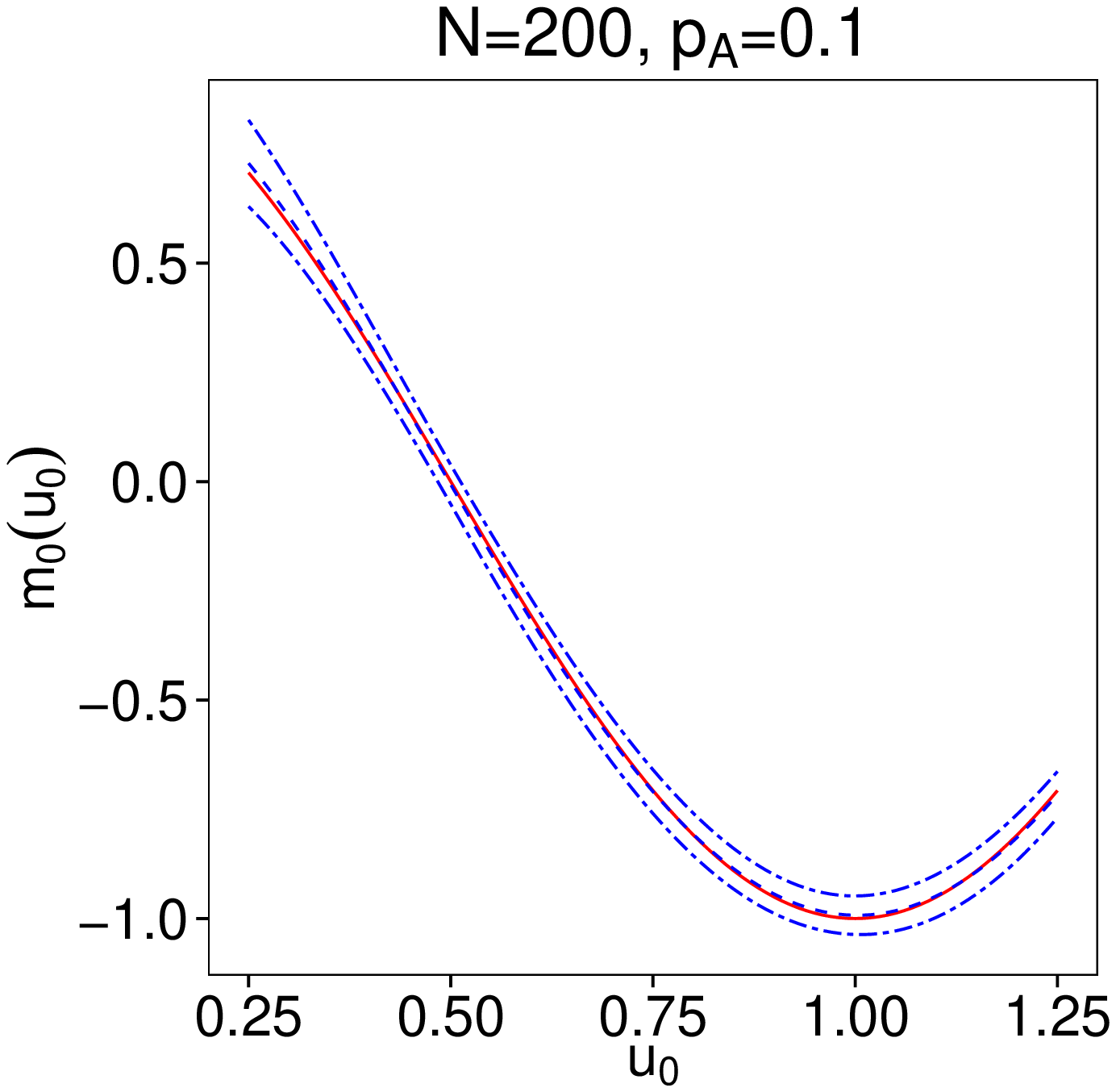}
  \end{minipage}
  \begin{minipage}[t]{0.32\textwidth}
        \includegraphics[width=5.5cm, height=4.5cm]{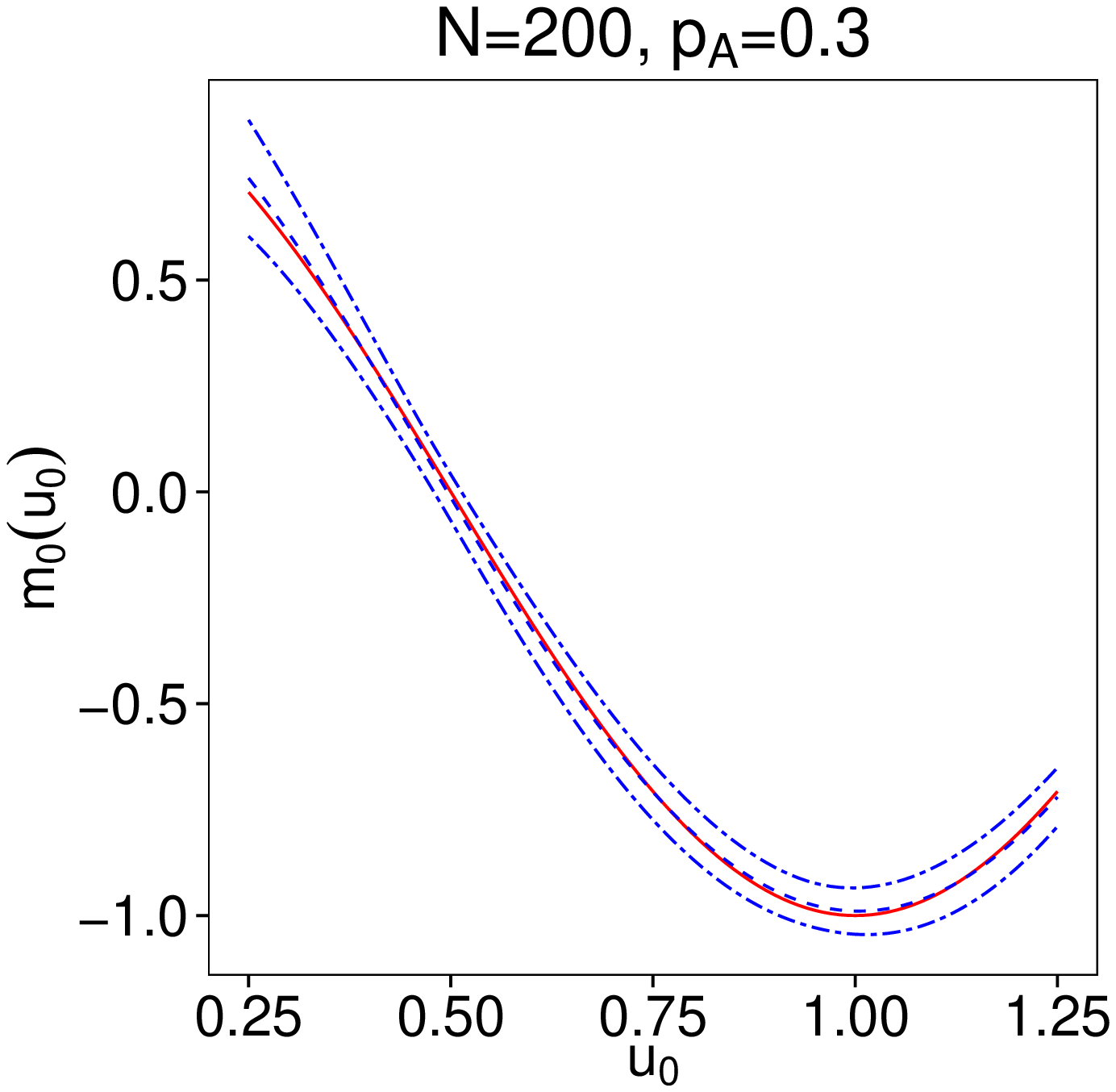}
  \end{minipage}
  \begin{minipage}[t]{0.32\textwidth}
        \includegraphics[width=5.5cm, height=4.5cm]{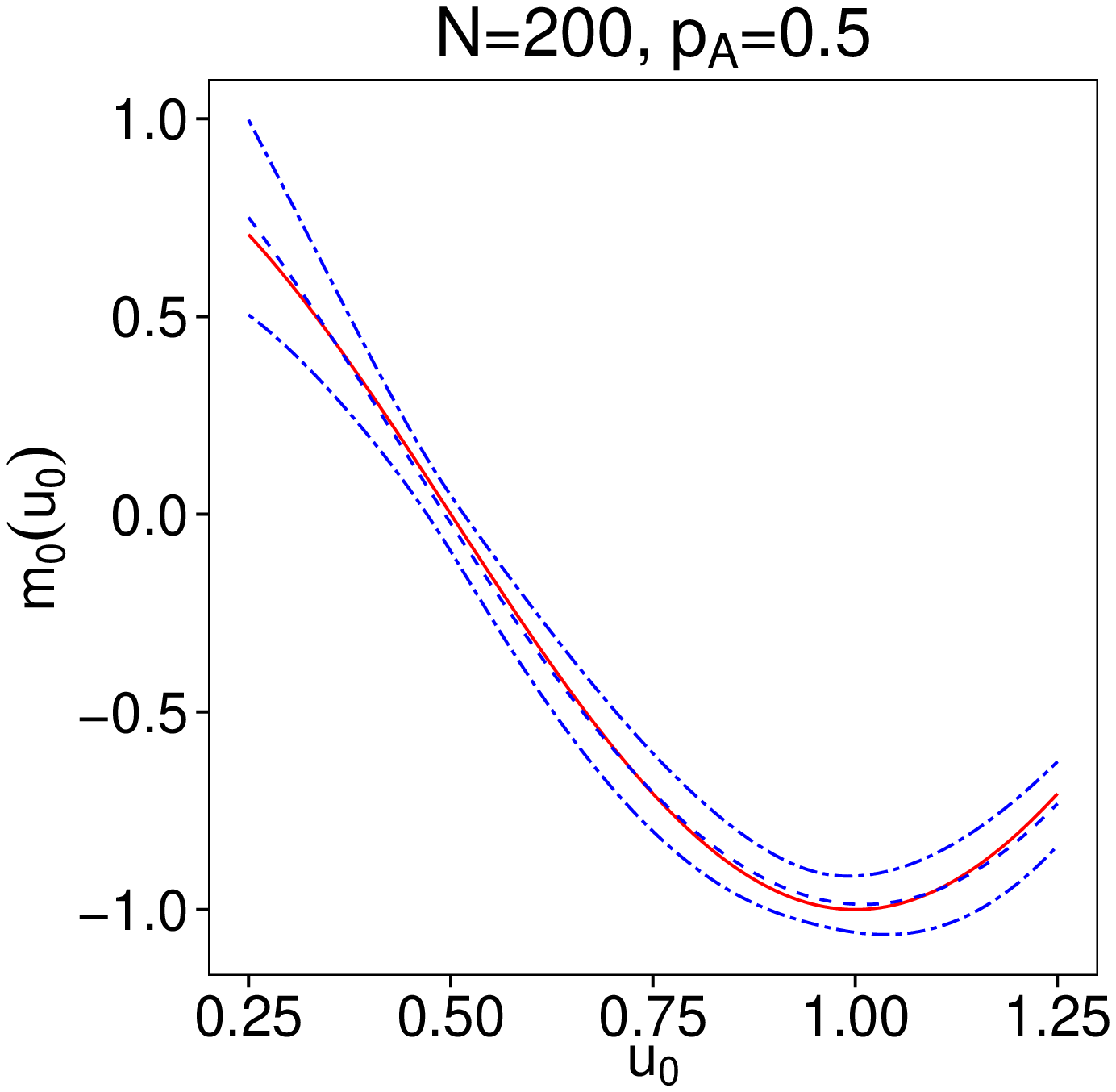}
  \end{minipage}
}
\subfigure{
  \begin{minipage}[t]{0.32\textwidth}
        \includegraphics[width=5.5cm, height=4.5cm]{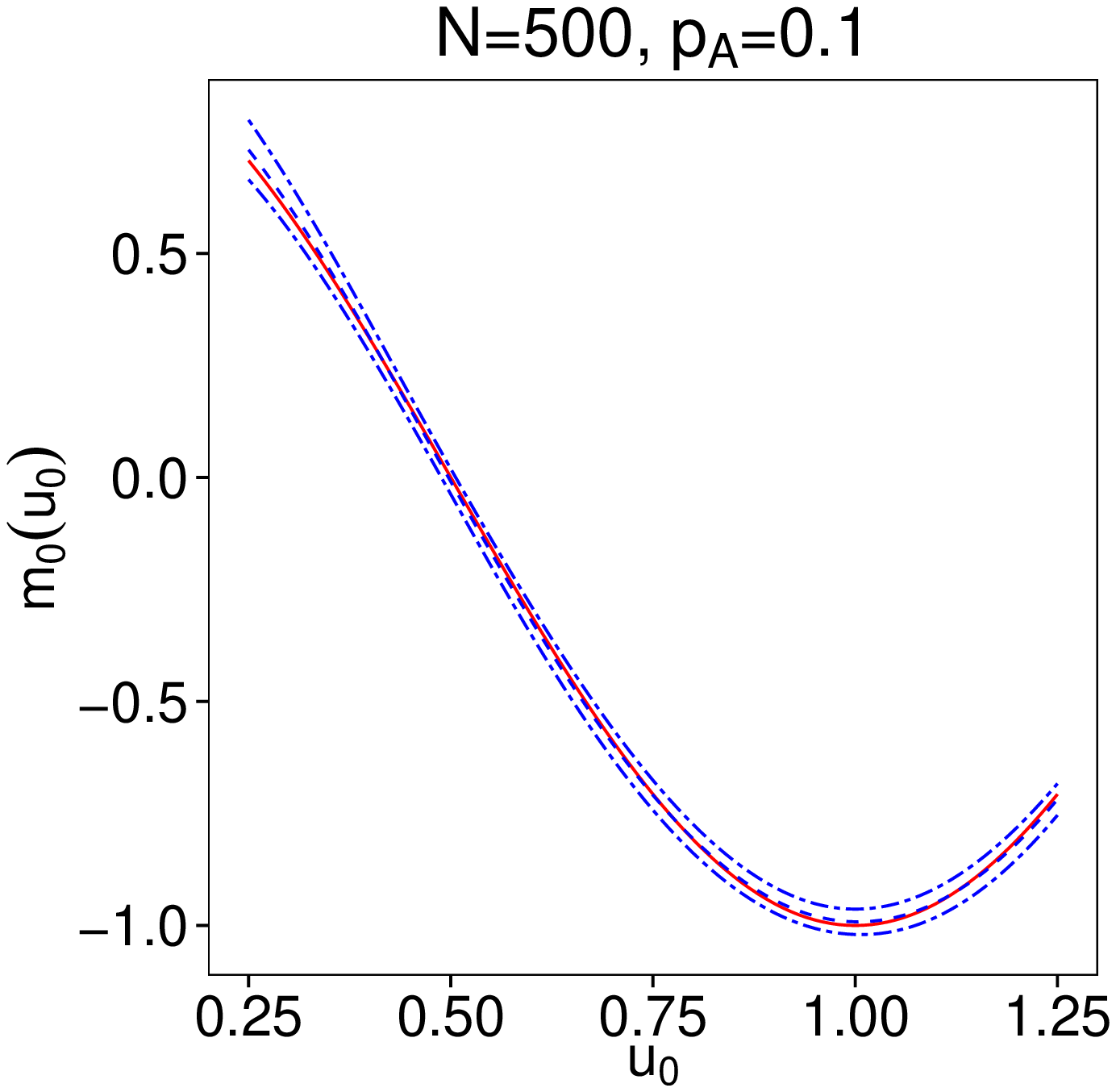}
  \end{minipage}
  \begin{minipage}[t]{0.32\textwidth}
        \includegraphics[width=5.5cm, height=4.5cm]{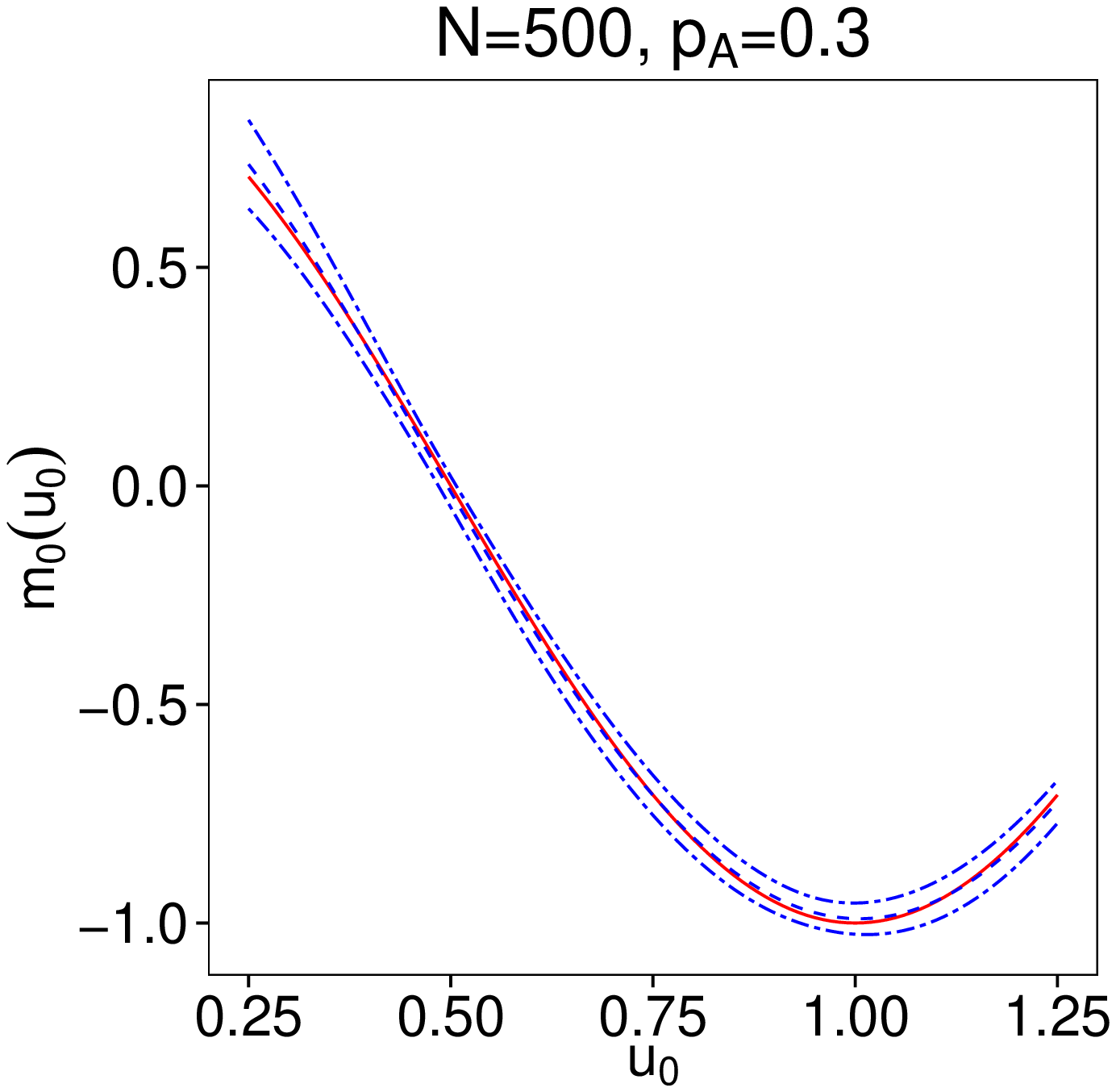}
  \end{minipage}
  \begin{minipage}[t]{0.32\textwidth}
        \includegraphics[width=5.5cm, height=4.5cm]{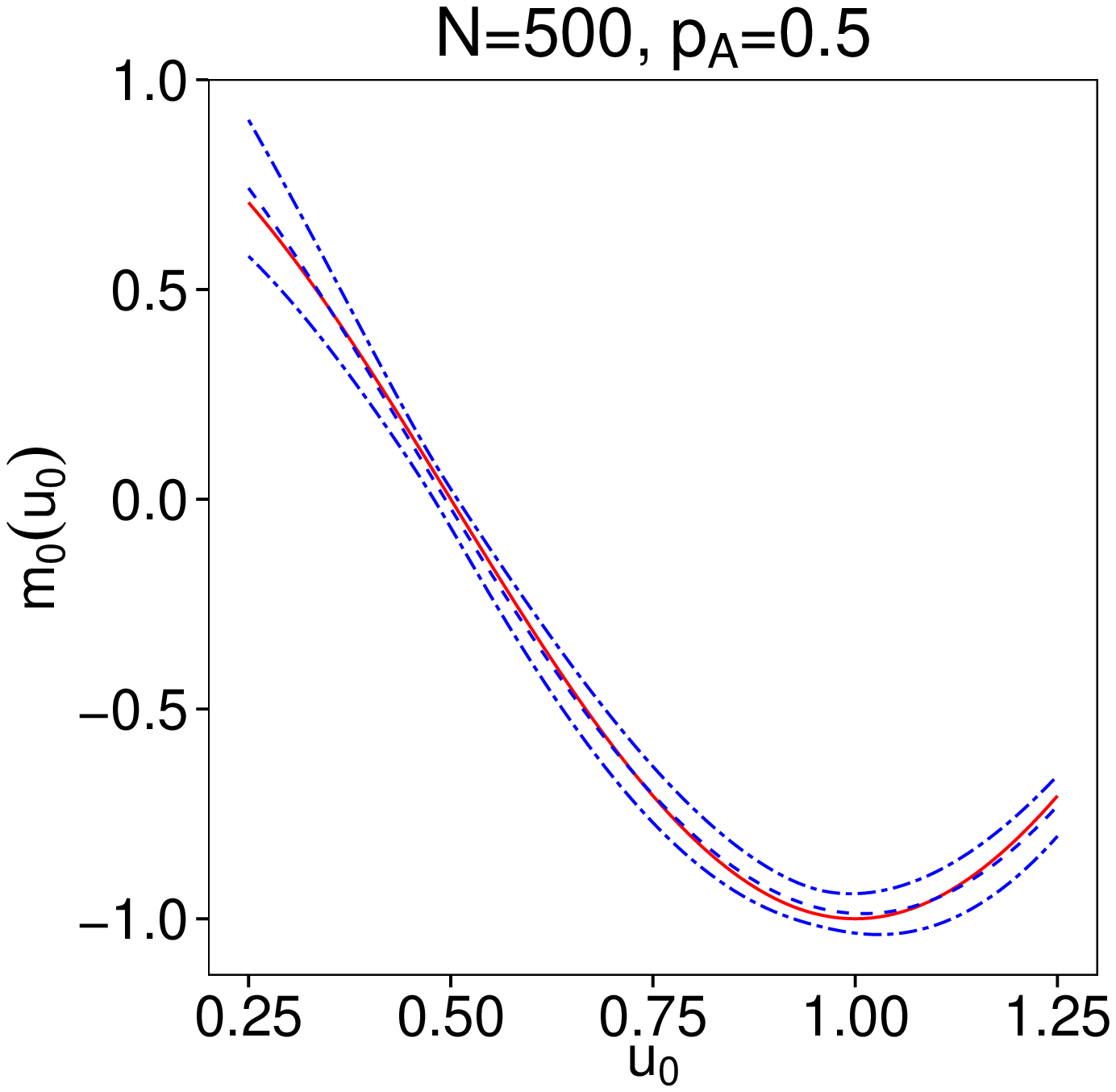}
  \end{minipage}
}

\caption{The estimation of function $m_{0}(\cdot)$ under different MAFs when $N$=200, 500 and $\rho$=0.8. The estimated and true functions are denoted by the solid and dashed
lines respectively. The 95\% confidence band is denoted by the dotted-dash line.}
\label{m0rho08}
\end{figure}

\begin{figure}[H]\centering
\centering
\subfigure{
  \begin{minipage}[t]{0.32\textwidth}
        \includegraphics[width=5.5cm, height=4.5cm]{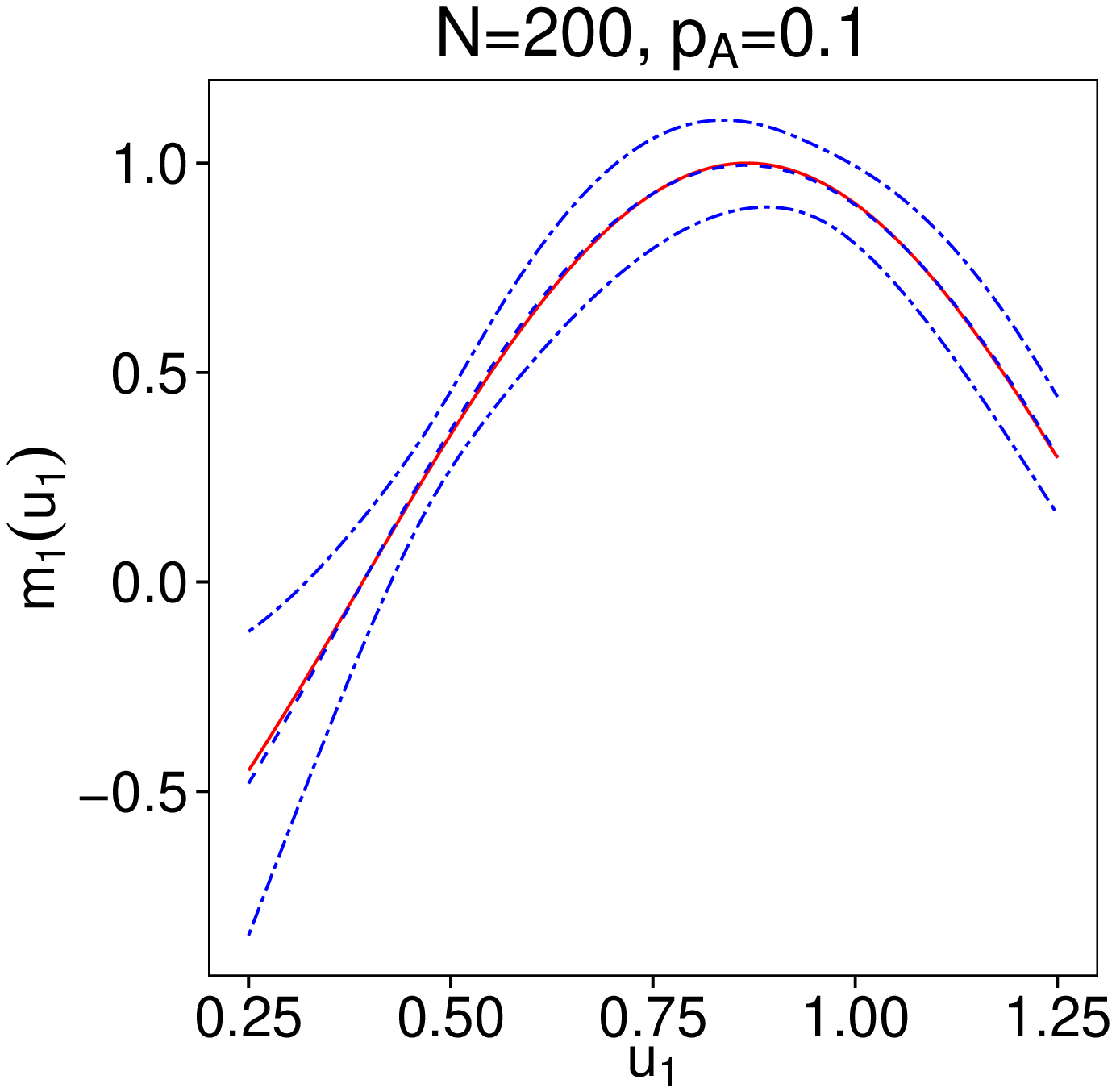}
  \end{minipage}
  \begin{minipage}[t]{0.32\textwidth}
        \includegraphics[width=5.5cm, height=4.5cm]{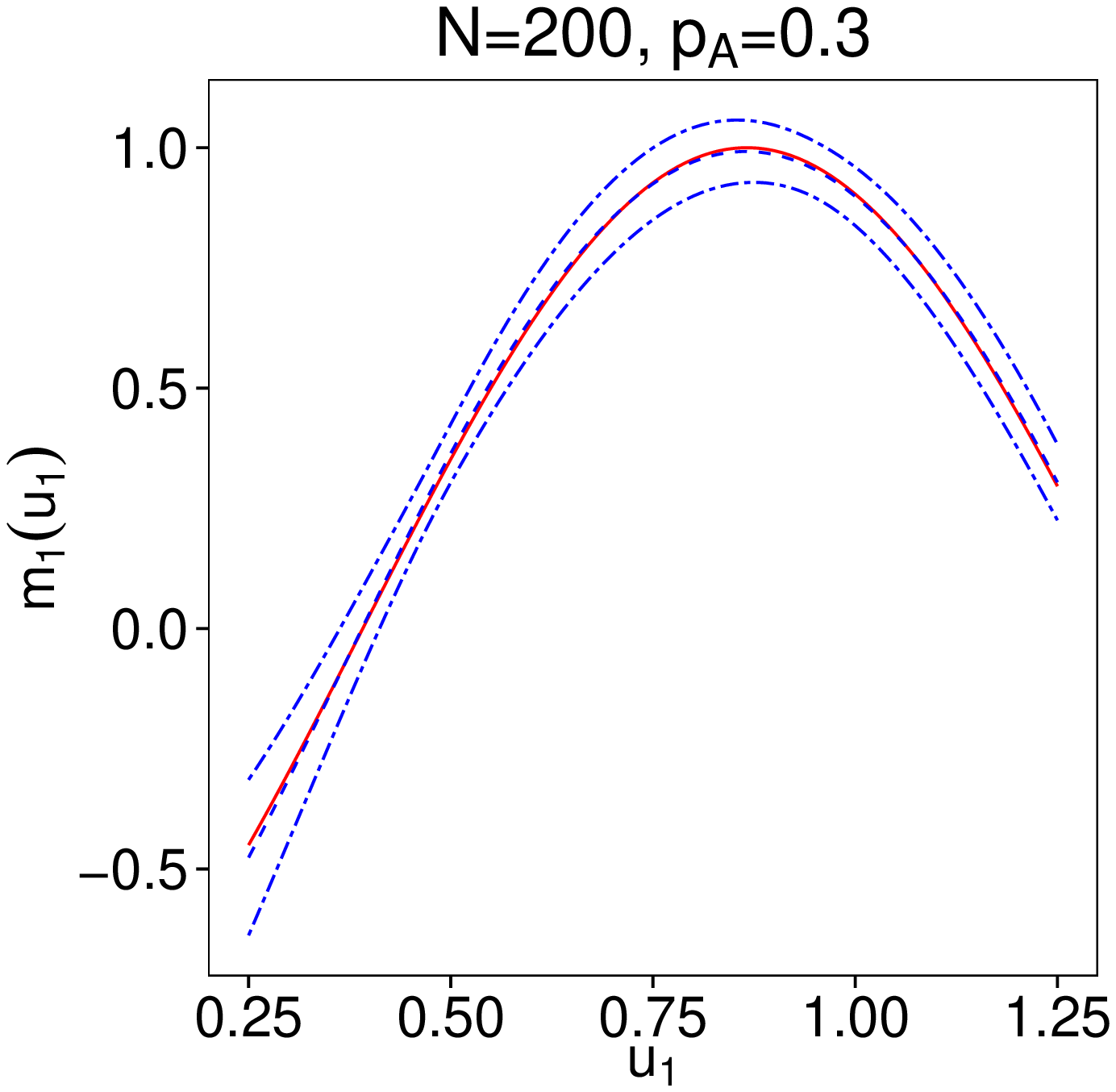}
  \end{minipage}
  \begin{minipage}[t]{0.32\textwidth}
        \includegraphics[width=5.5cm, height=4.5cm]{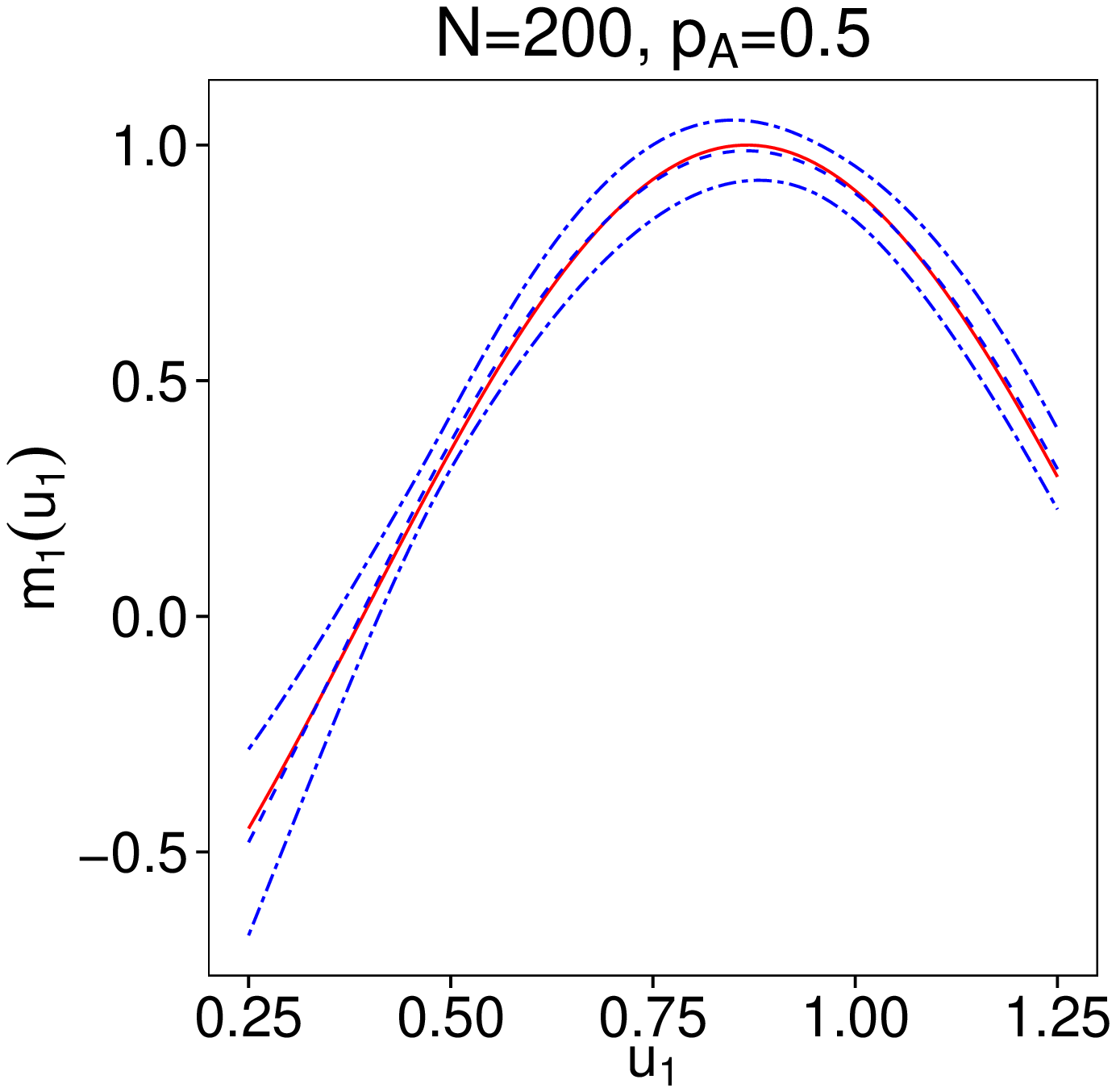}
  \end{minipage}
}
\subfigure{
  \begin{minipage}[t]{0.32\textwidth}
        \includegraphics[width=5.5cm, height=4.5cm]{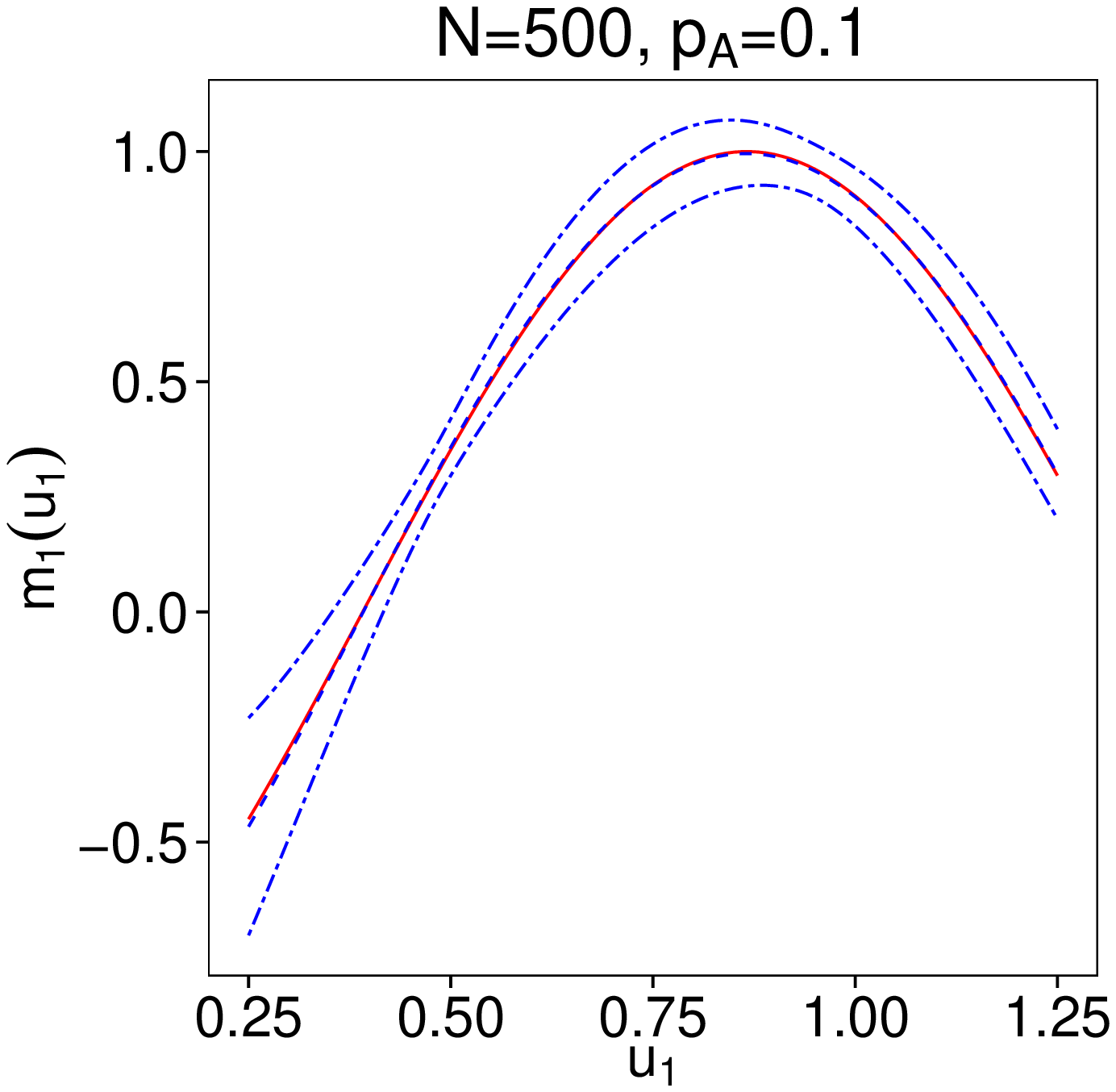}
  \end{minipage}
  \begin{minipage}[t]{0.32\textwidth}
        \includegraphics[width=5.5cm, height=4.5cm]{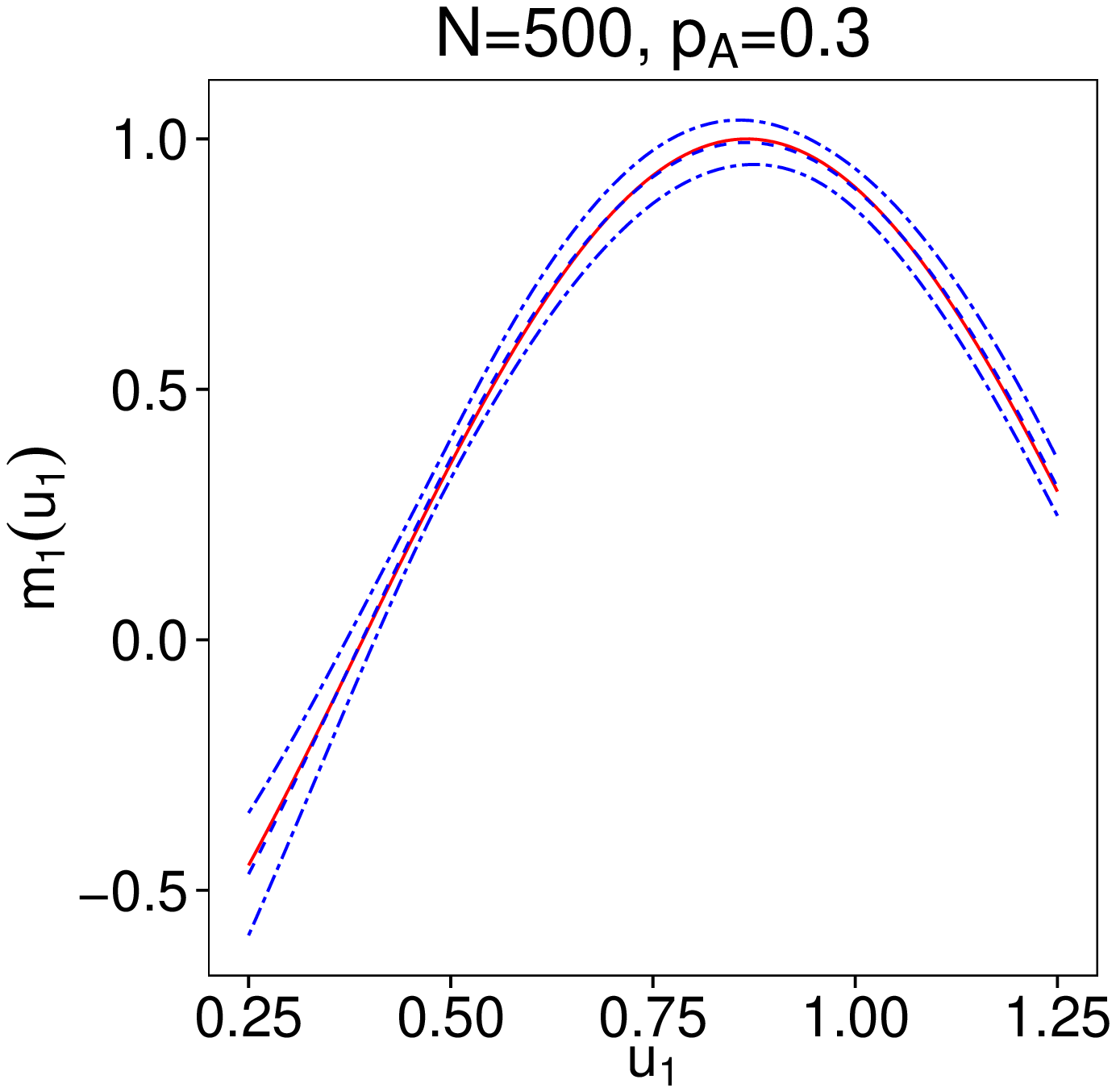}
  \end{minipage}
  \begin{minipage}[t]{0.32\textwidth}
        \includegraphics[width=5.5cm, height=4.5cm]{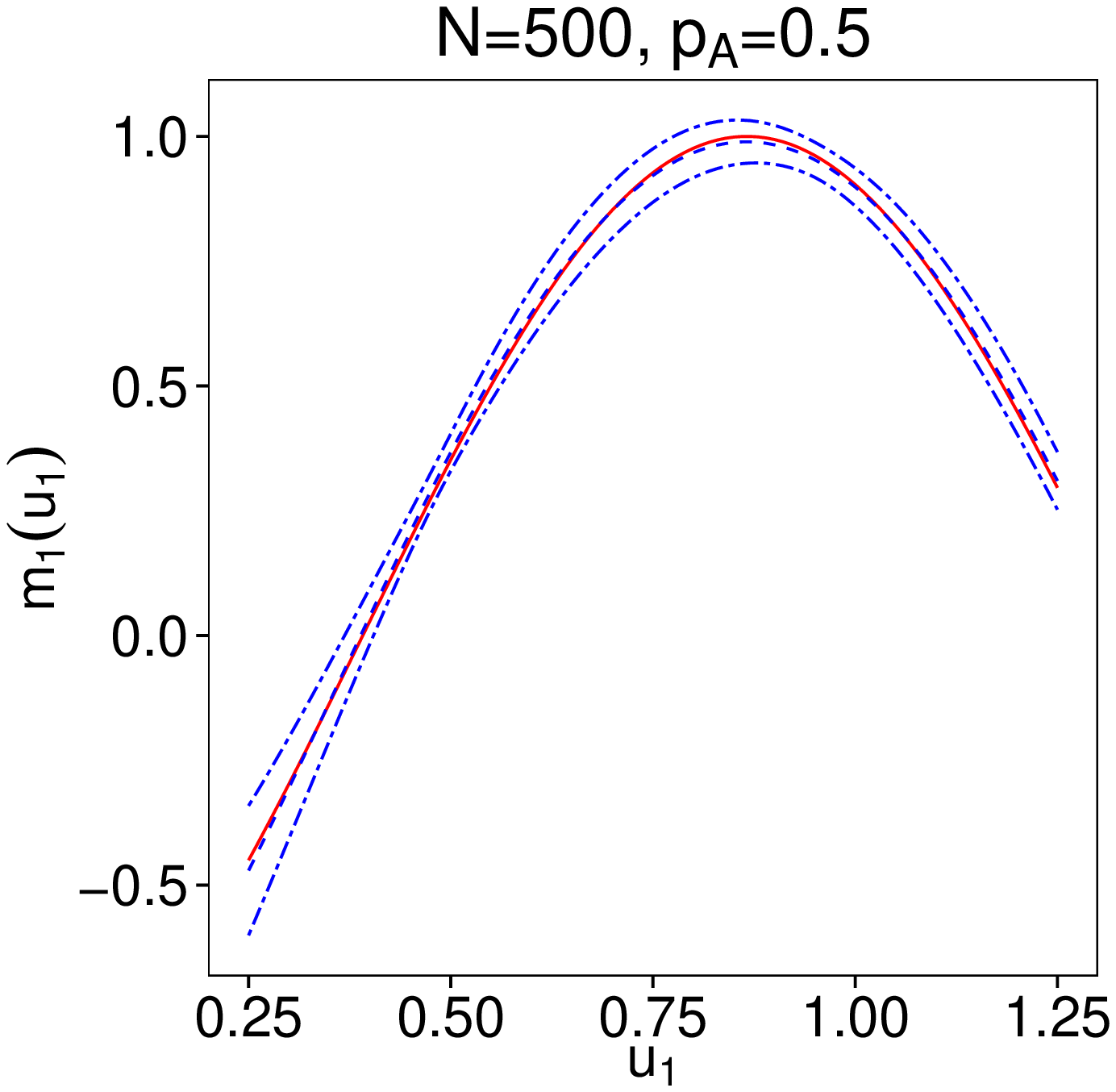}
  \end{minipage}
}
\caption{The estimation of function $m_{1}(\cdot)$ under different MAFs when $N$=200, 500 and $\rho$=0.8 The estimated and true functions are denoted by the solid and dashed
lines respectively. The 95\% confidence band is denoted by the dotted-dash line.}
\label{m1rho08}
\end{figure}

The performance of the estimation for $\rho=0.8$ is shown in Table \ref{rho08}, Figure \ref{m0rho08} and Figure \ref{m1rho08}. It is seen that the SD and SE are smaller when $\rho$ is larger compared to the results when $\rho=0.5$. The confidence bands are a little bit wider, especially for $m_{0}$ when $p_{A}$=0.5 and for $m_{1}$ when $p_{A}$=0.1 for larger $\rho$. In summary, the simulation results show that the estimation method performs reasonably well under different simulation settings in finite samples.

 \begin{table}[H]\centering
 \caption{Simulation results for $p_A = 0.1, 0.3, 0.5$ with sample size $N=200, 500$ and correlation $\rho$=0.8 }{
  \centering
  {\footnotesize \setlength{\tabcolsep}{1.0mm}
 \begin{tabular}{cccccccccccccccccccccccccccccccccccccc}
 \hline
  &&&  \multicolumn{4}{c}{$p_A=0.1$} &    &  \multicolumn{4}{c}{$p_A=0.3$} && \multicolumn{4}{c}{$p_A=0.5$}  \\
   \cline{5-8}\cline{10-13}\cline{15-18}
    $N$&Param& True& & Bias  & SD& SE& CP && Bias & SD& SE& CP && Bias & SD& SE& CP
          \\
 \hline
200&$\beta_{01}$ &    0.620 & & 4.4E-04&    0.005&    0.005&    95.8 & &   5.5E-04&    0.006&    0.006&    95.8& & -5.0E-06&    0.007&    0.007&     95.3   \\
&$\beta_{02}$ &    0.555 & & -2.2E-04&    0.006&    0.005&     92.3 & &   -3.2E-04&    0.007&    0.006&     91.8& & -2.8E-04&    0.008&    0.007&     92.7   \\
&$\beta_{03}$ &    0.555 & & -3.6E-04&    0.006&    0.005&    94.1 & &   -4.0E-04&    0.006&    0.006&     94.6& & 1.4E-04&    0.007&    0.007&      93.7   \\
&$\beta_{11}$ &    0.577 & & -6.7E-05&    0.014&    0.012&     90.3 & &   -2.4E-04&    0.007&    0.007&    94.3& & -7.7E-04&    0.006&    0.006&     92.7   \\
&$\beta_{12}$ &    0.577 & & -2.4E-04&    0.014&    0.012&    91.8 & &    -1.3E-04&    0.007&    0.007&     94.2& & 3.3E-05&    0.006&    0.006&      93.4   \\
&$\beta_{13}$ &    0.577 & & -1.8E-04&    0.014&    0.012&    89.9 & &   2.4E-04&    0.007&    0.007&    93.7& &6.4E-04&    0.006&    0.006&     93.5   \\\\
500 &$\beta_{01}$ &    0.620 & & 5.3E-04&    0.004&    0.003&    94.0 & &   5.8E-04&    0.004&    0.004&    95.4& &3.3E-04&    0.004&    0.005&     95.6   \\
&$\beta_{02}$ &    0.555 & & -4.0E-04&    0.003&    0.003&     93.8 & &   -4.2E-04&    0.004&    0.004&     94.8& &-1.3E-04&    0.005&    0.004&      95.0   \\
&$\beta_{03}$ &    0.555 & & -2.3E-04&    0.004&    0.003&     93.1 & &   -2.8E-04&    0.004&    0.004&     94.2& &-2.9E-04&    0.004&    0.004&      94.7   \\
&$\beta_{11}$ &    0.577 & & 2.5E-04&    0.008&    0.007&     94.0 & &   -1.5E-04&    0.004&    0.004&     95.3& &-6.8E-04&    0.004&    0.004&     93.9   \\
&$\beta_{12}$ &    0.577 & & -2.9E-04&    0.008&    0.007&    93.5 & &   4.4E-05&    0.004&    0.004&    95.7& &4.5E-04&    0.004&    0.004&      93.5   \\
&$\beta_{13}$ &    0.577 & & -1.1E-04&    0.007&    0.007&    95.4 & &   6.1E-05&    0.004&    0.004&    95.4& &1.9E-04&    0.004&    0.004&     95.2   \\

 \hline
 \end{tabular}}}
 \label{rho08}
 \end{table}

\subsection{Performance of hypothesis tests}

We evaluate the performance of the test for the nonparametric function under the null hypothesis $H_{0}: m_{1}(\cdot) = m_{1}^{0}(\cdot)$, where $m_{1}^{0}(u_{1})=\delta_{0}+\delta_{1}u_{1}$, $\delta_{0}$ and $\delta_{1}$ are some constants, which corresponds to a linear G$\times$E interaction. If we fail to reject the null, then a linear model can be fit to further assess the linear G$\times$E interaction. Otherwise, we conclude nonlinear G$\times$E interaction. Power is evaluated under a sequence of alternative models with different values of $\tau$, which is denoted by $H_{1}^{\tau}: m_{1}^{\tau}(\cdot)=m_{1}^{0}(\cdot)+\tau\{m_{1}(\cdot)-m_{1}^{0}(\cdot)\}$. When $\tau$ = 0, the corresponding power is the false positive rate.

\begin{figure}[H]\centering
\centering
\subfigure{
  \begin{minipage}[t]{0.4\textwidth}
        \includegraphics[width=6.5cm, height=6cm]{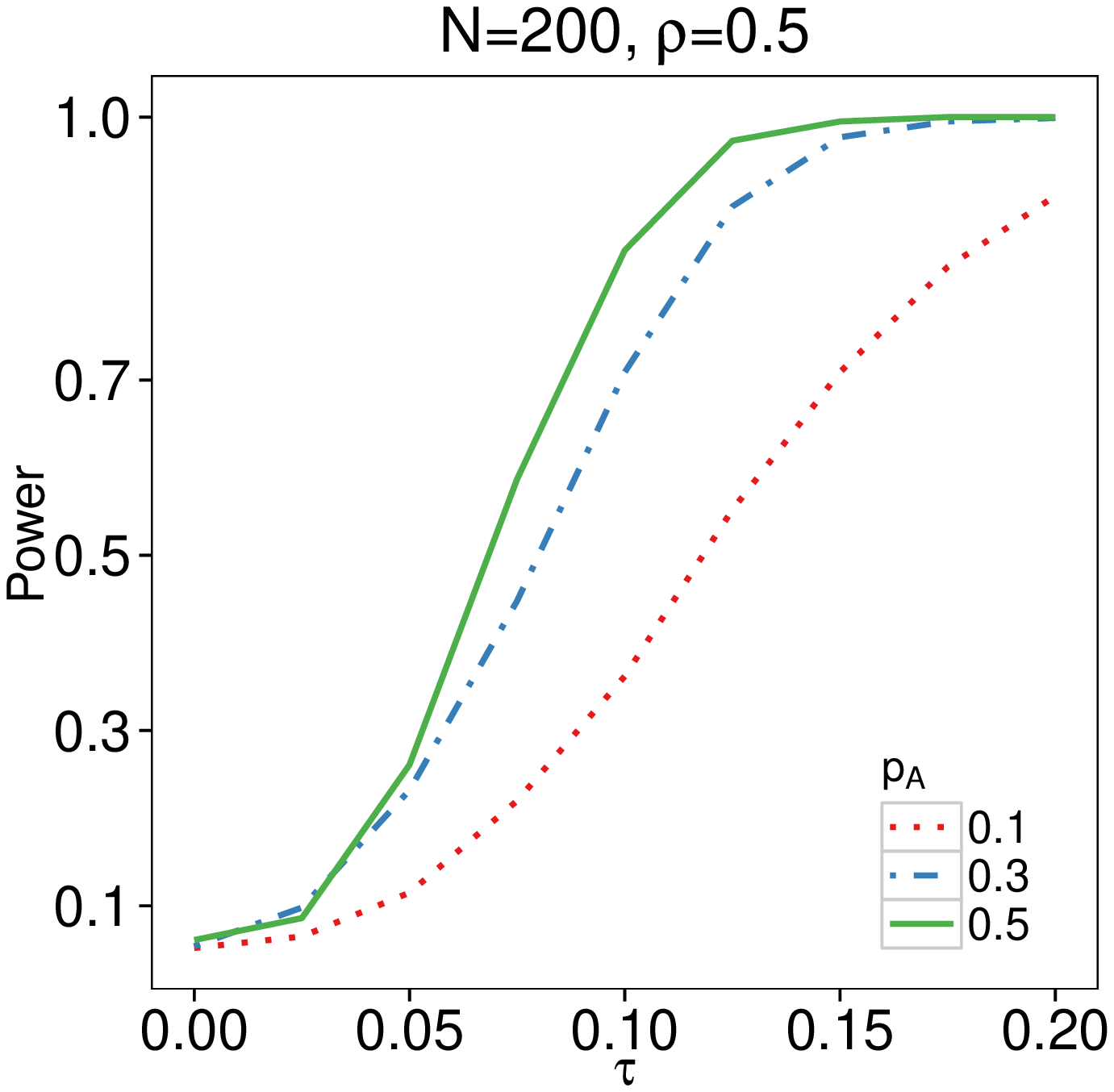}
  \end{minipage}
  ~~~~~~~~~~~
  \begin{minipage}[t]{0.4\textwidth}
        \includegraphics[width=6.5cm, height=6cm]{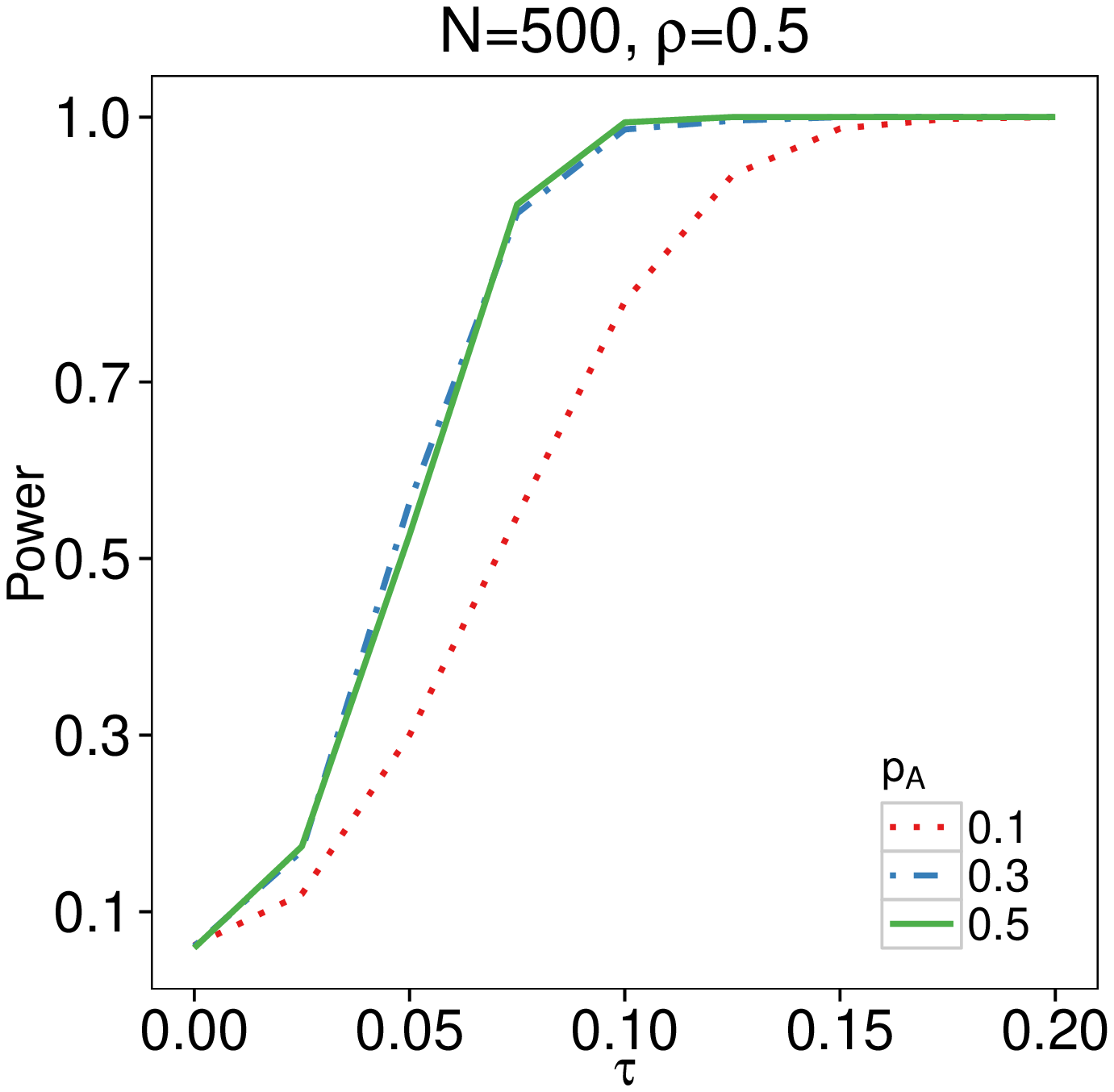}
  \end{minipage}
  }
\caption{The empirical size and power of testing the linearity of nonparametric function $m_{1}$ under different MAFs when $N$=200, 500 and $\rho$=0.5.}
\label{power05}
\end{figure}

\begin{figure}[H]\centering
\centering
\subfigure{
  \begin{minipage}[t]{0.4\textwidth}
        \includegraphics[width=6.5cm, height=6cm]{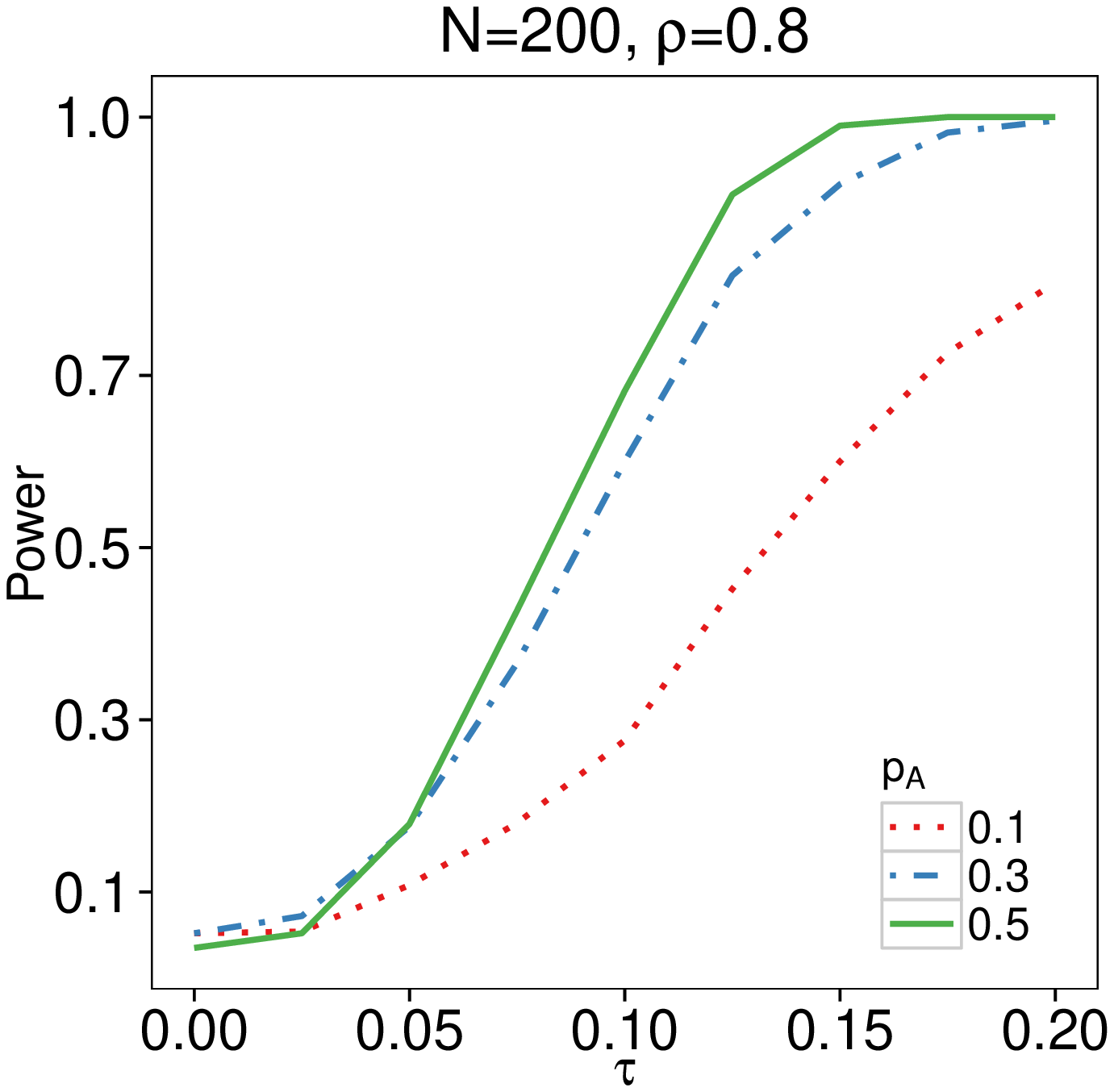}
  \end{minipage}
  ~~~~~~~~~~~~
  \begin{minipage}[t]{0.4\textwidth}
        \includegraphics[width=6.5cm, height=6cm]{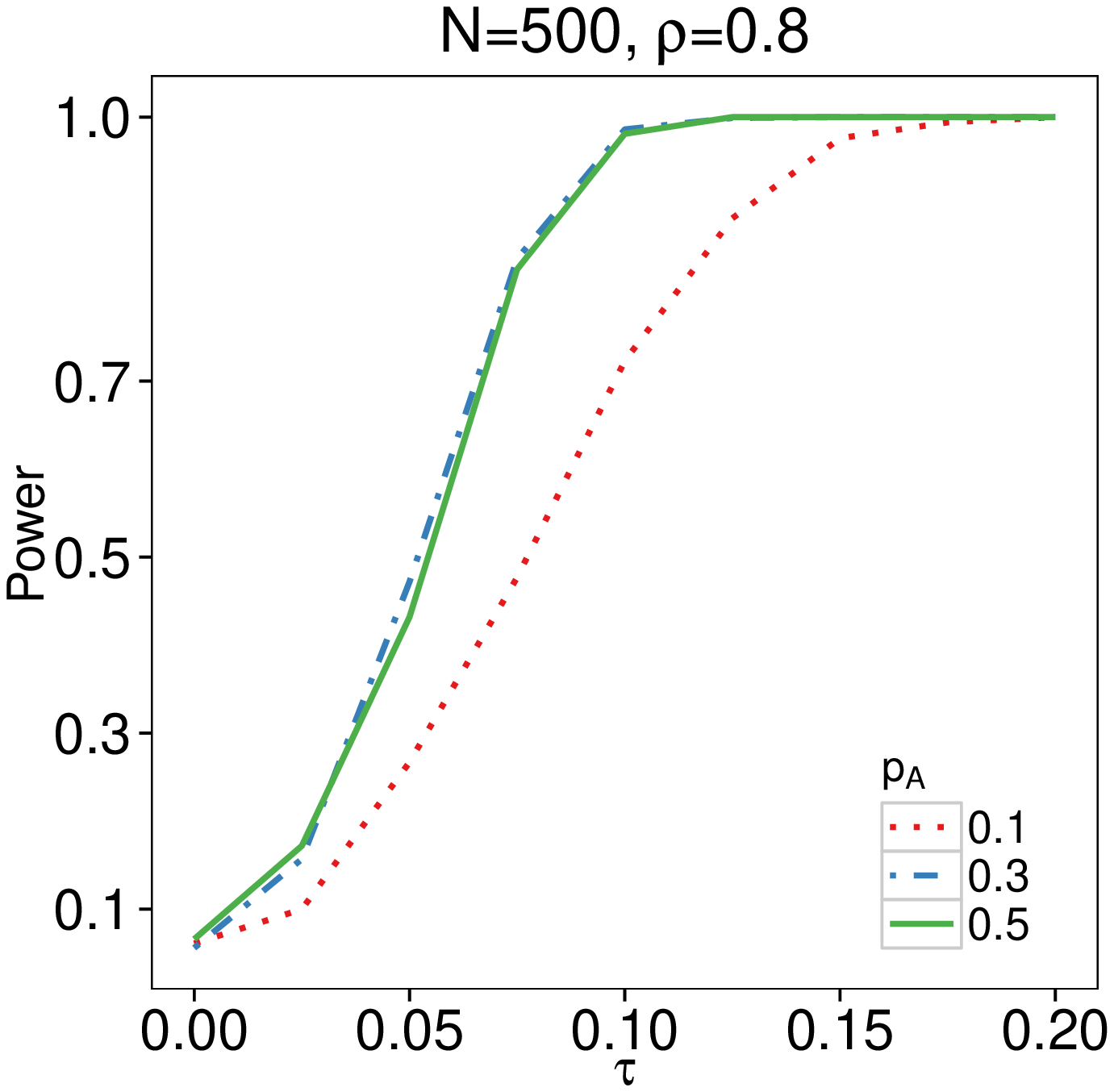}
  \end{minipage}
  }
\caption{The empirical size and power of testing the linearity of nonparametric function $m_{1}$ under different MAFs when $N$=200, 500 and $\rho$=0.8.}
\label{power08}
\end{figure}

Figure \ref{power05} shows the size (when $\tau$ = 0) and power (when $\tau>$ 0) at the 0.05 significance level. We obtain 1000 Monte Carlo simulations each with 5000 replications to access the null distribution of test statistic under sample sizes $N$ = 200, 500 with $\rho$ = 0.5. The empirical Type I error under three MAFs are very close to the nominal level 0.05 and the power increases dramatically when MAF increases from 0.1 to 0.3. Results for $\rho$ = 0.8 is presented in Figure \ref{power08}. Similarly, the empirical Type I error is close to 0.05 and the power increases rapidly when MAF increases from 0.1 to 0.3. Compared to the performance when $\rho$ = 0.5 shown in Figure \ref{power05}, the power increases a little bit slower when $\rho$ = 0.8. The results indicate that our method can reasonably control the false positive rates and has appropriate power to detect the genetic variation.

\section{Real data application}\label{realdata}

We applied the proposed FVICM model to a real data set from a study examining the association of A118G SNPs in an open reading frame of OPRM1 with experimental pain sensitivity (Jonson and Terra, 2002). A group of 163 men and women in ages from 32 to 86 years participated in the study. Systolic blood pressure (SBP), diastolic blood pressure (DBP) and heart rate (HR) were measured at 6 Dobutamine dosage levels for each subject. Dobutamine is a medication that is used to treat congestive heart failure by increasing heart rate and cardiac contractility. Dobutamine was injected into these subjects to investigate their response in heart rate and blood pressure to this drug, at different dosage levels: 0 (baseline), 5, 10, 20, 30 and 40mcg/min. In this study, dosage levels were treated as ``time" and measurements at different dosage levels were considered as longitudinal measurements. In addition to that, age and body mass index (BMI) were also recorded.

Five SNPs in genes $Beta_{1}AR$ and $Beta_{2}AR$ were genotyped, namely, $\textit{codon16}$, $\textit{codon27}$, $\textit{codon49}$, $\textit{codon389}$, and $\textit{codon492}$. We chose $X_{1}$= dosage level as the ``time-varying" variable, and $X_{2}=$ age and $X_{3}$= BMI as the ``time-invariant" variables. Our goal was to evaluate how the SNPs interact with age, BMI and dose level to affect SBP, DBP and HR. With the proposed FVICM model, we can model the dynamic gene effect on drug response under different dosage levels.

In this analysis, we tested whether any SNP was associated with the drug response in a linear fashion based on the hypothesis test $H_{0}: m_{1}(u_{1})$ = $\delta_{0}+\delta_{1}u_{1}$ with p-value denoted by $p_{m_{1}}$ in Table \ref{SBP} - \ref{HR}. We also reported the p-values for testing the significance of the index loading coefficients $\beta_{11}$, $\beta_{12}$ and $\beta_{13}$, which were labeled by $p_{\beta_{11}}$, $p_{\beta_{12}}$ and $p_{\beta_{13}}$, based on the asymptotic normality of the estimates. We also compared our proposed model to an additive varying-coefficient model (AVCM) $E(Y|\mathbf{X}, G)=\beta_{01}^{*}(X_{1})+\beta_{02}^{*}X_{2}+\beta_{03}^{*}X_{3}+\{\beta_{11}^{*}(X_{1})+\beta_{12}^{*}X_{2}+\beta_{13}^{*}X_{3}\}G$, where $\beta_{01}^{*}(\cdot)$ and $\beta_{11}^{*}(\cdot)$ are unknown functions of $X_{1}$. To see the relative gain by integrative analysis, we calculated the MSEs of both models. The p-values for testing $H_{0}: \beta_{11}^{*}(\cdot)=\beta_{12}^{*}=\beta_{13}^{*}=0$ for AVCM is also reported in the tables and denoted by $p_{AVCM}$.

\begin{figure}[H]\centering
 \centering
 \subfigure{
   \begin{minipage}[t]{0.32\textwidth}
         \includegraphics[width=5.5cm, height=5cm]{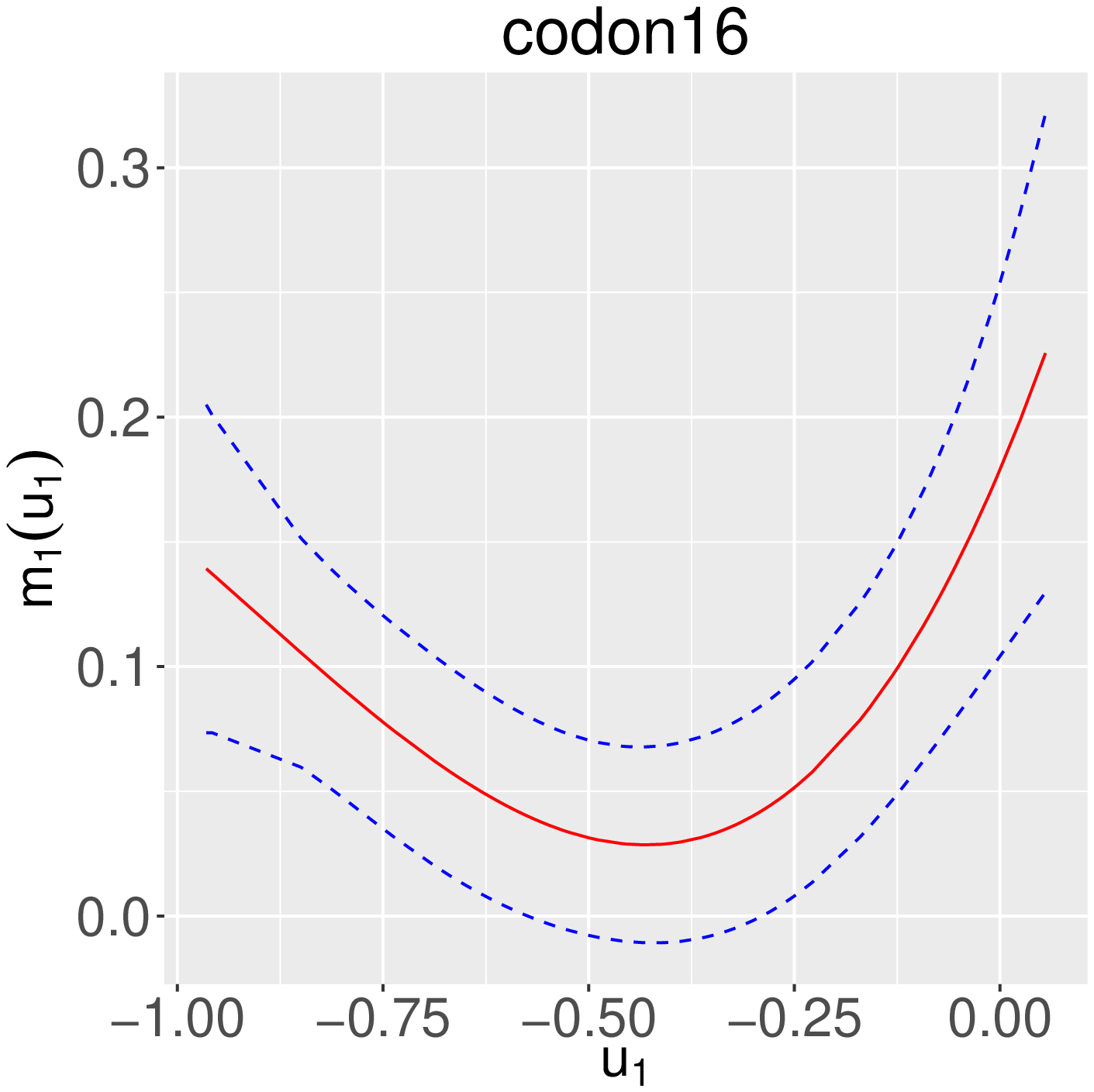}
   \end{minipage}
   \begin{minipage}[t]{0.32\textwidth}
         \includegraphics[width=5.5cm, height=5cm]{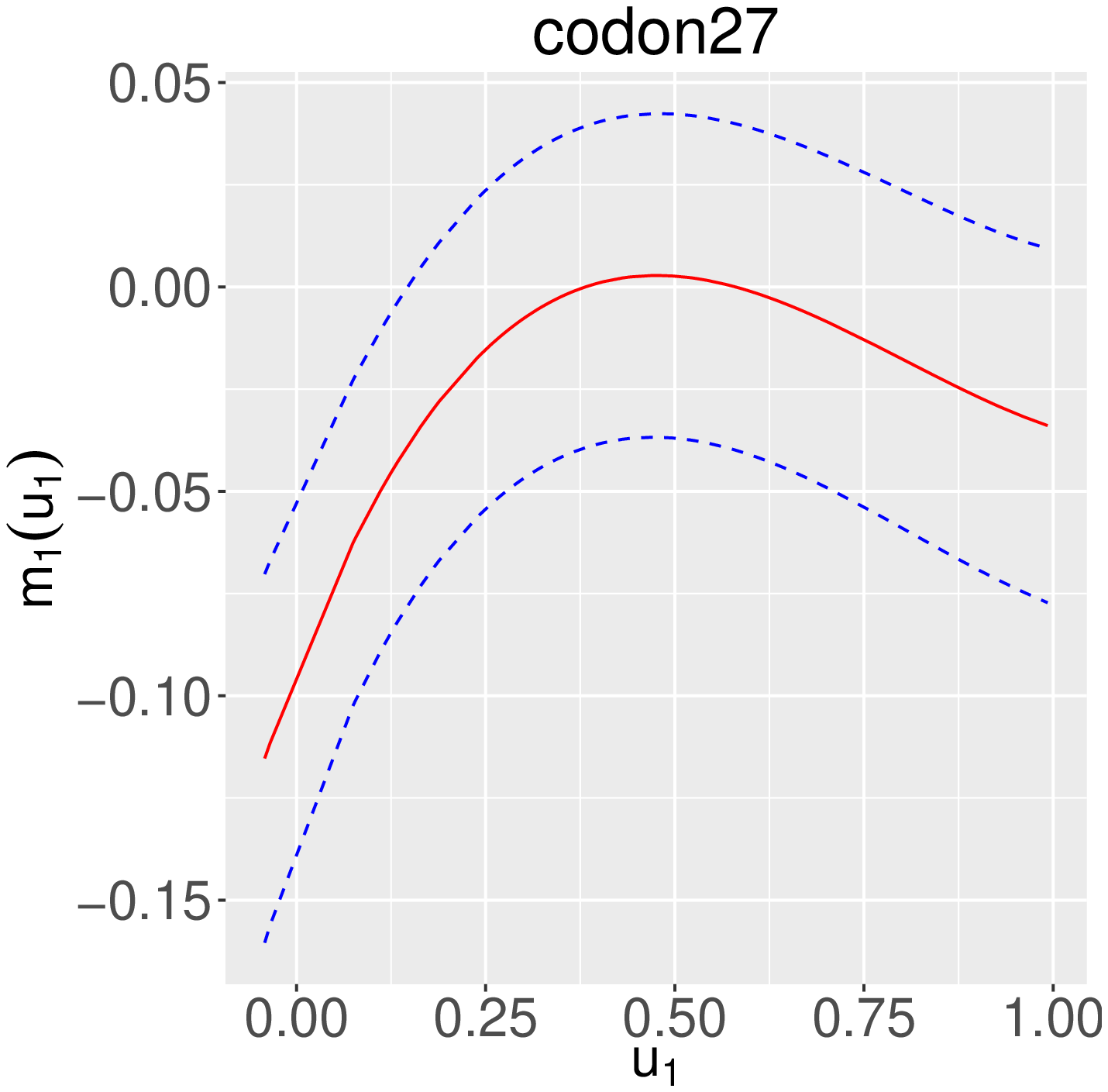}
   \end{minipage}
   \begin{minipage}[t]{0.32\textwidth}
         \includegraphics[width=5.5cm, height=5cm]{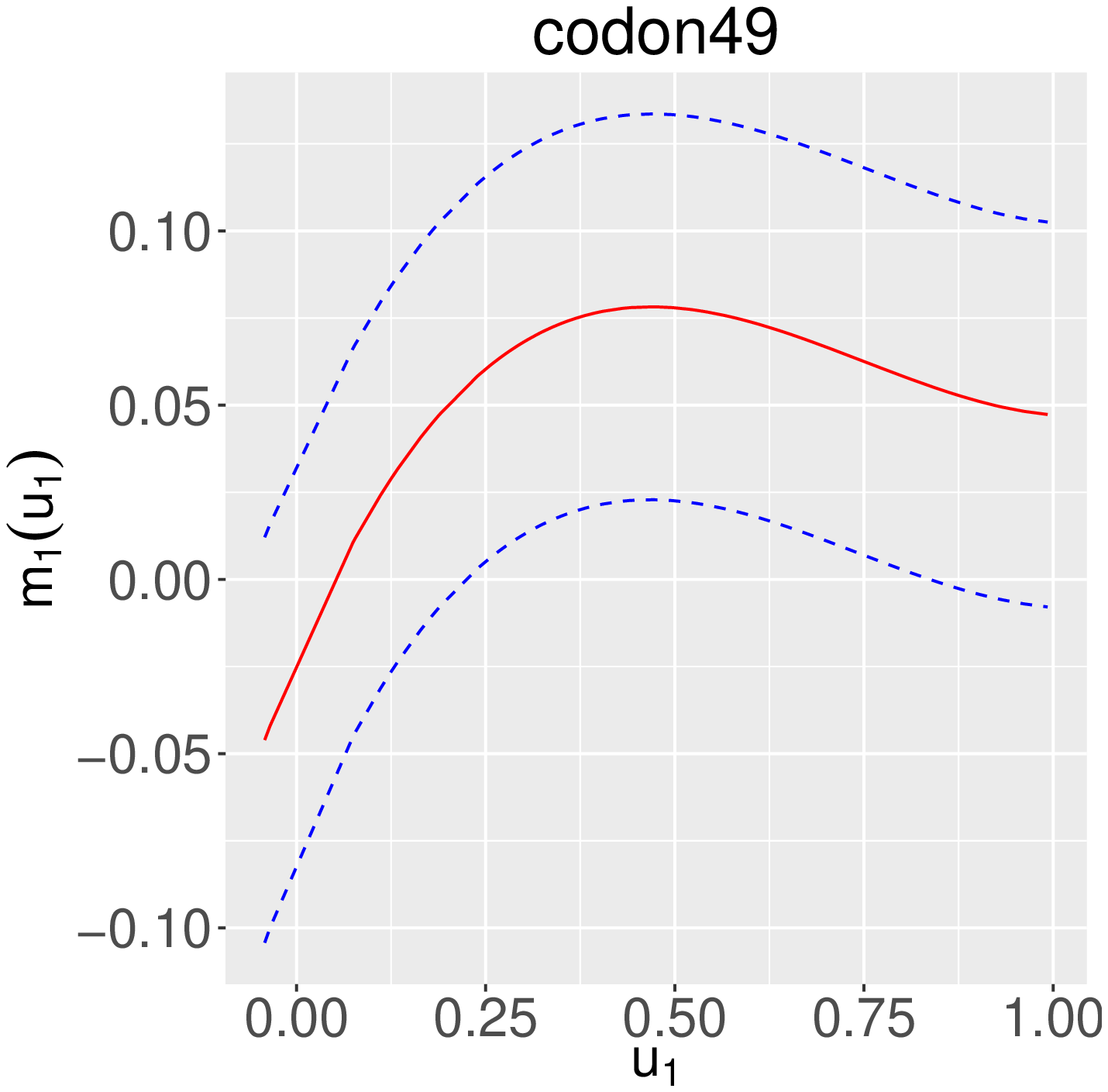}
   \end{minipage}
 }

 \subfigure{
   \begin{minipage}[t]{0.32\textwidth}
         \includegraphics[width=5.5cm, height=5cm]{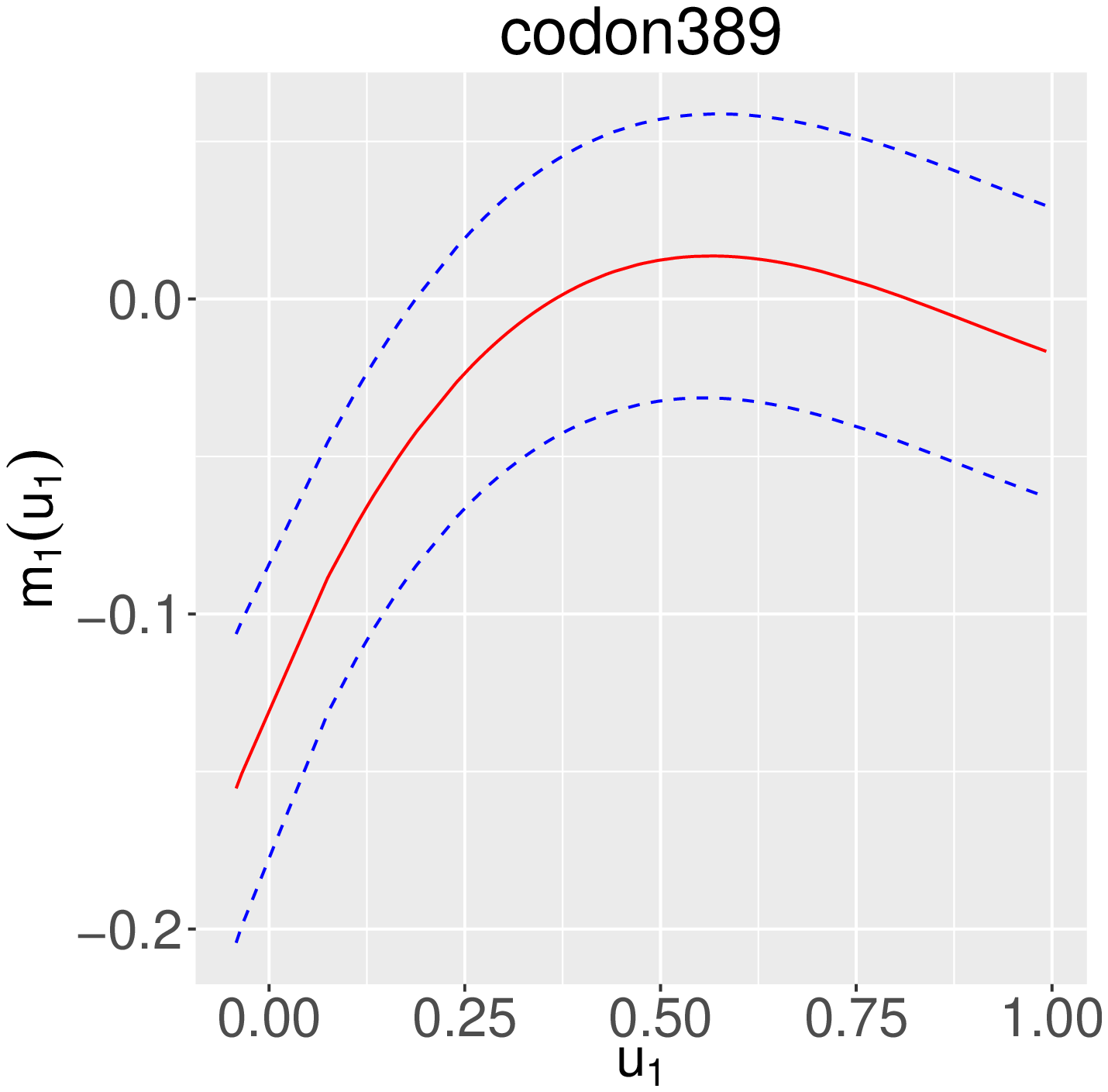}
   \end{minipage}
   \begin{minipage}[t]{0.32\textwidth}
         \includegraphics[width=5.4cm, height=5cm]{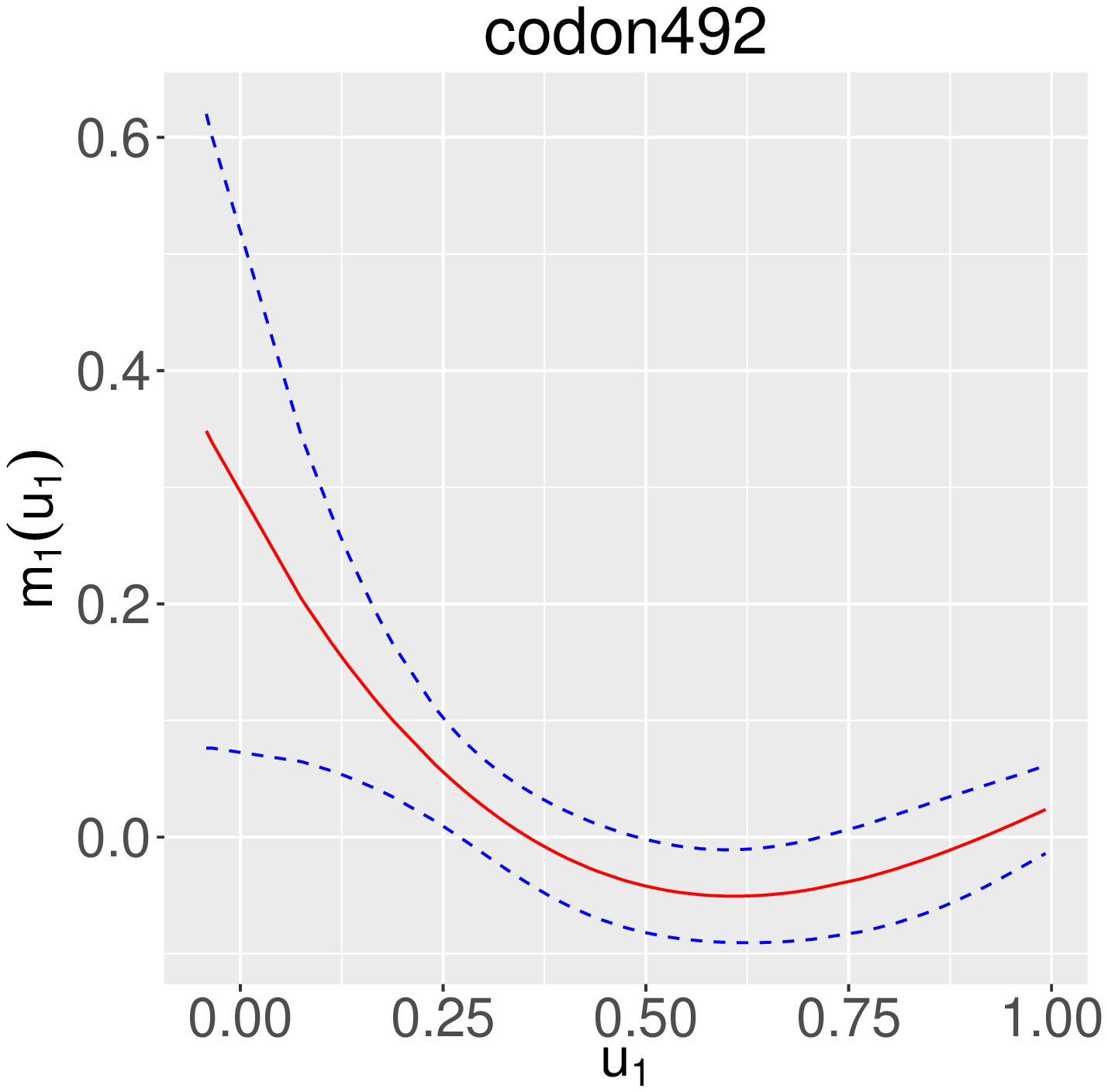}
   \end{minipage}
 }
 \caption{Plot of the estimate (solid curve) of the nonparametric function $m_{1}(u_{1})$ for SNPs codon16, codon27, codon49, codon389 and codon492. The 95\% confidence band is denoted by the dashed line. The response is SBP.}
 \label{SBPpic}
 \end{figure}

Table \ref{SBP} summarizes the performance of our method for response SBP. In the table, $p_{m_{1}}$ for all the 5 SNPs are smaller than the significance level 0.05, which implies the nonlinear function of the SNPs on SBP in response to the dosage level, age and BMI as a whole. The MSEs in the last two columns shows that FVICM fits the data better than AVCM, indicating the benefit of integrative analysis. Besides, the testing results for AVCM do not show significance of the coefficients, which further implies that the genetic effects of SNPs are nonlinearly modified by the mixture of these three variables. Figure \ref{SBPpic} shows the fitted nonlinear functions for each SNP, along with the 95\% confidence bands.

\begin{table}[H]
\centering
 \caption{List of SNPs with MAF, alleles, p-values under different hypotheses and MSE for SBP.}{
  {\small \setlength{\tabcolsep}{1.0mm}
 \begin{tabular}{ccccccccccccccc}
 \hline
  &&&  \multicolumn{6}{c}{p-value} &&  \multicolumn{2}{c}{MSE}  \\
   \cline{5-9}\cline{11-12}
    SNP ID& MAF & Alleles && $p_{m_{1}}$ & $p_{\beta_{11}}$ & $p_{\beta_{12}}$ & $p_{\beta_{13}}$&$p_{AVCM}$ && FVICM & AVCM
          \\
 \hline
 codon16 &    0.3990 & A/G & &   $<$1.0E-04&    0.0011&   $<$1.0E-04 &   0.0917&    0.5308& &   0.0403&    0.0421   \\
 codon27 &    0.4160 &  G/C& &   $<$1.0E-04&    $<$1.0E-04&     0.0027 &   0.1675&    0.6748& &   0.0388&     0.0415  \\
 codon49 &    0.1387 &  G/A& &   $<$1.0E-04&    $<$1.0E-04&    0.3614 &    0.8668&    0.2910& &   0.0398&     0.0410  \\
 codon389 &    0.3045 &  G/C& &  $<$1.0E-04&   $<$1.0E-04&     $<$1.0E-04 &    0.7552&    0.3927& &   0.0397&    0.0431   \\
 codon492 &    0.4250 &  T/C& &  $<$1.0E-04&    0.4102&    $<$1.0E-04 &     0.0182&    0.2990& &   0.0392&     0.0409 \\

\hline
 \end{tabular}}}\label{tab-4.1}
 \label{SBP}
 \end{table}

\begin{figure}[H]\centering
 \centering
 \subfigure{
   \begin{minipage}[t]{0.32\textwidth}
         \includegraphics[width=5.5cm, height=5cm]{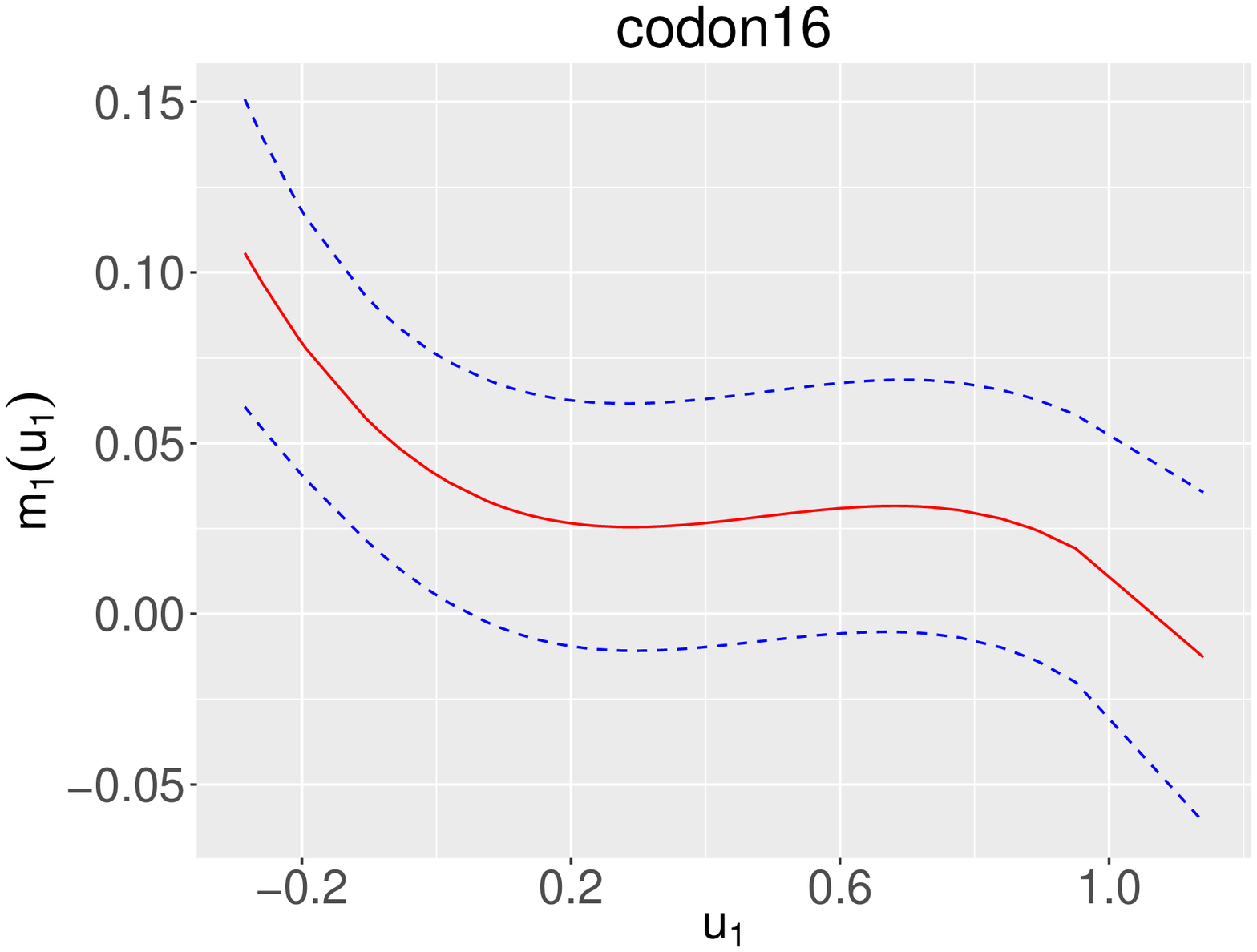}
   \end{minipage}
   \begin{minipage}[t]{0.32\textwidth}
         \includegraphics[width=5.5cm, height=5cm]{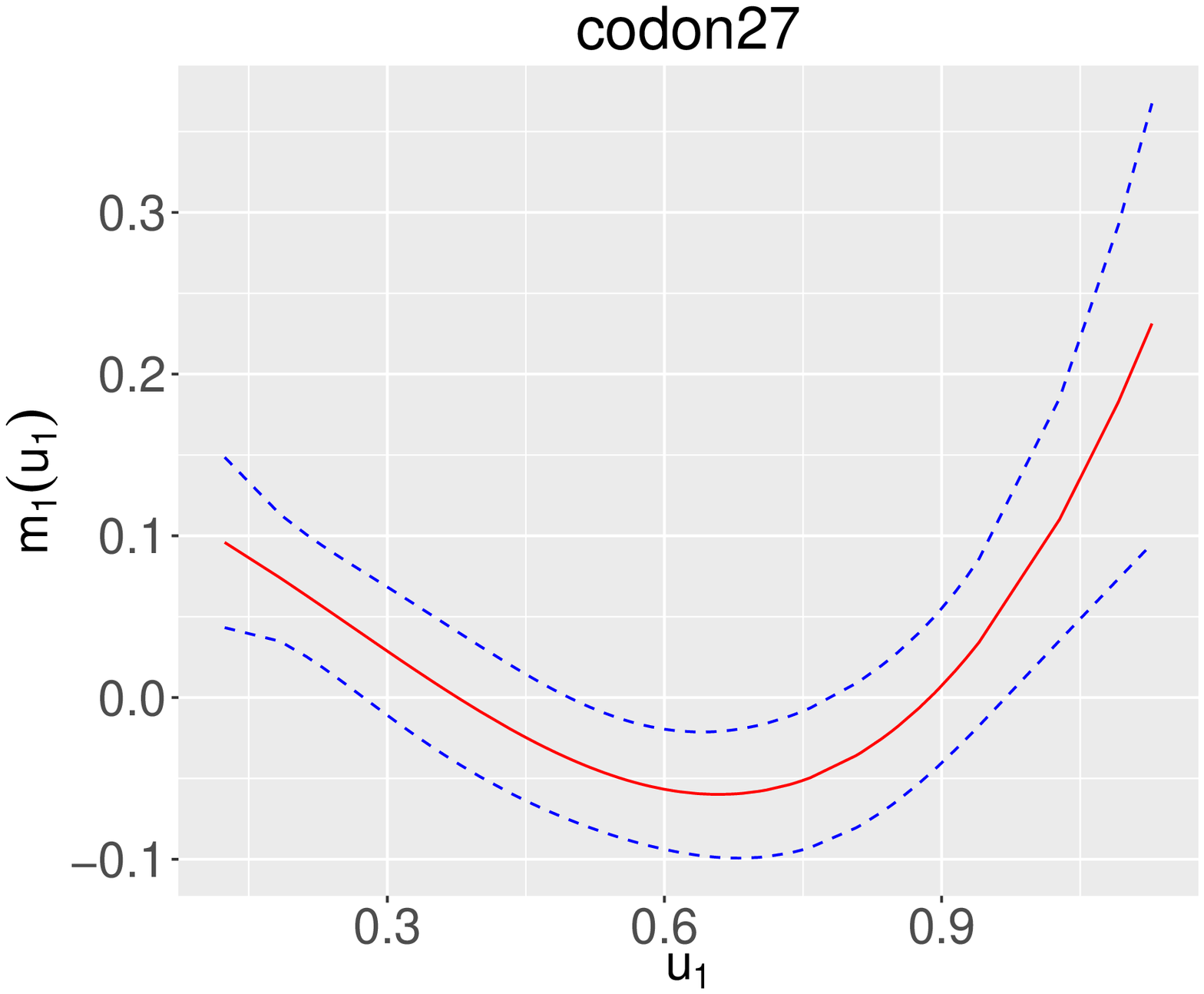}
   \end{minipage}
   \begin{minipage}[t]{0.32\textwidth}
         \includegraphics[width=5.5cm, height=5cm]{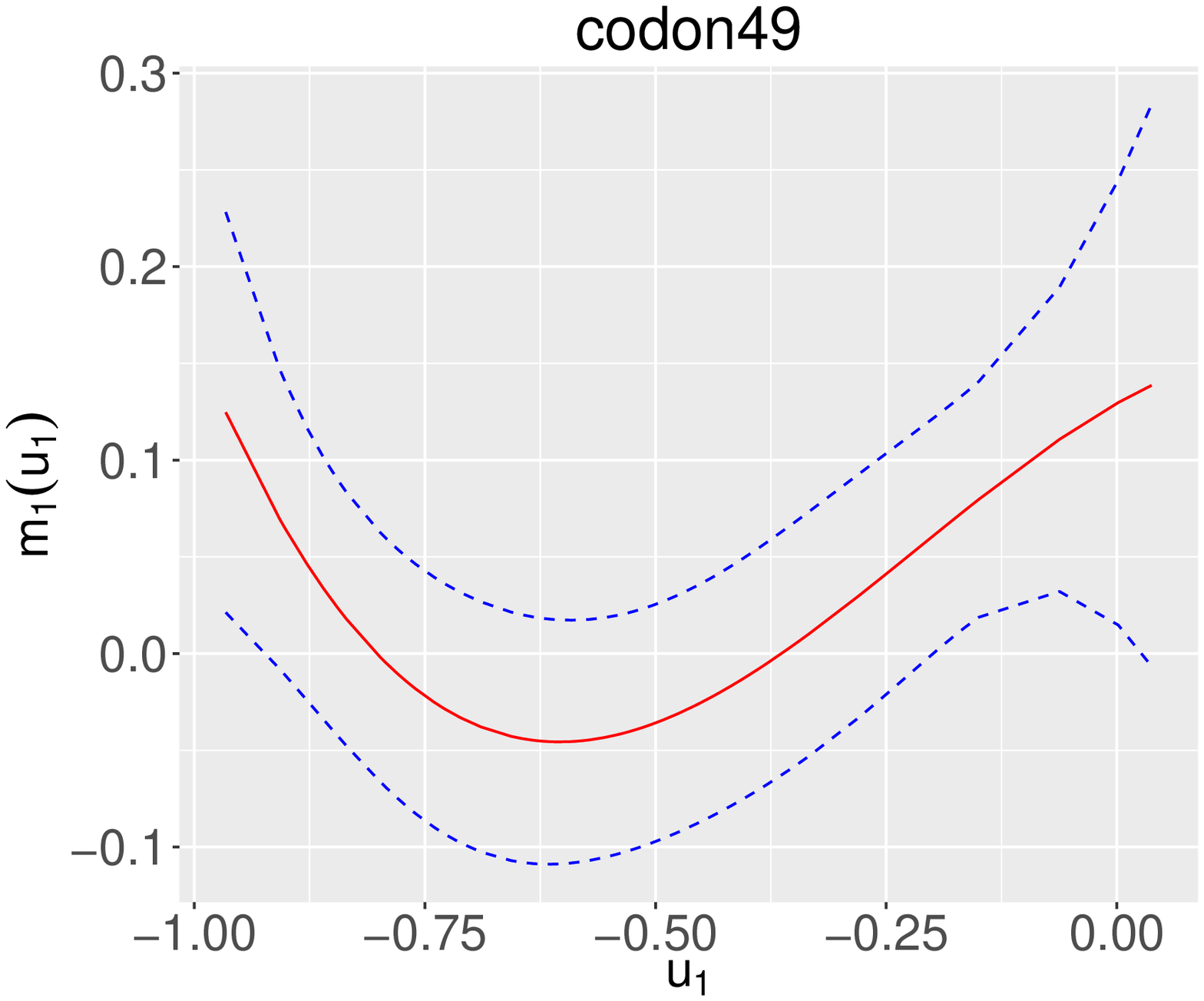}
   \end{minipage}
 }
 \subfigure{
   \begin{minipage}[t]{0.32\textwidth}
         \includegraphics[width=5.5cm, height=5cm]{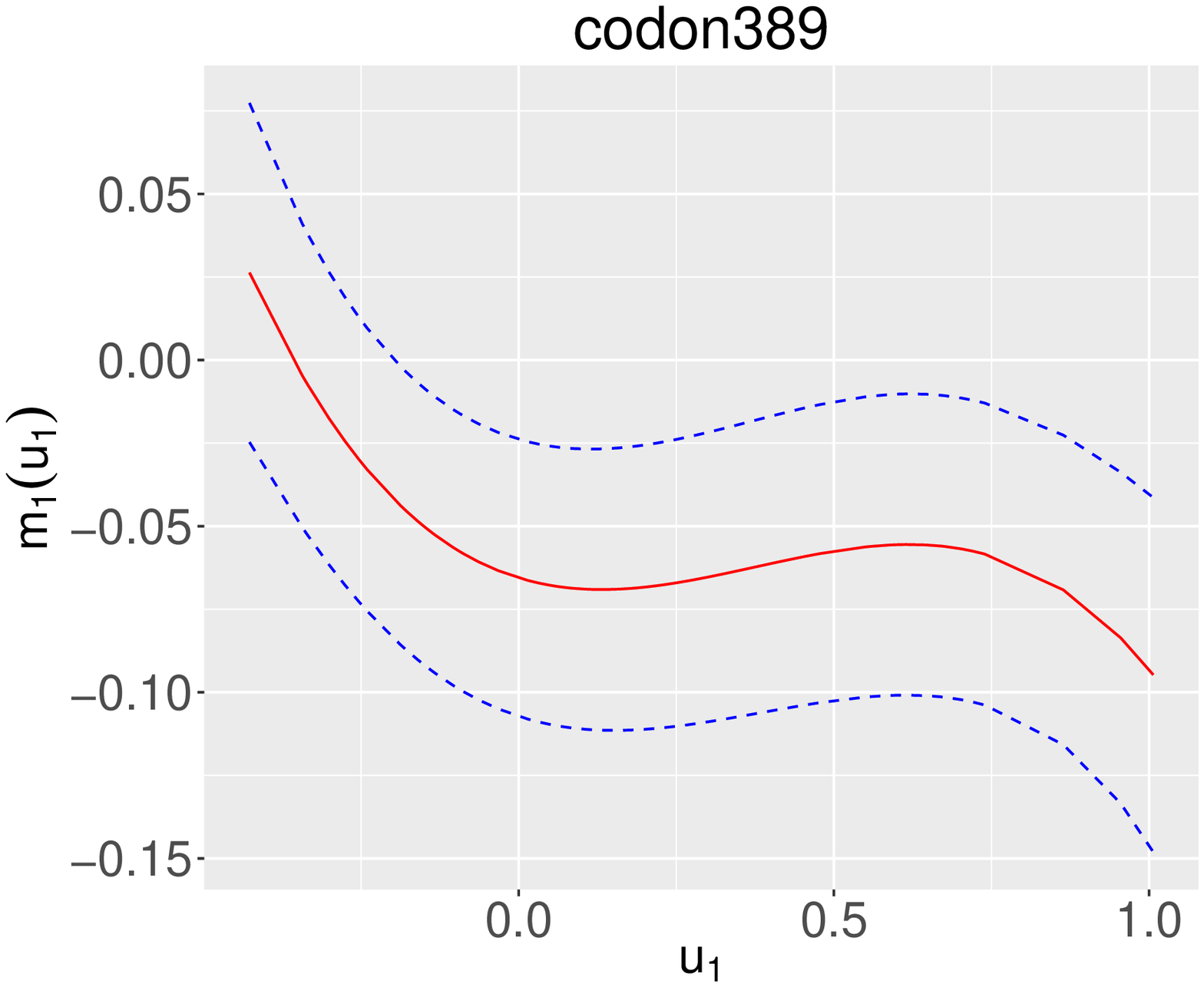}
   \end{minipage}
   \begin{minipage}[t]{0.32\textwidth}
         \includegraphics[width=5.4cm, height=5cm]{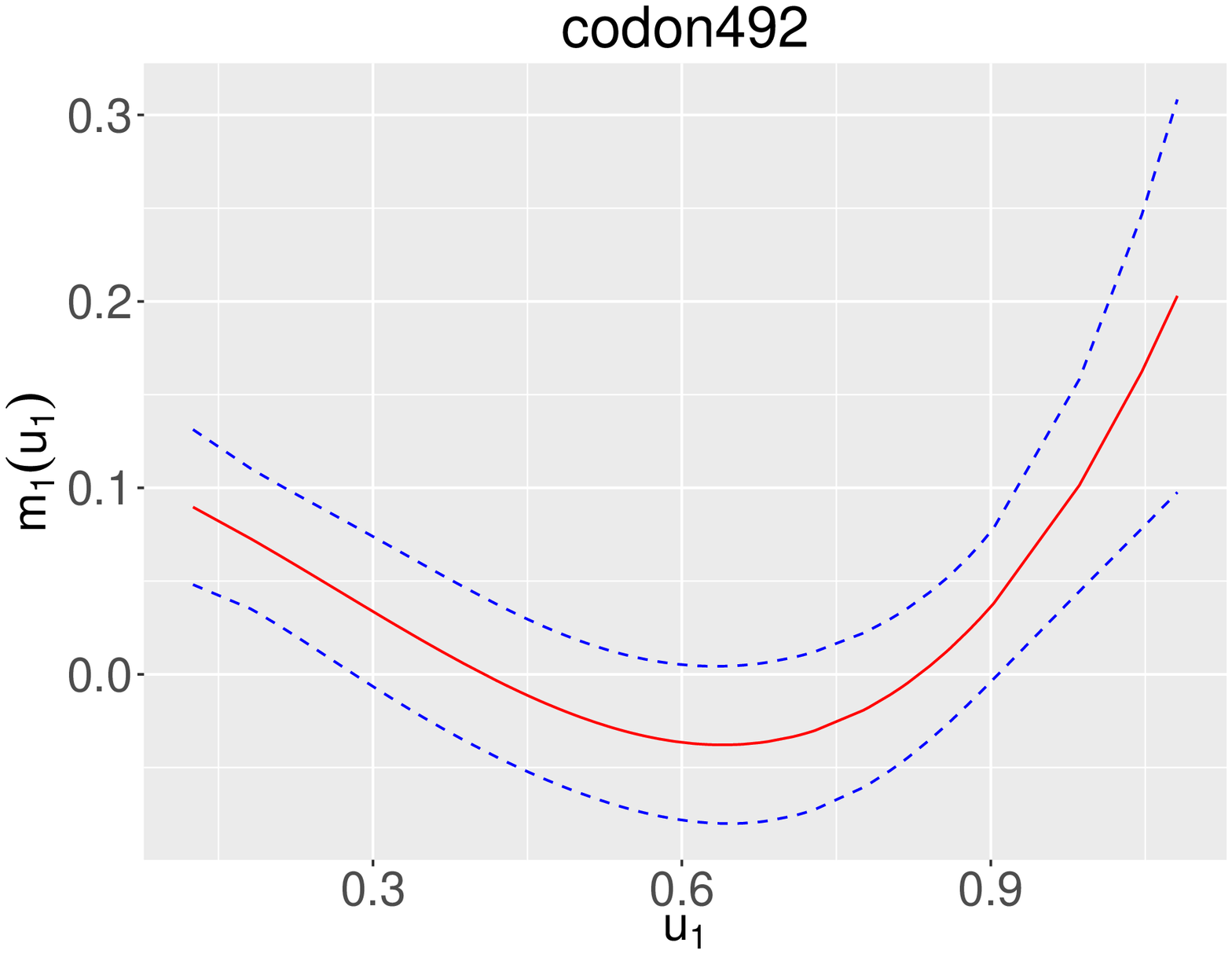}
   \end{minipage}

 }
 \caption{Plot of the estimate (solid curve) of the nonparametric function $m_{1}(u_{1})$ for SNPs codon16, codon27, codon49, codon389 and codon492. The 95\% confidence band is denoted by the dashed line. The response is DBP.}
 \label{DBPpic}
 \end{figure}

Table \ref{DBP} presents similar results for response DBP. The values of $p_{m_{1}}$ shows that the test results for all 5 SNPs are significant, indicating nonlinear interactions for all 5 SNPs, while no significance is shown for AVCM model. MSEs further support our method by showing smaller value for FVICM comparing with AVCM. The estimated interaction curves with 95\% confidence bands are shown in Figure \ref{DBPpic}.

\begin{table}[H]
\centering
 \caption{List of SNPs with MAF, alleles, p-values under different hypotheses and MSE for DBP.}{
  {\small \setlength{\tabcolsep}{1.0mm}
 \begin{tabular}{ccccccccccccccc}
 \hline
  &&&  \multicolumn{6}{c}{p-value} &&  \multicolumn{2}{c}{MSE}  \\
   \cline{5-9}\cline{11-12}
    SNP ID& MAF & Alleles && $p_{m_{1}}$ & $p_{\beta_{11}}$ & $p_{\beta_{12}}$ & $p_{\beta_{13}}$&$p_{AVCM}$ && FVICM & AVCM
          \\
 \hline
 codon16 &    0.3990 & A/G & &   0.0066&    $<$1.0E-04&   0.2834 &   0.0007&    0.3160& &   0.0366&    0.0372   \\
 codon27 &    0.4160 &  G/C& &   0.0004&    0.8431&     $<$1.0E-04 &   $<$1.0E-04&    0.0946& &   0.0360&     0.0386  \\
 codon49 &    0.1387 &  G/A& &   0.0003&    0.5750&    $<$1.0E-04 &    0.0042&    0.7986& &   0.0369&     0.0395 \\
 codon389 &    0.3045 &  G/C& &  0.0001&   $<$1.0E-04&     0.9675 &    $<$1.0E-04&    0.2615& &   0.0369&    0.0377   \\
 codon492 &    0.4250 &  T/C& &   0.0001&    0.7934&    $<$1.0E-04 &     $<$1.0E-04&    0.5837& &   0.0369&     0.0389 \\
\hline
 \end{tabular}}}\label{tab-4.1}
 \label{DBP}
 \end{table}

\begin{figure}[H]\centering
 \centering
 \subfigure{
   \begin{minipage}[t]{0.32\textwidth}
         \includegraphics[width=5.5cm, height=5cm]{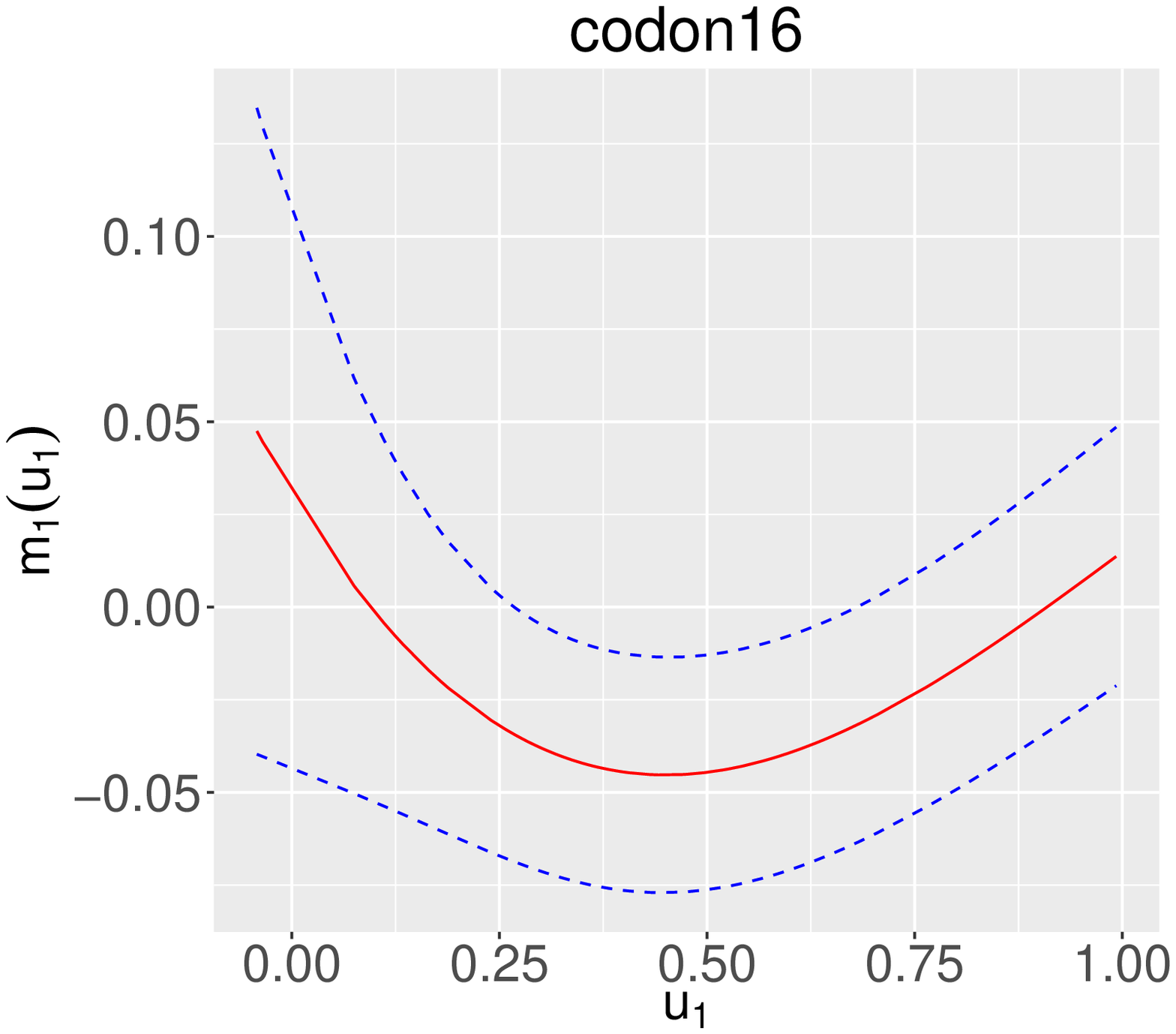}
   \end{minipage}
   \begin{minipage}[t]{0.32\textwidth}
         \includegraphics[width=5.5cm, height=5cm]{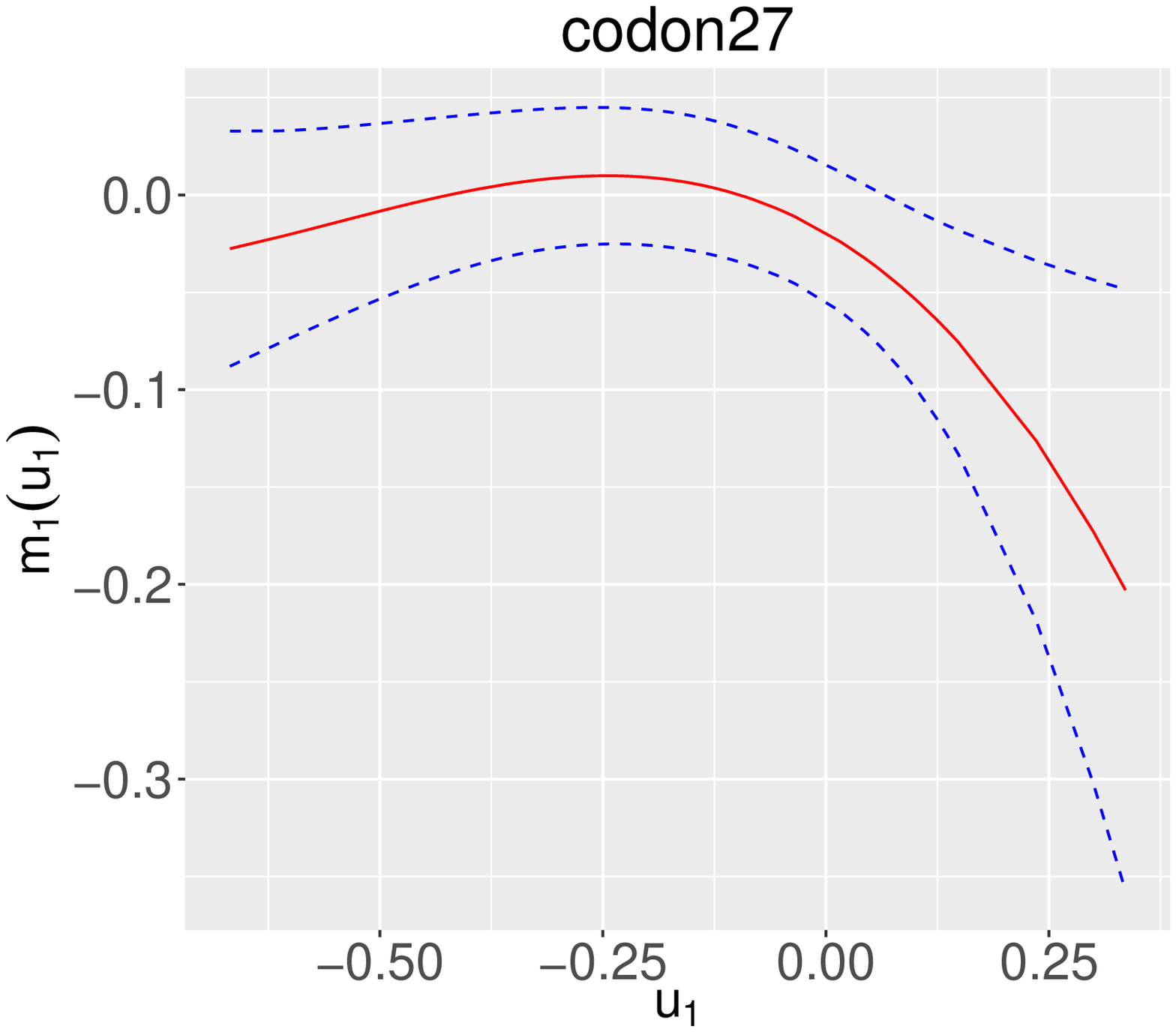}
   \end{minipage}
   \begin{minipage}[t]{0.32\textwidth}
         \includegraphics[width=5.5cm, height=5cm]{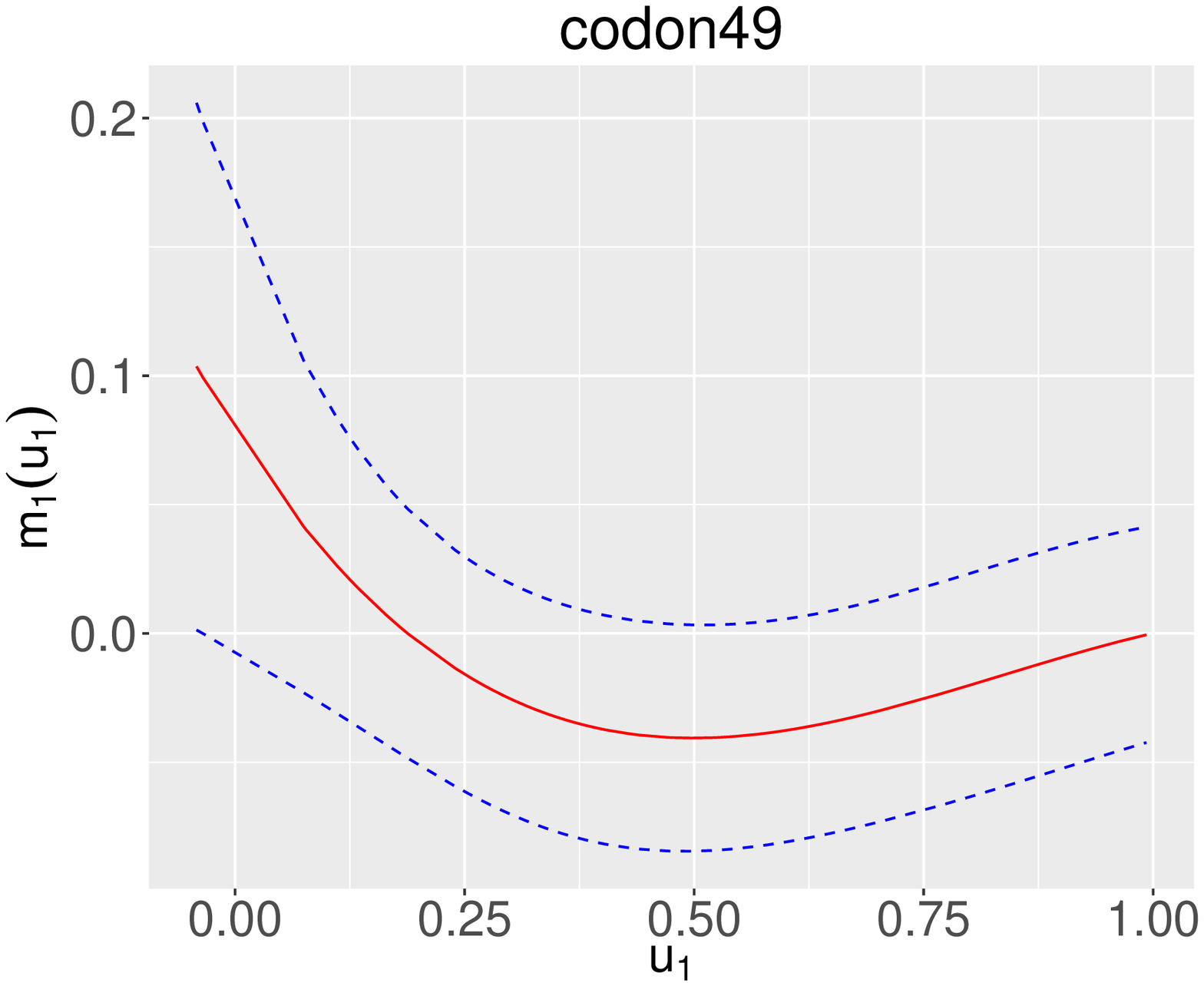}
   \end{minipage}
 }
 \subfigure{
   \begin{minipage}[t]{0.32\textwidth}
         \includegraphics[width=5.5cm, height=5cm]{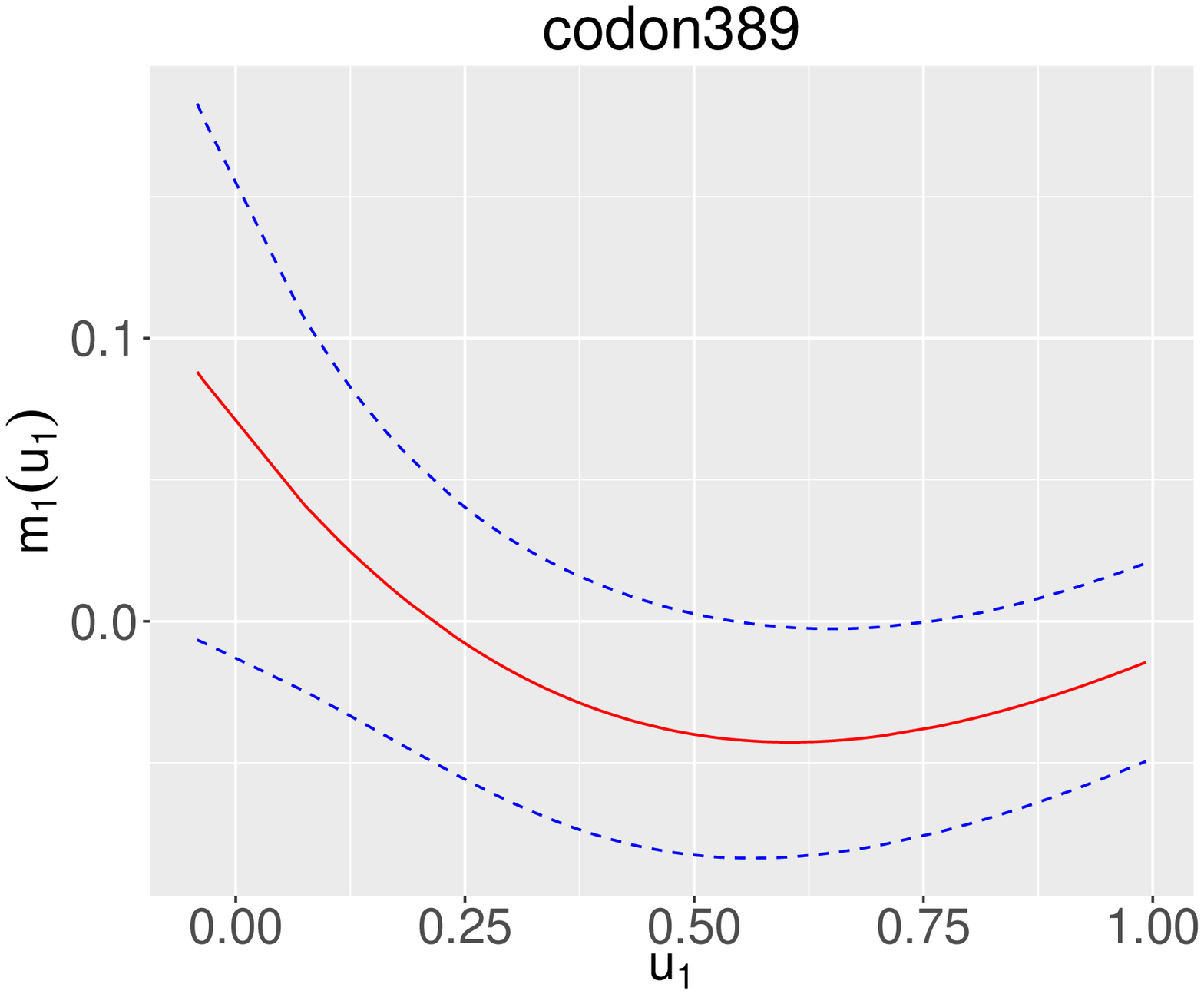}
   \end{minipage}
   \begin{minipage}[t]{0.32\textwidth}
         \includegraphics[width=5.5cm, height=5cm]{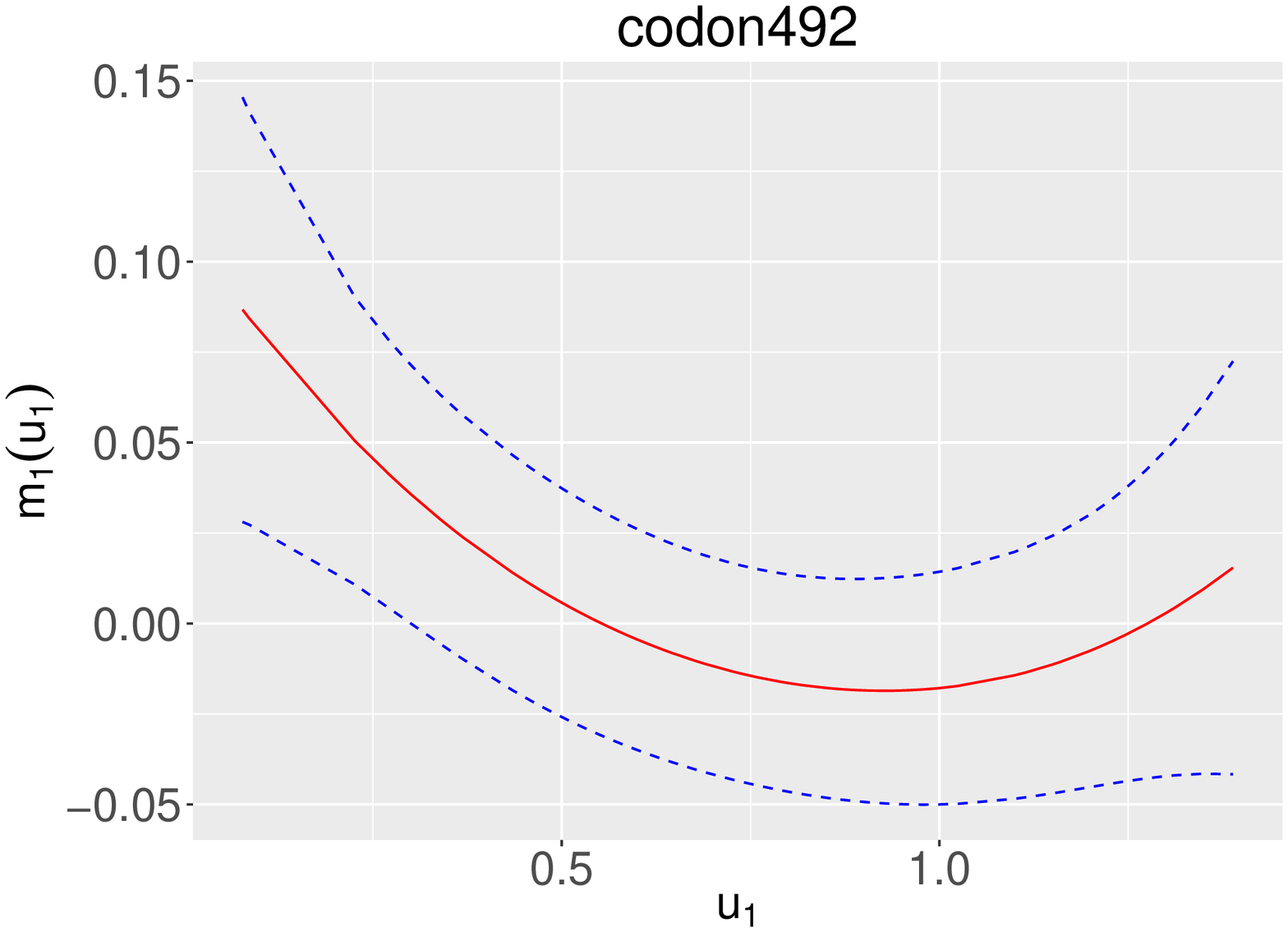}
   \end{minipage}
 }
 \caption{Plot of the estimate (solid curve) of the nonparametric function $m_{1}(u_{1})$ for SNPs codon16, codon27, codon49, codon389 and codon492. The 95\% confidence band is denoted by the dashed line. The response is HR.}
 \label{HRpic}
 \end{figure}

In Table \ref{HR}, the performance of our method for trait HR also leads to similar conclusion except for SNP \textit{codon16}, which shows (marginal) significant test results for both models. For all the other SNPs, FVICM outperforms AVCM in terms of MSE. Figure \ref{HRpic} displays the corresponding estimated nonlinear interaction curves.

\begin{table}[H]
\centering
 \caption{List of SNPs with MAF, alleles, p-values under different hypotheses and MSE for HR.}{
  {\small \setlength{\tabcolsep}{1.0mm}
 \begin{tabular}{ccccccccccccccc}
 \hline
  &&&  \multicolumn{6}{c}{p-value} &&  \multicolumn{2}{c}{MSE}  \\
   \cline{5-9}\cline{11-12}
    SNP ID& MAF & Alleles && $p_{m_{1}}$ & $p_{\beta_{11}}$ & $p_{\beta_{12}}$ & $p_{\beta_{13}}$&$p_{AVCM}$ && FVICM & AVCM
          \\
 \hline
 codon16 &    0.3990 & A/G & &   $<$1.0E-04&    $<$1.0E-04&   0.1158 &   0.0028&    0.0328& &   0.0309&    0.0308   \\
 codon27 &    0.4160 &  G/C& &   $<$1.0E-04&    0.0007&     0.6434 &   0.0001&    0.9620& &   0.0320&     0.0325  \\
 codon49 &    0.1387 &  G/A& &   0.0001&    0.0147&    0.0172 &    0.0133&    0.8371& &   0.0298&     0.0300 \\
 codon389 &    0.3045 &  G/C& &  $<$1.0E-04&   $<$1.0E-04&     0.0024 &    0.0021&    0.8959& &   0.0311&    0.0313   \\
 codon492 &    0.4250 &  T/C& &   0.0002&    $<$1.0E-04&    0.0011 &     0.0582&    0.3732& &   0.0315&     0.0316 \\
\hline
 \end{tabular}}}\label{tab-4.1}
 \label{HR}
 \end{table}

\section{Discussion}\label{discussion}
In this paper, we proposed a functional varying index coefficient model to study gene effects nonlinearly modified by a mixture of environmental variables in a longitudinal design. We implemented the quadratic inference function (QIF) method to estimate the index loading parameters and the spline coefficients. Furthermore, we applied the pseudo likelihood ratio test in a linear mixed model representation to test the linearity of the nonparametric coefficient function. Simulation studies were conducted to illustrate the estimation and testing procedures and confirmed the asymptotical property. Real analysis showed that our model outperforms the additive varying coefficient model, which considers the G$\times$E effect for each single environmental factor separately.

Our FVICM model is different from the varying coefficient model for longitudinal data. In fact, the varying coefficient model is a special case of our model when the dimension of the $X$ variable reduces to 1. FVICM is able to capture the effect of genes nonlinearly modified by the joint effect of multiple environmental variables as a whole. In addition, it can reduce multiple testing burden by treating multiple environmental variables as a single index variable. The advantage of modeling multiple variables as a single index and further assessing its effect via a nonparametric function has also been demonstrated by Ma and Song (2015) in a cross-sectional design. Our real data analysis results further confirmed the advantage in a longitudinal design.

We appled the model to a pain sensitivity study. Testing results indicated that all five SNPs have significant nonlinear interaction effects with environmental factors, which makes practical sense since these SNPs were genotyped from candidate genes. Our model was motivated by a practical need in G$\times$E study and offer additional insights that otherwise cannot be revealed by models with cross-sectional data. By checking the nonlinear effect function together with the confidence band, people can get a sense how genes respond to the combined change of the environmental factors over time to affect a response variable.

Our method can be applied to any longitudinal data in which the purpose is to model nonlinear interaction effects. For example, we can consider gene expressions in a pathway (denoted as $\bm{X}$) and model how they regulate downstream genes ($G$) to affect a disease trait. Both the trait and gene expressions can be measured over time. Thus, one can understand the dynamic effect of genes nonlinearly regulated by a pathway to affect a disease trait.

\section*{Acknowledgements}
This work was supported in part by the National Human Genome Research Institute of the National Institutes of Health (NIH) under award number R21HG010073 (to Y. Cui) and by the National Science Foundation under award number DMS-2212928 (to H. Wang). The content is solely the responsibility of the authors and does not necessarily represent the official views of NIH and NSF.

\section*{Appendix: Proofs}\label{appendix}
To establish the asymptotic properties for the estimator of $\bm{\theta}$, we need the following regularity conditions.
\begin{description}
  \item[(A1)] $\{n_{i}\}$ is a bounded sequence of integers.
  \item[(A2)]  The parameter space $\Omega$ is compact and $\bm{\theta}_{0}^{*}$ is an interior point of $\Omega$.
  \item[(A3)] The parameter $\bm{\theta}^{*}$ is identifiable, that is, there is a unique $\bm{\theta}_{0}^{*}\in \Omega$ satisfying the mean zero model assumption, i.e., $E[g(\bm{\theta}_{0}^{*})]=0$.
  \item[(A4)] $E[g(\bm{\theta})]$ is continuous in $\bm{\theta}$.
  \item[(A5)] $\bar{C}_{N}(\widehat{\bm{\theta}}^{*})=\frac{1}{N}\sum_{i=1}^{N}g_{i}(\widehat{\bm{\theta}}^{*})g_{i}(\widehat{\bm{\theta}}^{*})^{T}$ converges almost surely to $\textbf{C}_{0}$, which is a constant and invertible matrix.
  \item[(A6)] The first derivative of $\bar{g}_{N}$ exists and is continuous. $\frac{\partial \bar{g}_{N}}{\partial\bm{\theta}^{*}}(\widehat{\bm{\theta}}^{*})$ converges in probability to $\textbf{G}_{0}$ if $\widehat{\bm{\theta}}^{*}$ converges in probability to $\bm{\theta}_{0}^{*}$.
\end{description}

\textit{Proof of Theorem 1:}
If we can prove that $\widehat{\bm{\theta}}^{*}$ exists and converges to $\bm{\theta}^{*}_{0}$ almost surely, then we can prove the consistency of $\bm{\theta}$
directly.
$\widehat{\bm{\theta}}^{*}=\arg\min(N^{-1}Q_{N}(\bm{\theta}^{*})+ \lambda \bm{\theta}^{*T}\textbf{D}\bm{\theta}^{*})$ exists because (\ref{target}) has zero as a lower bound and the global minimum exists. To prove the consistency, first, the estimator $\widehat{\bm{\theta}}^{*}$ is obtained by minimizing $N^{-1}Q_{N}(\bm{\theta}^{*})+ \lambda \bm{\theta}^{*T}\textbf{D}\bm{\theta}^{*}$, then we have
\begin{equation}\label{unequal}
  \frac{1}{N}Q_{N}(\widehat{\bm{\theta}}^{*})+\lambda_{N}\widehat{\bm{\theta}}^{*T}D\widehat{\bm{\theta}}^{*}\leq  \frac{1}{N}Q_{N}(\bm{\theta}_{0}^{*})+\lambda_{N}\bm{\theta}_{0}^{*T}D\bm{\theta}_{0}^{*}.
\end{equation}
Since
\begin{equation*}
   \frac{1}{N}Q_{N}(\bm{\theta}_{0}^{*})= \bar{g}_{N}^{T}(\bm{\theta}_{0}^{*})\bar{C}_{N}^{-1}(\bm{\theta}_{0}^{*})\bar{g}_{N}(\bm{\theta}_{0}^{*})=o(1)
\end{equation*}
by the strong law of large number and (A5), and $\lambda_{N}=o(1)$,
\begin{equation*}
  \frac{1}{N}Q_{N}(\bm{\theta}_{0}^{*})+\lambda_{N}\bm{\theta}_{0}^{*T}D\bm{\theta}_{0}^{*}\xrightarrow{a.s.}0.
\end{equation*}
Thus, we can obtain from (\ref{unequal}) that
\begin{equation}\label{abc}
  \frac{1}{N}Q_{N}(\widehat{\bm{\theta}}^{*})=\bar{g}_{N}^{T}(\widehat{\bm{\theta}}^{*})\bar{C}_{N}^{-1}(\widehat{\bm{\theta}}^{*})\bar{g}_{N}(\widehat{\bm{\theta}}^{*})\xrightarrow{a.s.}0.
\end{equation}
Since the parameter space $\Omega$ is compact, by Glivenko-Cantelli theorem,
\begin{equation*}
  \sup_{\theta^{*}\in\Omega}\Big|\bar{g}_{N}(\bm{\theta}^{*})-E[g(\bm{\theta}^{*})] \Big|\xrightarrow{a.s.}0.
\end{equation*}
Hence, by (A5) and the continuous mapping theorem,
\begin{equation*}
  \left|\bar{g}_{N}^{T}(\widehat{\bm{\theta}}^{*})\bar{C}_{N}^{-1}(\widehat{\bm{\theta}}^{*})\bar{g}_{N}(\widehat{\bm{\theta}}^{*})-E[g(\widehat{\bm{\theta}}^{*})]^{T}\textbf{C}_{0}^{-1}E[g(\widehat{\bm{\theta}}^{*})]\right|\xrightarrow{a.s.}0.
\end{equation*}
Combined with (\ref{abc}), we get
\begin{equation}\label{abcd}
  E[g(\widehat{\bm{\theta}}^{*})]^{T}\textbf{C}_{0}^{-1}E[g(\widehat{\bm{\theta}}^{*})]\xrightarrow{a.s.}0.
\end{equation}
Next, we show that it is impossible that $\widehat{\bm{\theta}}^{*}$ remains outside of $U$, where $U$ is any neighborhood of the true parameter $\bm{\theta}_{0}^{*}$. Suppose there exists a neighborhood $U$ such that $\widehat{\bm{\theta}}^{*}\in U^{c}$. Since $E[g(\bm{\theta}^{*})]^{T}\textbf{C}_{0}^{-1}E[g(\bm{\theta}^{*})]$ is a continuous function and $U^{c}$ is compact, there exists a point $\widetilde{\bm{\theta}}^{*}\in U^{c}$ such that
  $E[g(\widetilde{\bm{\theta}}^{*})]^{T}\textbf{C}_{0}^{-1}E[g(\widetilde{\bm{\theta}}^{*})]$
achieves its minimum in $U^{c}$. By the identifiability of $\bm{\theta}^{*}$ in (A3), there is a unique $\bm{\theta}_{0}^{*}\in \Omega$ satisfying $E[g(\bm{\theta}_{0})]=0$, and we have
\begin{equation*}
  E[g(\bm{\theta}^{*})]^{T}\textbf{C}_{0}^{-1}E[g(\bm{\theta}^{*})]>0,
\end{equation*}
which contradicts ($\ref{abcd}$). Then we can prove that $\widehat{\bm{\theta}}^{*}$ converges almost surely to $\bm{\theta}^{*}$. Thus, $\widehat{\bm{\theta}}$ is a consistent estimator of $\bm{\theta}$.

\textit{Proof of Theorem 2:}
The estimate of $\bm{\theta}$ satisfies
\begin{equation*}
  0=\frac{1}{N}\frac{\partial Q_{N}}{\partial\bm{\theta}^{*}}(\widehat{\bm{\theta}}^{*})+2\lambda_{N}D\widehat{\bm{\theta}}^{*}.
\end{equation*}
By Taylor expansion, we obtain
\begin{equation*}
  0=\frac{1}{N}\frac{\partial Q_{N}}{\partial\bm{\theta}}^{*}(\bm{\theta}_{0}^{*})+2\lambda_{N}D\bm{\theta}_{0}^{*}+\Big(\frac{1}{N}\frac{\partial^{2} Q_{N}}{\partial\bm{\theta}^{*2}}(\widetilde{\bm{\theta}}^{*})+2\lambda_{N}D\Big)(\widehat{\bm{\theta}}^{*}-\bm{\theta}_{0}^{*}),
\end{equation*}
where $\widetilde{\bm{\theta}}^{*}$ is some value between $\widehat{\bm{\theta}}^{*}$ and $\bm{\theta}_{0}^{*}$.
Thus, we have
\begin{equation}\label{gammadiff}
  \widehat{\bm{\theta}}^{*}-\bm{\theta}_{0}^{*}=-\Big(\frac{1}{N}\frac{\partial^{2} Q_{N}}{\partial\bm{\theta}^{*2}}(\widetilde{\bm{\theta}}^{*})+2\lambda_{N}D\Big)^{-1}\Big(\frac{1}{N}\frac{\partial Q_{N}}{\partial\bm{\theta}^{*}}(\bm{\theta}_{0}^{*})+2\lambda_{N}D\bm{\theta}_{0}^{*}\Big).
\end{equation}
Since $\widehat{\bm{\theta}}^{*}$ converges to $\bm{\theta}_{0}^{*}$ in probability and $\widetilde{\bm{\theta}}^{*}$ is between $\widehat{\bm{\theta}}^{*}$ and $\bm{\theta}_{0}^{*}$, by (A5) and (A6) we can get
\begin{equation*}
  \frac{1}{N}\frac{\partial^{2} Q_{N}}{\partial\bm{\theta}^{*2}}(\widetilde{\bm{\theta}}^{*})=2\frac{\partial \bar{g}_{N}}{\partial\bm{\theta}^{*}}^{T}(\widetilde{\bm{\theta}}^{*})\bar{C}_{N}^{-1}(\widetilde{\bm{\theta}}^{*})\frac{\partial \bar{g}_{N}}{\partial\bm{\theta}^{*}}(\widetilde{\bm{\theta}}^{*})+o_{p}(1)\xrightarrow{~p~}2\textbf{G}_{0}^{T}\textbf{C}_{0}^{-1}\textbf{G}_{0}
\end{equation*}
When $\lambda_{N}=o(N^{-1/2})$,
\begin{equation*}
  \Big(\frac{1}{N}\frac{\partial^{2} Q_{N}}{\partial\bm{\theta}^{*2}}(\widetilde{\bm{\theta}}^{*})+2\lambda_{N}D\Big)^{-1}=\frac{1}{2}(\textbf{G}_{0}^{T}\textbf{C}_{0}^{-1}\textbf{G}_{0})^{-1}+o_{p}(N^{-1/2}).
\end{equation*}
Similarly, since
\begin{equation*}
  \frac{1}{N}\frac{\partial Q_{N}}{\partial\bm{\theta}^{*}}(\bm{\theta}_{0}^{*})=\frac{\partial \bar{g}_{N}}{\partial\bm{\theta}^{*}}^{T}(\bm{\theta}_{0}^{*})\bar{C}_{N}^{-1}(\bm{\theta}_{0}^{*})\bar{g}_{N}(\bm{\theta}_{0}^{*})
\end{equation*}
and $\lambda_{N}=o(N^{-1/2})$, we have
\begin{equation*}
  \frac{1}{N}\frac{\partial Q_{N}}{\partial\bm{\theta}^{*}}(\bm{\theta}_{0}^{*})+2\lambda_{N}D\bm{\theta}_{0}^{*}=G_{0}^{T}C_{0}^{-1}\bar{g}_{N}(\bm{\theta}_{0}^{*})+o(N^{-1/2}).
\end{equation*}
Therefore, (\ref{gammadiff}) can be written as
\begin{equation}\label{aa}
  \sqrt{N}(\widehat{\bm{\theta}}^{*}-\bm{\theta}_{0}^{*})=-\sqrt{N}(\textbf{G}_{0}^{T}\textbf{C}_{0}^{-1}\textbf{G}_{0})^{-1}\textbf{G}_{0}^{T}\textbf{C}_{0}^{-1}\bar{g}_{N}(\bm{\theta}_{0}^{*})+o_{p}(1).
\end{equation}
By Central Limit Theorem,
\begin{equation}\label{g}
  \sqrt{N}\bar{g}_{N}(\bm{\theta}_{0}^{*})\xrightarrow{d}N(\textbf{0},\textbf{C}_{0}).
\end{equation}
Using (\ref{aa}) and (\ref{g}), we obtain
\begin{equation*}
  \sqrt{N}(\widehat{\bm{\theta}}^{*}-\bm{\theta}_{0}^{*})\xrightarrow{d}N(\textbf{0},(\textbf{G}_{0}^{T}\textbf{C}_{0}^{-1}\textbf{G}_{0})^{-1}),
\end{equation*}
and directly,
\begin{equation*}
  \sqrt{N}(\widehat{\bm{\theta}}-\bm{\theta}_{0})\xrightarrow{d}N(\textbf{0},\textbf{J}(\textbf{G}_{0}^{T}\textbf{C}_{0}^{-1}\textbf{G}_{0})^{-1}\textbf{J}^{T}).
\end{equation*}

\end{document}